\documentclass[aps, prd, twocolumn, amsmath, floats,floatfix, superscriptaddress, nofootinbib]{revtex4}

\usepackage{natbib}
\usepackage{graphicx,color}
\usepackage{amsmath,amssymb}
\usepackage{verbatim}
\usepackage{float}
\usepackage{wasysym}
\usepackage{amssymb,graphicx,subfigure}
\usepackage{epsfig}
\usepackage{psfrag}
\usepackage{dsfont}
\usepackage{amsfonts}
\usepackage{mathrsfs}
\usepackage{multirow}
\usepackage{times}
\usepackage{bm}
\usepackage{hyperref}
\usepackage{xspace}
\usepackage[export]{adjustbox}
\usepackage{orcidlink}
\hypersetup{
  colorlinks=true,        
  linkcolor=blue,         
  citecolor=cyan,         
  }
\usepackage{pifont}
\usepackage[normalem]{ulem}

\begin{document}

\definecolor{orange}{rgb}{0.9,0.45,0}
\newcommand{\re}{\mbox{Re}}
\newcommand{\im}{\mbox{Im}}
\newcommand{\miq}[1]{\textcolor{orange}{#1}}
\newcommand{\red}[1]{{\color{red} #1}}
\newcommand{\pcd}[1]{{\color{teal} #1}}
\newcommand{\tf}[1]{\textcolor{blue}{#1}}
\newcommand{\dg}[1]{\textcolor{teal}{#1}}
\newcommand{\mr}[1]{{\color{magenta} #1}}
\def\CovDev{D}
\def\Res{{\mathcal R}}
\def\Gammaflat{\hat \Gamma}
\def\metricflat{\hat \gamma}
\def\Dflat{\hat {\mathcal D}}
\def\part_n{\partial_\perp}

\def\Lie{\mathcal{L}}
\def\A{\mathcal{X}}
\def\Aphi{\A_{\phi}}
\def\hAphi{\hat{\A}_{\phi}}
\def\E{\mathcal{E}}
\def\Ham{\mathcal{H}}
\def\M{\mathcal{M}}
\def\R{\mathcal{R}}
\def\p{\partial}

\def\hg{\hat{\gamma}}
\def\hA{\hat{A}}
\def\hD{\hat{D}}
\def\hE{\hat{E}}
\def\hR{\hat{R}}
\def\hcA{\hat{\mathcal{A}}}
\def\hDelt{\hat{\triangle}}

\def\na{\nabla}
\def\dif{{\rm{d}}}
\def\non{\nonumber}
\newcommand{\erf}{\textrm{erf}}
\renewcommand{\t}{\times}
\long\def\symbolfootnote[#1]#2{\begingroup%
\def\thefootnote{\fnsymbol{footnote}}\footnote[#1]{#2}\endgroup}


\title{Identifying thermal effects in neutron star merger remnants with model-agnostic waveform reconstructions and third-generation detectors}

\author{Miquel Miravet-Tenés \orcidlink{0000-0002-8766-1156}}\thanks{E-mail: m.miravet-tenes@soton.ac.uk}
\affiliation{Departament d'Astronomia i Astrofísica, Universitat de València, C/ Dr Moliner 50, 46100, Burjassot (València), Spain} 
\affiliation{Mathematical Sciences and STAG Research Centre, University of Southampton, Southampton SO17 1BJ, UK}
\author{Davide Guerra \orcidlink{0000-0003-0029-5390}}
\affiliation{Departament d'Astronomia i Astrofísica, Universitat de València, C/ Dr Moliner 50, 46100, Burjassot (València), Spain}
\author{Milton Ruiz
\orcidlink{0000-0002-7532-4144}}
\affiliation{Departament d'Astronomia i Astrofísica, Universitat de València, C/ Dr Moliner 50, 46100, Burjassot (València), Spain} 
\author{Pablo Cerdá-Durán
\orcidlink{0000-0003-4293-340X}}
\affiliation{Departament d'Astronomia i Astrofísica, Universitat de València, C/ Dr Moliner 50, 46100, Burjassot (València), Spain} 
\affiliation{Observatori Astronòmic, Universitat de València, C/ Catedrático José Beltrán 2, 46980, Paterna (València), Spain}
\author{José A. Font
\orcidlink{0000-0001-6650-2634}}
\affiliation{Departament d'Astronomia i Astrofísica, Universitat de València, C/ Dr Moliner 50, 46100, Burjassot (València), Spain} 
\affiliation{Observatori Astronòmic, Universitat de València, C/ Catedrático José Beltrán 2, 46980, Paterna (València), Spain}


\date{\today}

\begin{abstract}
We probe the intrinsic differences in simulated gravitational-wave signals from binary neutron star (BNS) mergers, arising from varying approaches to incorporating thermal effects in numerical-relativity modeling. We consider a {\it hybrid} approach in which the equation of state (EOS) comprises a cold, zero-temperature, piecewise-polytropic part and a thermal part described by an ideal gas, and a {\it tabulated} approach based on self-consistent, microphysical, finite-temperature EOS. We use time-domain waveforms corresponding to BNS merger simulations with four different EOSs. Those are injected into Gaussian noise given by the sensitivity of the third-generation detector Einstein Telescope and reconstructed using {\tt BayesWave}, a Bayesian data-analysis algorithm that recovers the signals through a model-agnostic approach.  The two representations of thermal effects result in different dominant peak frequencies in the spectra of the postmerger signals, for both the quadrupole fundamental mode and the late-time inertial modes. For some of the EOSs investigated those differences are large enough to be told apart, especially in the early postmerger phase when the signal amplitude is the loudest. Our results suggest that a self-consistent treatment of thermal effects in BNS postmerger modeling is essential to prevent significant
parameter biases in upcoming gravitational-wave detections.

\end{abstract}
\maketitle

\section{\label{sec:sec1}Introduction}
We are in a golden era of astrophysics where a plethora of new gravitational-wave (GW) observations is changing our
understanding of the Universe at an unprecedented rate~\cite{GWTC-1,GWTC-2,GWTC-2.1,GWTC-3}.  In particular, the observations of GWs from the first binary neutron  star (BNS) merger -- event GW170817 -- along with its postmerger emission of electromagnetic radiation,  spurred the era of multimessenger astronomy~\citep{LIGOScientific:2017pwl,GBM:2017lvd,LIGOScientific:2017zic}.
This single event provided:
(i) the most direct evidence that stellar compact mergers, where at least one of the binary companions is a neutron star (NS), are progenitors of the central engines that power short gamma-ray bursts; 
(ii) strong observational support to theoretical proposals linking BNS mergers with production sites for $r$-process nucleosynthesis and  kilonovae~\citep{Li:1998bw,Metzger:2016pju, Troja:2017,Kasen:2017};	
(iii) an independent measure for the expansion of the Universe~\citep{lvk_hubble:2017,Dietrich:2020efo};
and (iv)	tight constraints on the equation of state (EOS) of matter at supranuclear densities~\citep{Rezzolla:2017aly,Ruiz:2017due,Shibata:2017xdx,Margalit:2017dij,lvk_eos:2018,lvk_eos:2019}.  

From the GW observation point of view,
most of the results inferred from the analysis of GW170817 are based on the late inspiral part of the waveform. 
During this period, tidal forces transfer energy and angular momentum from the orbit to the NS.
The energy transferred is primarily converted into gravitational radiation and deformation of the NS structure.
However, the internal heating due to these forces is minimal. This is why NSs are often considered to have 
effectively zero temperature during this phase, as the tidal energy transfer does not significantly raise the 
internal temperature of the stars.
Following merger, however, shock heating rises up the temperature of the remnant to $\gtrsim 10$ MeV. Such high temperatures provide an additional pressure support that may change the internal structure of the remnant and its subsequent evolution. It is expected that observations of postmerger GWs, presently at the limit of the technology used in second-generation detectors~\cite{lvk_pm:2017}, will yield 
new insights on the nuclear EOS of NSs at finite temperature.

GW searches of BNS mergers and inference of source parameters rely on accurate waveform models for the inspiral signal. Those are based on analytical relativity (computing waveform approximants using post-Newtonian expansions or the effective-one-body approach~\cite{Lackey:2016krb,Narikawa:2022saj}) and numerical relativity, the full-fledged numerical solution of Einstein's field equations coupled to the equations describing NS matter and  radiation processes~(see,~e.g.,~\cite{Sun:2022vri,Foucart:2022kon,Foucart:2020qjb,Gieg:2022mut,Hayashi:2021oxy,Radice:2021jtw}). In numerical simulations of BNS mergers thermal effects are incorporated using two alternative approaches. The first one is a ``hybrid approach" which assumes that the pressure and the internal energy have two contributions, namely, a cold, zero-temperature part  described by a polytropic EOS (or a family of piecewise polytropes) and a thermal part described by an ideal-gas-like EOS~\citep{1993A&A...268..360J,Dimmelmeier:2002bk,shibata:2005ss}. The latter is given by   $P_{\rm th} =\rho_0\,\epsilon_{\rm th} (\Gamma_{\rm th}-1)$, with $\rho_0$ the rest-mass density, and $P_{\rm th}$ and $\epsilon_{\rm th}$ the thermal pressure and thermal energy density, respectively, and $\Gamma_{\rm th}$ the adiabatic index, a constant that lays in the range $1\lesssim \Gamma_{\rm th} \lesssim 2$ 
for causality constraints,  but that in typical BNS merger simulations is set  between $1.6$ and $2$~(see,~e.g.,~\citep{Constantinou:2015mna,Takami:2014tva}). The second approach employs tabulated representations of microphysical finite-temperature EOSs, providing a self-consistent method to probe the impact of thermal effects in the merger dynamics. 
Although the hybrid approach is computationally preferred,
it has some limitations. In particular, it has been shown that the value of the thermal adiabatic index $\Gamma_{\rm th}$ above half saturation density  strongly depends on the nucleon effective mass~\citep{Lim:2019ozm}. Therefore, it is likely that this approach overestimates the thermal pressure by a few orders of magnitude~\citep{Raithel:2021hye}, which may induce significant changes in the GW frequencies~\citep{Bauswein:2010dn,Figura:2021bcn}. 
As the tabulated approach incorporates the temperature self-consistently the above issue is not present. BNS merger simulations based on tabulated EOSs, while computationally more challenging than those based on the hybrid approach, are becoming increasingly more common~
\citep{Oechslin:2006uk,Bauswein:2010dn,Sekiguchi:2011zd,Fields:2023bhs,Espino:2022mtb,Werneck:2022exo,Guerra:2023}. These two alternative ways of including thermal effects in the numerical simulations result in measurable differences in the GW signal, especially in the postmerger part, significantly affecting the frequency spectra (see e.g.~\cite{Bauswein:2010dn,Guerra:2023}). In this work we explore whether these biases significantly impact the detectability of the GW signals.

During the first~$\sim 5\,\rm ms$ after merger, nonaxisymmetric deformations of the remnant are accompanied by the emission of high-frequency GWs. 
The frequency spectra are characterized by the presence of distinctive peaks associated with oscillation modes due to nonlinear interactions between the quadrupole and quasiradial modes and the rotation of the nonaxisymmetric binary remnant. These peaks are typically denoted as $f_{2\pm0}$, $f_{\rm spiral}$, and $f_2$ (or $f_{\rm peak}$)~\cite{Stergioulas:2011,Hotokezaka:2013,Bauswein:2015,Takami:2015,Bauswein:2016,Bauswein:2019}. As pointed out in~\cite{Rezzolla:2016} the frequency of the $f_2$ mode changes by around $\sim 5\%$ in time. These (initial) frequency values are denoted as $f_{2,i}$ to distinguish them from the  value of $f_2$  reached during the quasistationary evolution of the GW signal. Through the analysis of these peaks, inference on NS properties may be possible.  In particular, it has been shown that their frequencies are  related quasiuniversally with the tidal deformability of the stars, and the maximum-mass of nonrotating configurations~\citep{Read:2013,Bauswein:2015,Takami:2015,Rezzolla:2016,Guerra:2023,Topolski:2023}. 

Long-term simulations of the postmerger remnant extending beyond $\sim 50$ ms have also revealed the appearance of inertial modes~\citep{DePietri:2018,DePietri:2020}. Their GWs dominate over those associated with the initial $f_2$ mode at late postmerger times but have lower frequencies and amplitudes. As inertial modes depend on the rotation rate of the star and on its thermal stratification, their detection in GWs would provide a unique opportunity to probe the rotational and thermal states of the merger remnant (see,~e.g.,~\cite{Kastaun:2008}).

Recently, long-term simulations of BNS mergers exploring the influence of the treatment of the thermal part of the EOS by comparing models using hybrid and tabulated approaches have been reported in~\cite{Guerra:2023}. The differences found in the dynamics and GW emission can be used to gauge the importance of the numerical treatment of thermal effects in the EOS, which has observational implications. In this work, we investigate the identification of such differences in BNS merger remnants by reconstructing the GW signals of~\cite{Guerra:2023} using \texttt{BayesWave}\footnote{\href{https://git.ligo.org/lscsoft/bayeswave}{https://git.ligo.org/lscsoft/bayeswave}} \cite{Cornish:2015,Littenberg:2015}, a Bayesian data-analysis algorithm that recovers the postmerger signal through a morphology-independent approach using series of sine-Gaussian wavelets. We aim to study the uncertainties in the peak frequency distributions from several oscillation modes of the merger remnant, and see whether these uncertainties arising from the signal reconstruction are large enough to make the tabulated and hybrid approaches compatible. We stress that our simulations focus on the possible identification of a single difference in the postmerger remnant -- the implementation of thermal effects in the EOS -- and, thus, assume that potential effects from neglected ingredients in the modeling (e.g.~magnetic fields, viscosity, neutrinos, or the knowledge of the underlying nuclear interaction) would be identical in both setups.
Notice that magnetic viscosity and/or cooling processes, such as neutrino emissions, can modify the characteristic GW frequencies of the binary remnant. In particular,  the GRMHD simulations of BNS mergers reported in~\cite{Ruiz:2021qmm} found that magnetic viscosity tends to shift the $f_2$ mode frequency to  lower frequencies by around $\sim 30-150\,\rm Hz$ depending on the stiffness of the EOS as well as on the numerical resolution. Neutrino processes, on the other hand, tend to slightly increase its frequency (see,~ e.g.,~\cite{Sun:2022vri,Foucart:2015gaa}). In addition, numerical simulations of BNS mergers suggested that a sufficiently strong first-order phase transition can alter the postmerger GW signal, leading to a significant  shift frequency of $\sim 500\,\rm Hz$~\cite{Blacker:2020nlq}.  Such deviations have also been observed in BNS mergers subject to nonconvex dynamics~\cite{Rivieccio:2024sfm}.

This investigation is a follow-up of our recent work in~\cite{Miravet:2023}, where we first employed \texttt{BayesWave} to analyze the detectability prospects of the inertial modes computed in the simulations of~\cite{DePietri:2018,DePietri:2020} employing only hybrid BNS models. Moreover, we  further extend the analysis of \cite{Miravet:2023} by studying  the identification of differences in the treatment of thermal effects across the {\it entire} postmerger signal, i.e.,~both in the early part where the $f_2$ mode dominates and in the late part where inertial modes are excited. As done in \cite{Miravet:2023}, we perform waveform injections corresponding to a set of EOSs, each with a hybrid and a tabulated version, into the noise of  the third-generation GW detector Einstein Telescope (ET)~\cite{ET:2010, Hild:2011,Science_case_ET,COBA_study} from BNS sources at different distances. The posterior distributions of the recovered waveforms give us distributions of the peak frequencies that can be related to physical properties of the merger remnant via empirical relations. 

Our analysis is complementary to the recent work reported in~\cite{Calderon-Bustillo:2023} where Bayesian model selection was used to explore differences between the hybrid and the tabulated approaches with the same set of GW signals from~\cite{Guerra:2023}. This is a completely different approach to our model-agnostic reconstructions. The findings reported in~\cite{Calderon-Bustillo:2023}, where differences between tabulated and hybrid treatments of thermal effects were found to lead to differences in the postmerger GW observable by third-generation detectors at source distances $\leq 50$ Mpc, are consistent with what is reported here. We find that differences in the distribution of the main frequency peaks in the postmerger GW spectra in hybrid and tabulated models can be resolved in third-generation detectors up to distances similar to those reported in~\cite{Calderon-Bustillo:2023}. Recently, the studies in~\cite{Raithel:2023} showed that finite-temperature effects included through different prescriptions can be distinguished with future detectors if the cold EOS is well constrained. These results suggest that, depending on the choice of the EOS, the use of a self-consistent treatment of thermal effects through the tabulated approach is crucial to obtain realistic GW signals from numerical simulations. The use of a hybrid approach, which is an approximation of the ``real" EOS, may produce large enough changes in the simulated GW signal with respect to an actual postmerger signal.

The paper is organized as follows: we summarize the setup of the BNS merger simulations of~\cite{Guerra:2023} in Sec.~\ref{sec:Numerical_set}. Next, in Sec.~\ref{sec:reconstruction}, we briefly present the {\tt BayesWave} algorithm and introduce the quantities we use to assess the waveform reconstructions. Our main results are discussed in Sec.~\ref{sec:results}. We divide this section in two parts: The first one is focused on the early postmerger signal, where we also discuss different EOS-insensitive fits that relate the $f_{2,i}$ and $f_2$ modes with the tidal deformability of NSs. Then, in the second part, we consider the late postmerger phase and study the differences in the frequencies of the inertial modes for our set of EOSs. The conclusions of our work are presented in  Sec.~\ref{sec:discussion}. Finally,  Appendix~\ref{sec:appendixA} contains a brief summary of our findings for source inclinations and sky locations different than those considered in the main body of the paper where optimal orientation and sky location is assumed, and in Appendix~\ref{sec::appendixB} we relate the distance to the GW sources with the corresponding signal-to-noise ratio.

%
\section{Summary of the BNS mergers setup}
\label{sec:Numerical_set}

The gravitational waveforms employed in our analysis were computed in the numerical-relativity simulations of BNS mergers recently conducted by~\cite{Guerra:2023}. The initial data for those simulations consist of two equal-mass, irrotational NSs modeled by finite-temperature (tabulated) microphysical EOSs, namely {\tt DD2}~\cite{Typel:2009sy}, {\tt HShen}~\cite{Shen:2011qu}, {\tt LS220}~\cite{Lattimer91}, and {\tt SLy4}~\cite{chabanat98}. These initial data were built using {\tt LORENE}~\cite{Gourgoulhon:2000nn,tg02,Lorene}.  The EOS tables are obtained following the work of Schneider \textit{et al.}~\cite{Schneider:2017tfi} and are freely
available at~\cite{stellarcollapse}. The initial temperature is fixed to $T = 0.01\,\rm{MeV}$, the lowest value on the tables. These EOSs span a reasonable range of central densities, radii, and maximum gravitational masses for irrotational NSs. The initial separation of the two stars is $44.3\,\rm km$ and the rest-mass of each star is $M_0=1.4\,M_\odot$. Their properties are summarized in Table~\ref{table:Iparamenters}. For comparison purposes we also consider waveforms obtained in simulations of BNS mergers based on hybrid EOSs, consisting of a cold and a thermal part. The cold component of each EOS is made of  piecewise polytropic representations of the above EOSs using a piecewise regression as in~\cite{2021JOSS....6.3859P} with seven pieces~\cite{Read:2008iy}. Correspondingly, the thermal component is based on a $\Gamma$-law EOS with a constant adiabatic index $\Gamma_{\rm th} = 1.8$. 

\begin{figure}[ht]
\centering
   \includegraphics[width=1\linewidth]{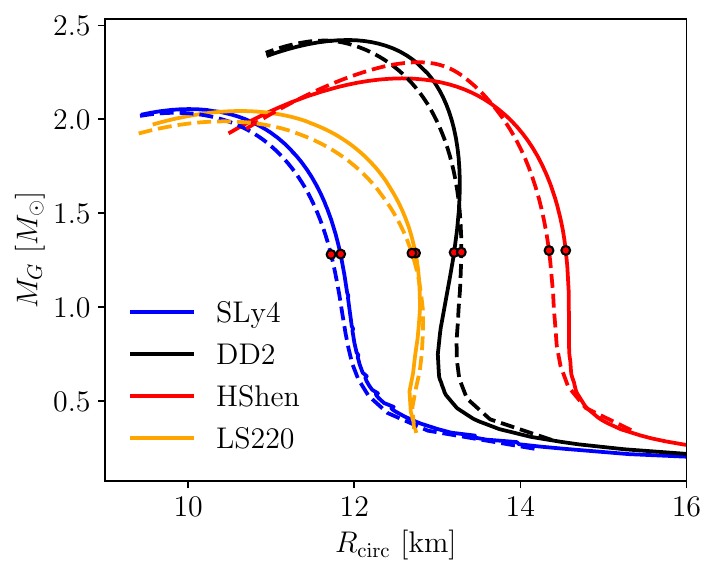}
   \caption{Gravitational mass versus circumferential radius for the tabulated (solid lines) and the hybrid (dotted lines) EOSs used in this work. The selected NS configurations are depicted with red dots.}
\label{fig:MR}
\end{figure}

The two types of BNS models we use -- hybrid and tabulated -- are built as similar as possible, to minimize the effects of differences that the initial data may have in our study. However, there are intrinsic differences in the way the models are built (e.g.,~the lowest value of the temperature in the tables is $T=0.01$ MeV, which affects the density distribution at low densities) that make it not possible to build the exact same stars. The gravitational mass versus circumferential radius for the two types of configurations and for the four EOSs used in this paper are displayed in Fig.~\ref{fig:MR} with solid red circles. The unavoidable small discrepancy visible in this plot leads to a difference of $\lesssim 10\%$ between the values of the tidal deformability $\Lambda$ of the tabulated and the hybrid EOSs listed in Table~\ref{table:Iparamenters}. We  refer the reader to~\cite{Guerra:2023} for further details.

\begin{center}
  \begin{table}[t!]
    \caption{Summary of the initial properties of the BNS configurations.
      We list the EOS, the temperature $T[\rm MeV]$, the gravitational mass $M[M_\odot]$,
      and the compactness $\mathcal{C}\equiv M/R_{\rm eq}$ and the tidal deformability
      $\Lambda = (2/3)\kappa_2\,\mathcal{C}^{-5}$ for each individual star.
      Here $R_{\rm eq}$ is the equatorial coordinate radius toward the companion
      of each star, and $\kappa_2$ is the second Love number. The ADM mass $M_{\rm ADM}[M_\odot]$, the ADM angular momentum $J_{\rm ADM}[M_\odot^2]$ and the angular velocity $\Omega [\rm krad/s]$, for an initial binary coordinate separation of $\sim 44.3\,\rm km$. In all cases
      the NS has a rest-mass $M_0=1.4M_\odot$. The first (last) four rows correspond to the BNSs modeled through a fully tabulated (piecewise polytropic) EOS. Center dots denote “not applicable.”
      \label{table:Iparamenters}}
    \begin{tabular}{cccccccc}
      \hline
      \hline
       EOS & $T$     & $M$      & $\mathcal{C}$    &  $\Lambda$   &$M_{\rm ADM}$  & $J_{\rm ADM}$ & $\Omega$\\
      \hline
        {\tt SLy4}       & 0.01     &  1.28   & 0.16   & 536.00   & 2.54          & 6.63          & 1.77 \\
        {\tt DD2}        & 0.01     &  1.29   & 0.14   & 1098.68   & 2.56          & 6.73          & 1.78 \\
        {\tt HShen}      & 0.01     &  1.30   & 0.13   & 1804.67  & 2.58          & 6.82          & 1.78 \\
        {\tt LS220}      & 0.01     &  1.29   & 0.15   & 851.72   & 2.55          & 6.68          & 1.77 \\
       \hline
       \hline
        {\tt SLy4}       & ...        &  1.28   & 0.16   & 511.70   & 2.54          & 6.62          & 1.77 \\
        {\tt DD2}        & ...        &  1.29   & 0.14   & 1113.92   & 2.56          & 6.73          & 1.78 \\
        {\tt HShen}      & ...        &  1.30   & 0.13   & 1633.24  & 2.58          & 6.82          & 1.78 \\
        {\tt LS220}      &  ...       &  1.29   & 0.15   & 899.05   & 2.55          & 6.69          & 1.77 \\ \hline\hline
    \end{tabular}
  \end{table}
\end{center}

The initial data were evolved in~\cite{Guerra:2023} using the  {\tt IllinoisGRMHD} code~\cite{Werneck:2022exo,Etienne:2015cea}  embedded in the {\tt Einstein Toolkit} infrastructure~\cite{Loffler:2011ay}. Much of the numerical infrastructure has been extensively discussed in~\cite{Werneck:2022exo,Etienne:2015cea,Noble:2005gf,Guerra:2023} to which the interested reader is addressed for details. As a summary, we mention only that the code evolves the Baumgarte–Shapiro–Shibata–Nakamura  equations~\cite{Baumgarte:1998,Shibata:1995} coupled to the puncture gauge conditions using fourth-order  spatial differentiation. In all cases the damping coefficient appearing in the shift condition was set to 1/$M$, where $M$ is the Arnowitt-Deser-Misner (ADM) mass of the system. Moreover,  {\tt IllinoisGRMHD} adopts the Valencia formalism for the general relativistic hydrodynamics equations~\cite{Banyuls:1997, Font:2008} which are integrated with a state-of-the-art finite-volume algorithm. Time integration is performed using the method of lines with a fourth-order Runge-Kutta integration scheme with a Courant-Friedrichs-Lewy factor of 0.5.

Some of the evolutions reported in~\cite{Guerra:2023} extend for over $t-t_{0}\sim 150\,\rm ms$ after merger. This permits one to identify the imprint of thermal effects on the postmerger GW signals and in the frequency spectra. In particular, such long-term simulations allow us to study the potential dependence on the treatment of thermal effects for both the frequencies associated with the fundamental quadrupolar mode, excited about some 5 ms after merger, along with those of inertial modes, typically appearing at significantly longer postmerger times~\cite{DePietri:2018,DePietri:2020,Guerra:2023}.

\section{\label{sec:reconstruction}Waveform reconstruction}

\subsection{\label{sec:BAYESWAVE} The {\tt BayesWave} algorithm}

The tool we employ to assess possible observational differences in the treatment of thermal effects in the postmerger GW signal is {\tt BayesWave} \cite{Cornish:2015,Littenberg:2015}, a Bayesian algorithm that uses sine-Gaussian wavelets to reconstruct unmodeled signals with minimal assumptions \cite{BECSY:2017}. There are some parameters of the wavelets that the user can fix to optimize the reconstruction. Those appear in the time-domain expression of the wavelets for the ``+" and ``$\times$" polarizations:
\begin{align}
    h_+(t) & = A e^{-(2\pi f_0(t-t_0)/Q)^2}\cos{[2\pi f_0(t-t_0)+\phi_0]}\,, \\
    h_{\times}(t) & = \epsilon h_+(t)e^{i\pi/2}\,. \label{e_eq}
\end{align}
In these equations $Q$ is the quality factor, indicating how damped a wavelet is. The algorithm uses a transdimensional reversible jump Markov chain Monte Carlo (RJMCMC) to sample joint posteriors of other parameters of the wavelets, such as their amplitude $A$, their central frequency $f_0$, the offset phase $\phi_0$, and the ellipticity $\epsilon$. {\tt BayesWave} also chooses an optimal number of wavelets for the reconstruction, $N_W$. The RJMCMC method derives the posterior {distribution} of the reconstructed waveform and, using the waveform samples, posteriors of quantities that can be derived from the signals are obtained. For the extrinsic
parameters of the model, i.e., the sky location, polarisation angle and ellipticity,
uniform priors are adopted (see~\cite{Cornish:2015}).

\subsection{\label{sec:overlap_fpeak} Overlap and peak frequency}

We employ the \textit{overlap} function to study the similarity between the recovered model from {\tt BayesWave}, $h_r$, and the injected signal, $h_i$:
\begin{equation}\label{overlap}
    \mathcal{O} = \frac{\langle h_i, h_r \rangle}{\sqrt{\langle h_i,h_i\rangle}\sqrt{\langle h_r,h_r\rangle}}\,.
\end{equation}

This expression involves the inner product of two complex quantities, which is defined as 
\begin{equation}\label{inner_prod}
    \langle a,b \rangle \equiv 2\int_0^{\infty}\frac{a(f)b^*(f)+a^*(f)b(f)}{S_h(f)}df\,,
\end{equation}
where $S_h(f)$ refers to the one-sided noise power spectral density (PSD) of the detector. The asterisk denotes complex conjugation. 

The overlap function can take values from -1 to +1. A perfect match between the signals will result in $\mathcal{O} = +1$, and a perfect anticorrelation will give $\mathcal{O} = -1$. If there is no similarity between the signals, the overlap will be 0. Equation~(\ref{overlap}) is valid for a single-detector measurement. The expression for the weighted overlap of a network of $N$ detectors is
\begin{equation}\label{net_overlap}
    \mathcal{O}_{\rm network} = \frac{\sum^N_{k=1}\langle h_i^{(k)}, h_r^{(k)} \rangle}{\sqrt{\sum^N_{k=1}\langle h_i^{(k)},h_i^{(k)}\rangle}\sqrt{\sum^N_{k=1}\langle h_r^{(k)},h_r^{(k)}\rangle}}\,,
\end{equation}
where index $k$ stands for the $k$th detector.

\begin{figure*}[t]
    \centering 
    \includegraphics[width=0.49\textwidth]{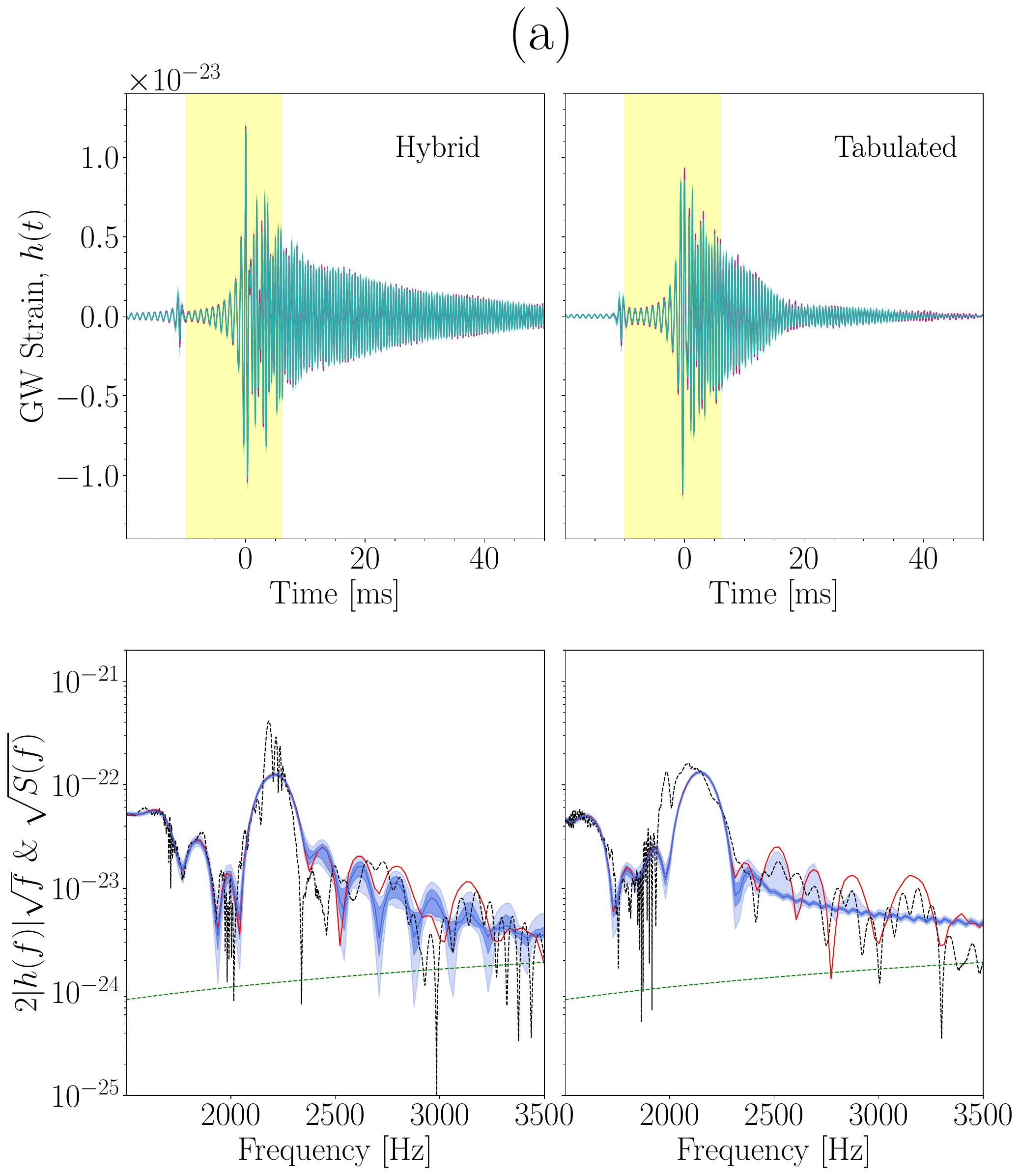}
    \includegraphics[width=0.49\textwidth]{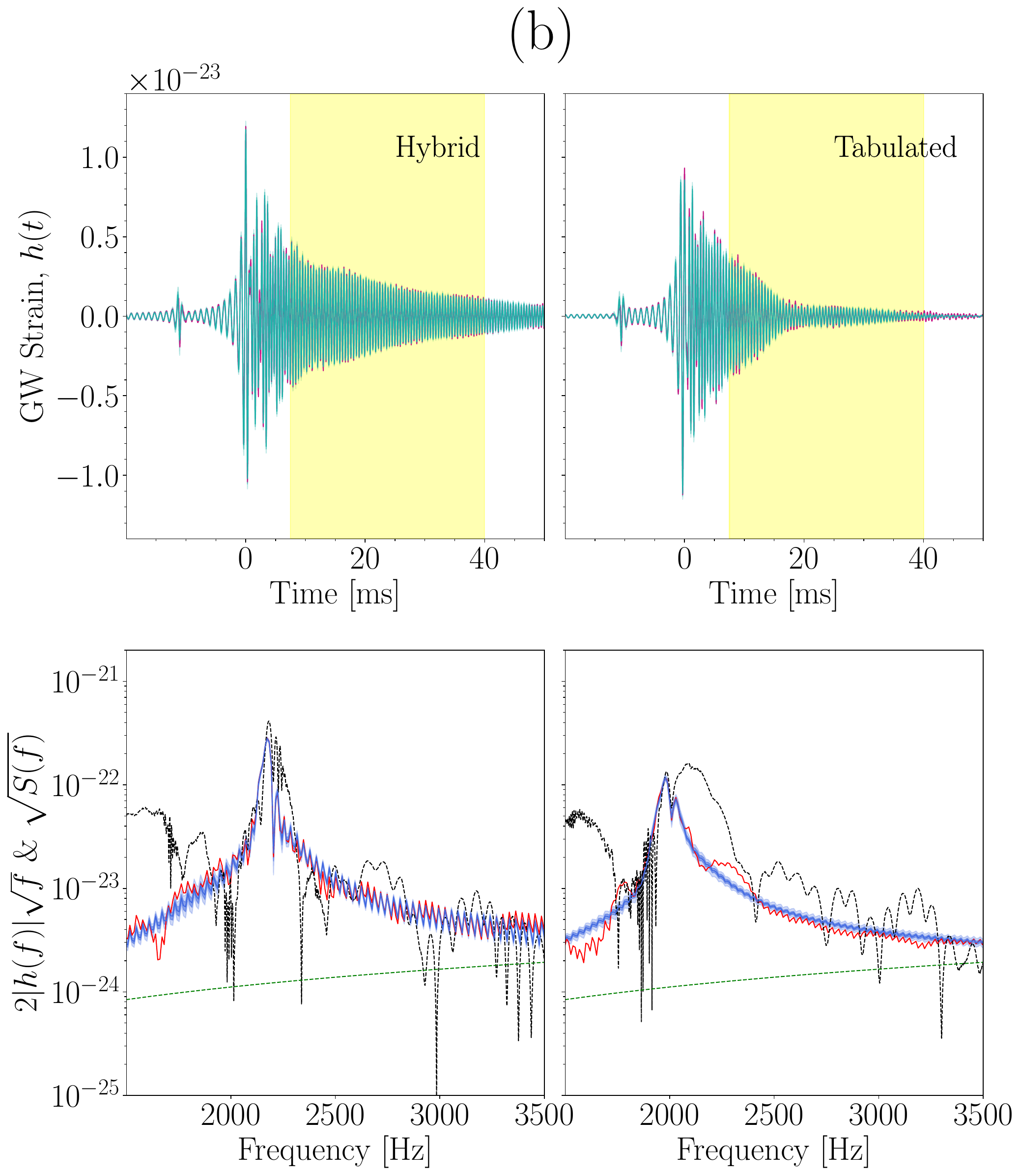}
    \caption{Top row: injected (red) and reconstructed (blue) time-domain waveforms from BNS mergers with the {\tt HShen} EOS. Bottom row: corresponding ASD, computed either using the complete waveforms (dotted line) or at the time windows depicted in yellow in the strain plots in (a) and (b) (see Table \ref{table:time_windows}).  The left (right) column of each of the two panels corresponds to the hybrid (tabulated) version of the EOS. The source is located at $D=20$ Mpc. The signals are injected into the ET-D configuration of the  ET detector, whose sensitivity curve is shown by the dashed green curve within the frequency range depicted.} 
    \label{fig:rec_HShen}
\end{figure*}

From the GW spectra, we can analyze the frequency peaks that arise due to the excitation of certain modes of oscillation in the merger remnant. Since the output of {\tt BayesWave} is time-domain signals, we need to apply the fast Fourier transform (FFT)~{\citep{Cooley:1965}} with a certain time window to study the part of the postmerger phase we are interested in. The FFT is computed using {\tt PyCactus}~\cite{pycactus:2021}, a {\tt Python} package that contains tools for postprocessing data from numerical simulations. Once the spectra are obtained, we look for their frequency peaks. Given the posterior distribution of reconstructed signals, we will end up with a posterior distribution of frequency peaks. Those may be connected to some physical parameters of the merger remnant via empirical relations. We expect the peak frequencies to be located in the range $f\in [1500,4000]$ Hz \cite{Chatz:2017,DePietri:2020}. We use this range to set the low-frequency and high-frequency cutoffs for the computation of the overlap and the frequency peaks.

\begin{figure*}[t]
\centering
     \includegraphics[width=0.49\textwidth]{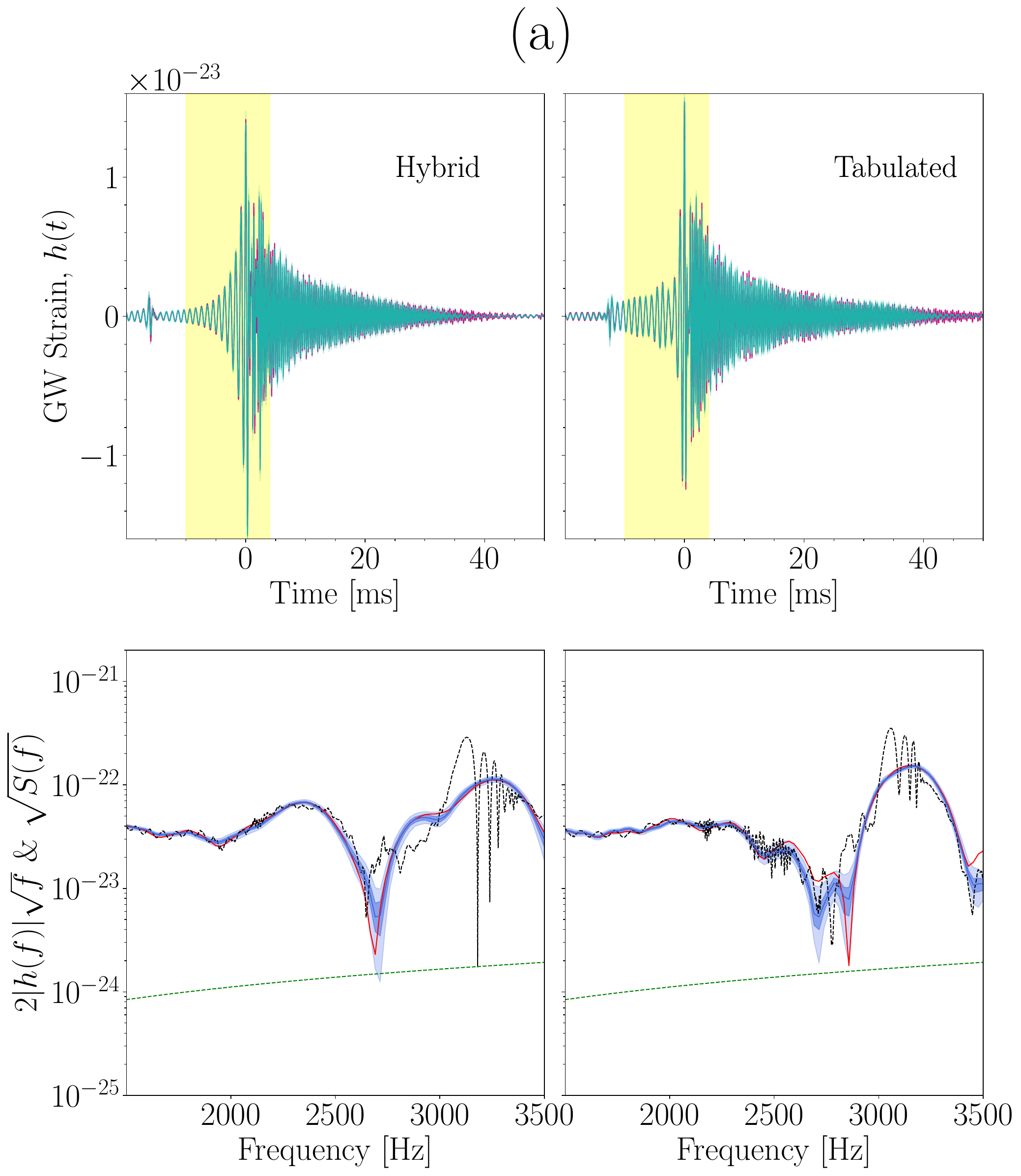}
   \includegraphics[width=0.49\textwidth]{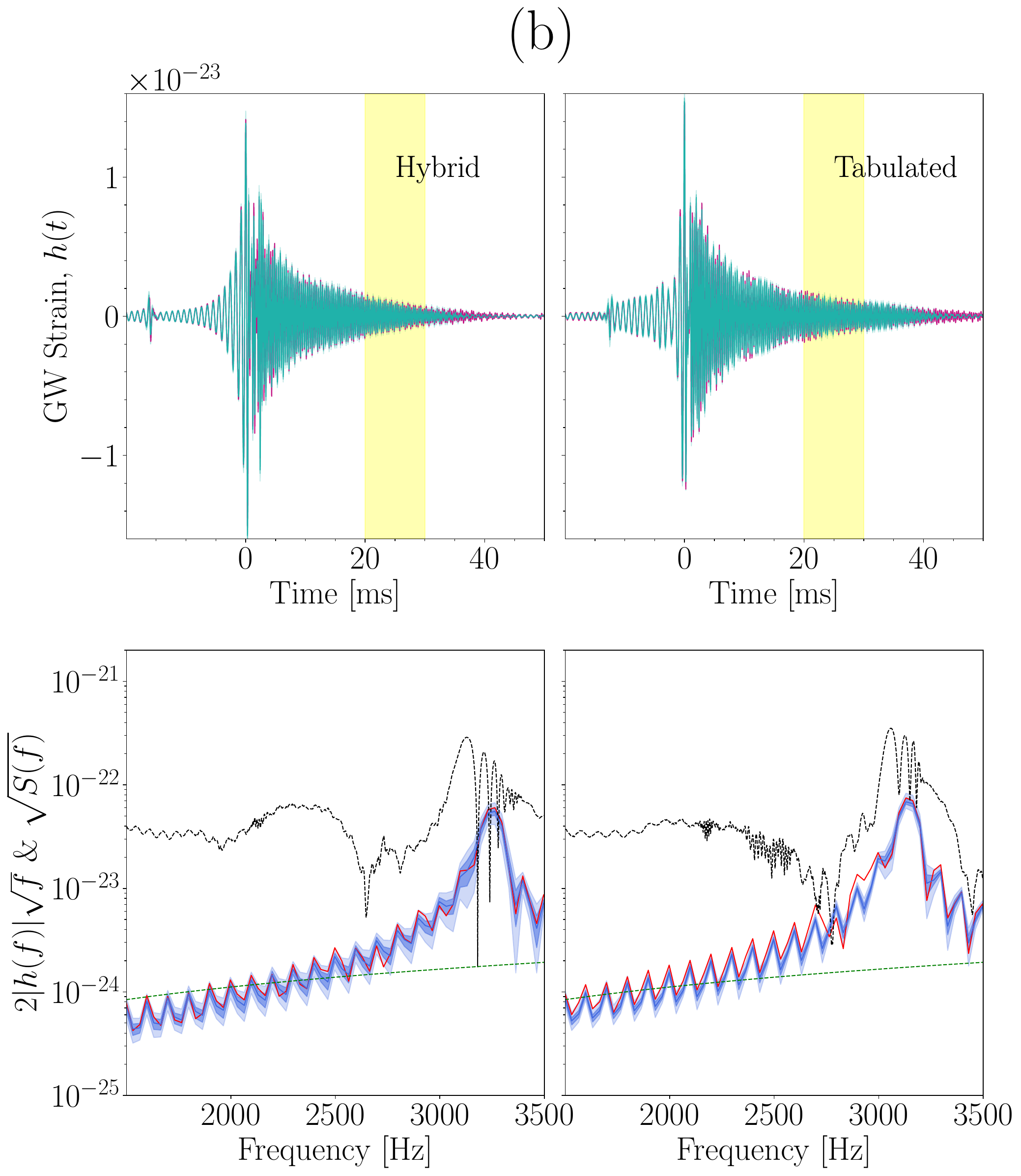}
   \caption{Top row: injected (red) and reconstructed (blue) time-domain waveforms from BNS mergers with the {\tt SLy4} EOS. Bottom row: corresponding ASD, computed either using the complete waveforms (dotted line) or at the time windows depicted in yellow in the strain plots in (a) and (b) (see Table~\ref{table:time_windows}). The left (right) column of each of the two panels corresponds to the hybrid (tabulated) version of the EOS. The source is located at $D =20$ Mpc. The signals are injected into the ET-D configuration of the ET detector, whose sensitivity curve is shown by the dashed green curve within the frequency range depicted.}
\label{fig:rec_SLy4}
\end{figure*}

\section{\label{sec:results}Results}

For all of our injections, we use the ET-D configuration from \cite{Hild:2011} as the sensitivity curve of the ET,  which is formed by a three-detector network on the same site. Our conclusions should also broadly hold for Cosmic Explorer~\cite{Cosmic_Explorer}, as its detection capabilities are similar to those of ET. For simplicity and as we did in \cite{Miravet:2023}, we consider Gaussian noise \citep{Blackburn:2008,Abbott:2009,Aasi:2012} (colored by the PSD of the detector) and no sources of noise or glitches are added. The waveforms are injected at different distances, which result in different signal-to-noise ratios (SNRs). We also assume that the source is optimally oriented with respect to the detector. For completeness, in  Appendix~\ref{sec:appendixA}, we discuss the differences in the overlap function for  non-optimal orientation and sky location. 

We set a maximum number of wavelets of $N_W^{\rm max} = 200$, a maximum quality factor of {$Q^{\rm max} = 200$}, $n=2\times 10^6$ iterations, and a sampling rate of 8192 Hz, resulting in the same setup used in \cite{Miravet:2023}.  The offset phase parameters are taken to be uniform in the range $\phi_0 = [0, 2\pi]$, and the signal wavelet amplitude prior is presented and discussed in~\cite{Cornish:2015}.

\subsection{\label{sec:early_pm} Early postmerger phase}

We begin by focusing on the first milliseconds after merger. During this early phase, strong nonaxisymmetric deformations and nonlinear oscillations are present, namely, combinations of oscillation modes and spiral deformations, leading to the emission of GW signals with frequencies around a few kilohertz. Since the amplitude of these signals is considerably larger than in the late postmerger phase, we also consider correspondingly larger distances, from 1 up to 200 Mpc.  

\subsubsection{\label{subsec:reco_early} Waveform reconstructions}
\label{sub_sec:early_wf}

Figures~\ref{fig:rec_HShen} and \ref{fig:rec_SLy4} show the nonwhitened, time-domain signal (top panels) and the amplitude spectral density (ASD, bottom panels) of the injected (red) and reconstructed (blue) GW signals (with the detector ASD), at a fixed distance of 20 Mpc. Figure~\ref{fig:rec_HShen} corresponds to the {\tt HShen} EOS and Fig.~\ref{fig:rec_SLy4} to the {\tt SLy4} EOS, respectively. Panels (a) and (b) in both figures differ by the time window used to compute the ASD, highlighted in yellow in the time-domain waveform plots. We adapt the time windows to each particular model and phase of the waveform. The time windows employed for each EOS and phase of the simulation (characterized by a dominant oscillation mode) are summarized in Table~\ref{table:time_windows}. The blue-shaded regions in the ASD plots in both figures show the 50\% and 90\% credible intervals (CIs) of the distribution of the recovered waveforms. These intervals are given by values of the percentiles 25th/75th and 5th/95th, respectively. 

The windows used in panel (a) of Figs.~\ref{fig:rec_HShen}
and \ref{fig:rec_SLy4} corresponds to the time interval $t \in [-10,6] $ ms and $t \in [-10,4] $ ms, respectively, $t=0$ being the time of merger. During this phase, the $f_{2,i}$ modes are excited, and they exhibit the frequency peaks shown in the bottom rows. The left column of the panels in both figures shows the reconstructions of the hybrid version of the EOS, whereas the right column depicts the tabulated version. For the case of {\tt HShen}, in Fig.~\ref{fig:rec_HShen}(a), the $f_{2,i}$ peaks are located around 2200 Hz. The ASD of the hybrid and tabulated version of the EOS are similar, but the tabulated one produces $f_{2,i}$ modes with a slightly lower frequency. Regarding {\tt SLy4}, in Fig.~\ref{fig:rec_SLy4}(a), the peaks are located around 3250 Hz, and the tabulated version of the EOS also has the peaks at slightly lower frequencies than the hybrid version. The differences between the hybrid and tabulated versions of the EOS are almost negligible, with a difference between the frequency peaks of $\approx 4\%$. For the {\tt DD2} and {\tt LS220} EOS the mismatch is even smaller.

\begin{center}
  \begin{table}[th]
  \caption{Time windows employed to capture the different oscillation modes appearing  during postmerger. Times are expressed in milliseconds and $t=0$ ms is the merger time.
      \label{table:time_windows}}
    \begin{tabular}{cccc}
      \hline
      \hline
       EOS & & Mode & \\
       \hline
       & $f_{2,i}$     & $f_2$      & Inertial  \\
      \hline
        {\tt SLy4}       &  [-10, 4]    &  [20, 30]   & [75, 140]      \\
        {\tt DD2}        & [-10, 4]       &  [7, 17]     & [70, 140]  \\
        {\tt HShen}      & [-10, 6]      &  $[7.5,40]$  & [105, 140]  \\
        {\tt LS220}      & [-10, 6]      &  [8, 15]     & [80, 140]   \\
       \hline
    \end{tabular}
  \end{table}
\end{center}

\begin{figure*}[t]
    \centering
    \includegraphics[width=0.49\textwidth]{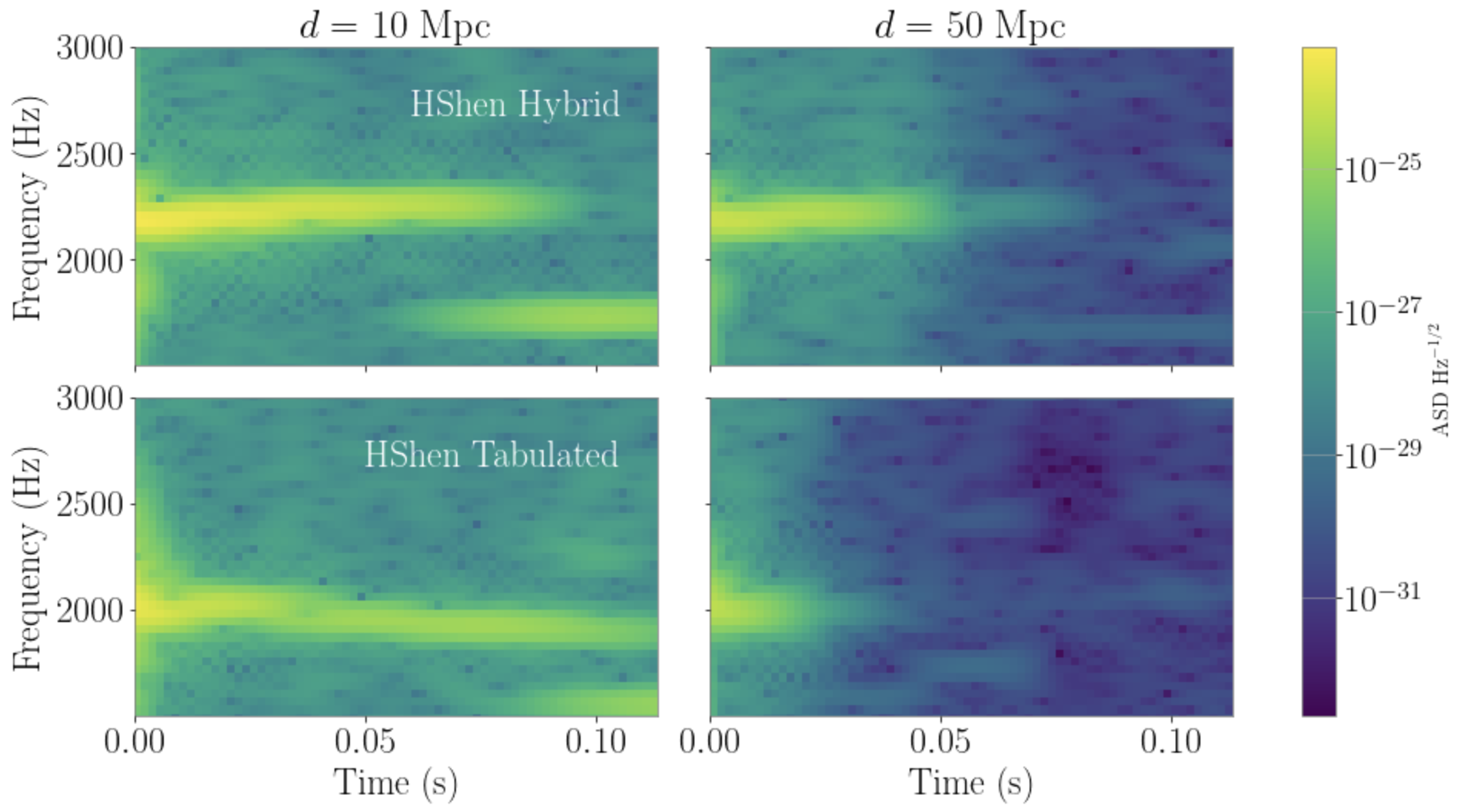}
    \includegraphics[width=0.49\textwidth]{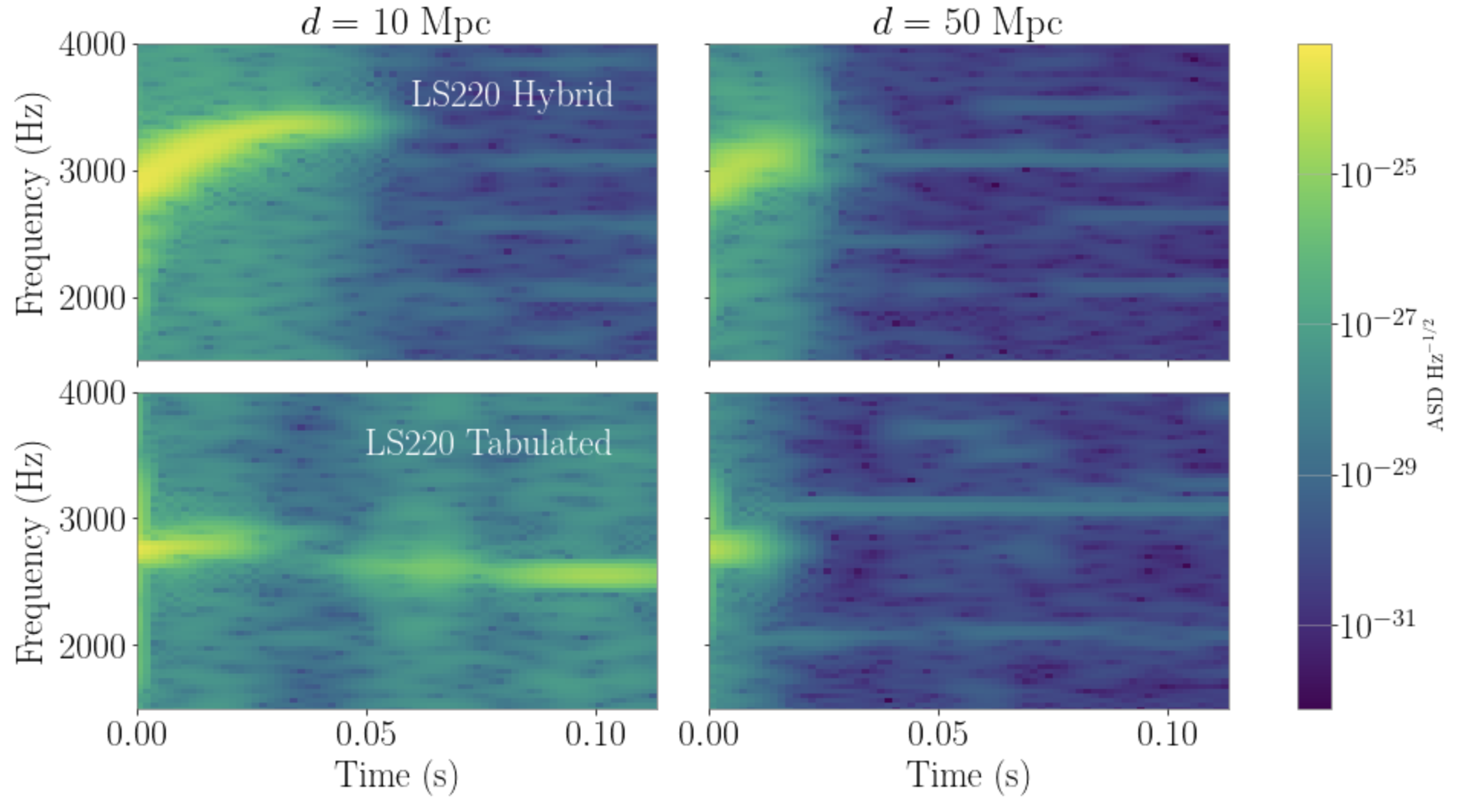}
    \caption{Spectrograms of the reconstructed GW signals at two source distances, $D = 10$ and 50 Mpc. The left (right) panels correspond to the {\tt HShen} ({\tt LS220}) EOS. Plots in the upper row depict the hybrid version of the EOS and those in the lower row the tabulated model. As the distance to the source increases, it becomes more difficult to capture the time evolution of the frequency of the signal at late times.}
    \label{fig:spectrograms}
\end{figure*}

In panel (b) in Figs.~\ref{fig:rec_HShen} and \ref{fig:rec_SLy4} we depict again the time domain and the spectra of the injected and recovered signals for the same two EOSs, but the time window is applied now for the intervals $t\in [7.5,40]$ ms ({\tt HShen}) and $t\in [20,30]$ ms ({\tt SLy4}). Therefore, the ASD of the bottom panels shows the appearance of the $f_2$ modes. For the {\tt SLy4} EOS, the amplitude of the $f_2$ modes is lower than that of the $f_{2,i}$ modes. For the {\tt HShen} EOS, there are more noticeable differences in the position of the peaks between the hybrid and tabulated models than for the {\tt SLy4} EOS. For both cases, the peaks of the $f_2$ modes appear at lower frequencies than for the $f_{2,i}$ modes.

Spectrograms of the median of the reconstructed signals are shown in Fig.~\ref{fig:spectrograms} for the case of {\tt HShen} (left) and {\tt LS220} (right). In both panels, the plots in the upper row refer to the hybrid version of the EOS and the lower row to the tabulated version. This figure shows the spectrograms for two difference source distances, 10 and 50 Mpc. For a distance of 10 Mpc all the stages of the postmerger signal are visible in the spectrogram of the {\tt HShen} EOS: There is an initial part where the signal is louder followed by a decrease in frequency and amplitude, visible up to more than 100 ms after merger. This trend occurs for both hybrid and tabulated versions of the {\tt HShen} EOS. As the distance increases, however, the last part of the signal cannot be reconstructed. Beyond 50 Mpc, the signal is visible only up to $t \sim 50 $ ms. Note that, for the hybrid version of the {\tt HShen} EOS, the signal is detectable for longer times. 

The four plots in the right panel of Fig.~\ref{fig:spectrograms} depict two completely different behaviors between the hybrid and tabulated versions of the {\tt LS220} EOS. In the first case (upper row), the frequency of the signal increases with time up to $t\sim 40$ ms for sources located at short distances (10 Mpc). At that point, the signal disappears because the remnant collapses to a black hole (BH) (at $t=66.2$ ms after merger)~\cite{Guerra:2023}. However, the tabulated version of the {\tt LS220} EOS (lower row) shows that a stable remnant evolves for more than 100 ms after merger. In this case, the GW signal decreases in amplitude, reaching frequencies of about 2 kHz. At larger source distances (50 Mpc), {\tt BayesWave} recovers only the inspiral phase and the very early stages after  merger, not capturing the collapse of the remnant for the hybrid version of the EOS.

\subsubsection{\label{subsec:fpeaks_early} Frequency peaks of the $f_{2,i}$ and $f_{2}$ modes}
\label{subsec:early_fp}

From the posterior distributions of the GW signals that {\tt BayesWave} provides,  one can compute the ASD via the FFT of the time-domain signal using a certain time window. This yields posterior distributions of the frequency peaks of the spectra. We start considering time windows spanning from $t\approx 10$ ms before merger to a few tens of milliseconds after merger (depending on the EOS; see Table~\ref{table:time_windows}). The size and position of the time windows are chosen to distinguish the frequency peaks related to the $f_{2,i}$ and $f_2$ modes.

\begin{figure*}
    \centering
    \includegraphics[width=\textwidth]{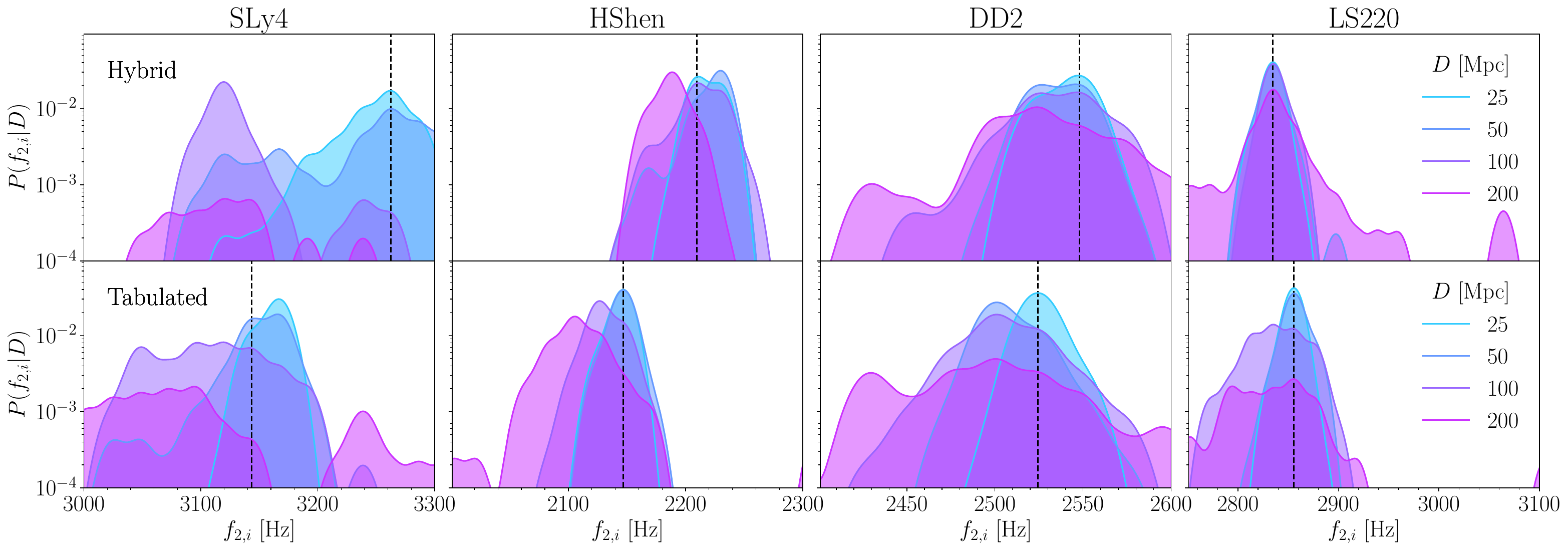}
    \includegraphics[width=\textwidth]{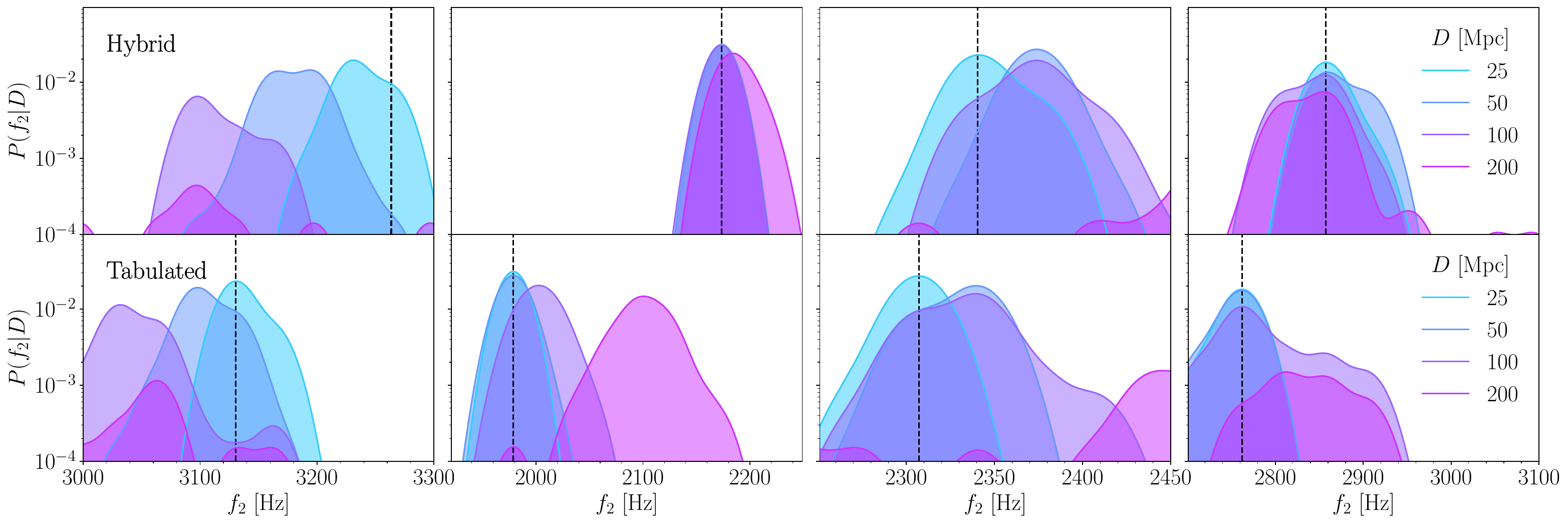}
    \includegraphics[width=0.765\textwidth,left]{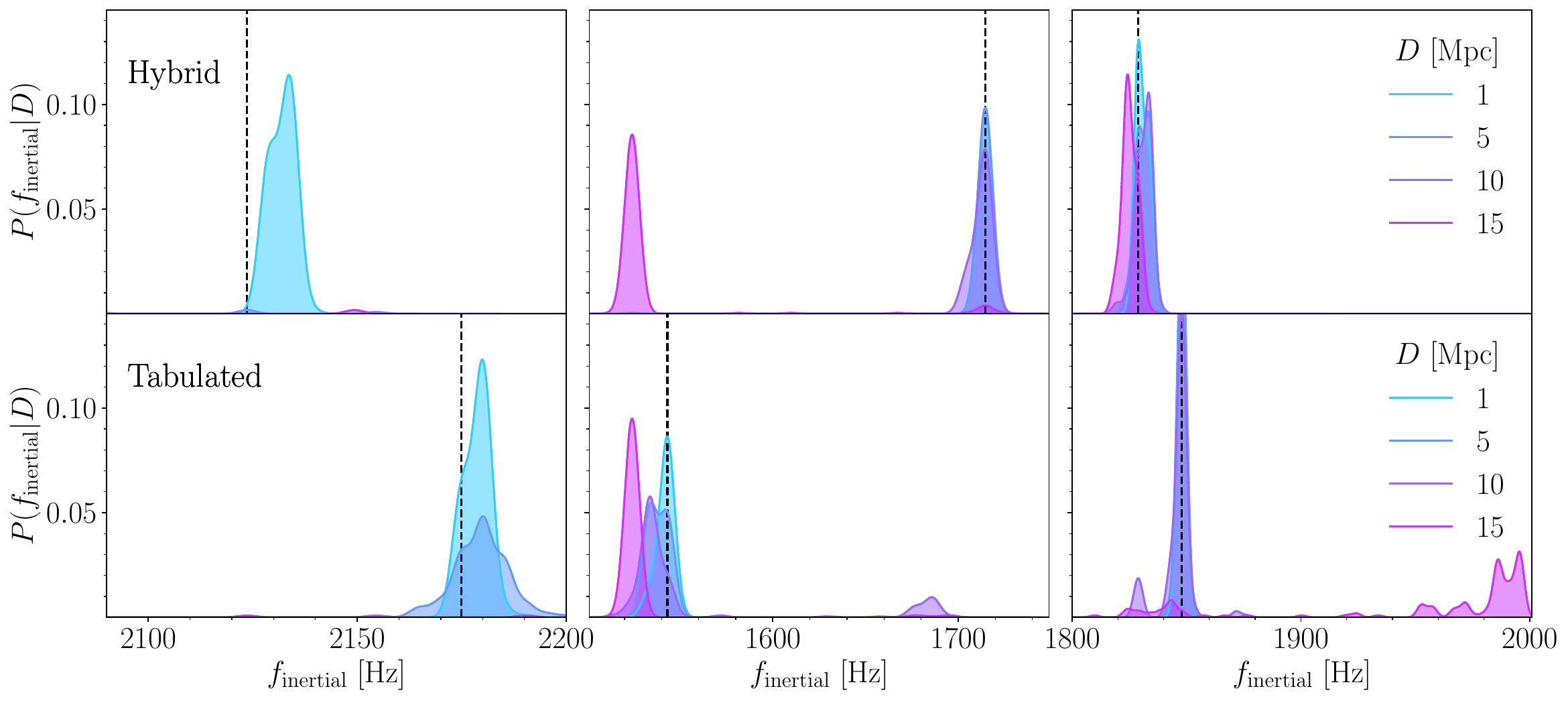}
    \caption{Posterior distributions of the frequency peaks for the $f_{2,i}$ modes (top), $f_2$ modes (middle), and $f_{\rm inertial}$ modes (bottom). Each column corresponds to a different EOS ({\tt SLy4}, {\tt HShen}, {\tt DD2}, and {\tt LS220}, from left to right). The upper (lower) rows are the hybrid (tabulated) versions of the corresponding EOS. Each color in the posterior distributions corresponds to a different distance to the source, indicated in the legends. The vertical black dashed lines are the frequency peaks of the injected signals. As expected, the shorter the distance, the narrower the distributions and the closer they are to the injected values. We do not show the peaks for {\tt LS220} in the bottom row because the hybrid version collapses to a BH and there is no late postmerger signal.}
    \label{fig:fpeaks}
\end{figure*}

The top and middle panels in Fig.~\ref{fig:fpeaks} show the posterior distributions of the frequency peaks for the $f_{2,i}$ and the $f_2$ modes, respectively. (The bottom panel in this figure will be discussed below.) The posterior distributions are constructed using a Gaussian kernel density estimator  and setting the bandwidth equal to the frequency resolution given by the FFT (which will be different depending on the time window considered). Each column corresponds to a certain EOS, from left to right {\tt SLy4}, {\tt HShen}, {\tt DD2}, and {\tt LS220}. The upper row shows the hybrid version of each EOS and the lower one the tabulated version. The different colors refer to several distances to the source, that range from $D =25$ Mpc to $D =200$ Mpc. For the $f_{2,i}$ mode, shown in the top panel of Fig.~\ref{fig:fpeaks}, differences between the treatment of thermal effects in the EOS are observed\footnote{We consider that the differences in the peak frequencies are ``detectable" when the posterior distributions for the frequency peaks do not overlap. The distributions overlap when their full widths at half maximum do so.}  for only {\tt SLy4} (first column) and {\tt HShen} (second column). These two EOSs show the largest deviation in the frequency peaks as a result of the distinct consideration of thermal effects. The frequency peak of {\tt SLy4} is not well recovered for $D > 50$ Mpc, with a $p$-value\footnote{We set a threshold of $p> 0.05$ (corresponding to 2$\sigma$) to consider as null hypothesis that the mean of the distribution is equal to the frequency peak of the injected signal. For distributions with $p \gtrsim 0.05$, we assume that the frequency peaks are well recovered, i.e., detectable.} of 0.05 and 0.13 for the hybrid and tabulated cases, respectively, at $D=200$ Mpc. This results in large enough differences between both versions of the EOS only for very close sources. In the case of {\tt DD2} and {\tt LS220}, the curves of the posterior distributions overlap for all distances, and no differences between the tabulated and hybrid EOSs might be seen. The range of detectability is almost the same for all EOSs but {\tt SLy4} (first column). For the other cases, the peaks are detectable up to $D\sim 200$ Mpc, with $p$-values over 0.15 for all distances. 

\begin{figure*}[t]
    \centering
    \includegraphics[width=\textwidth]{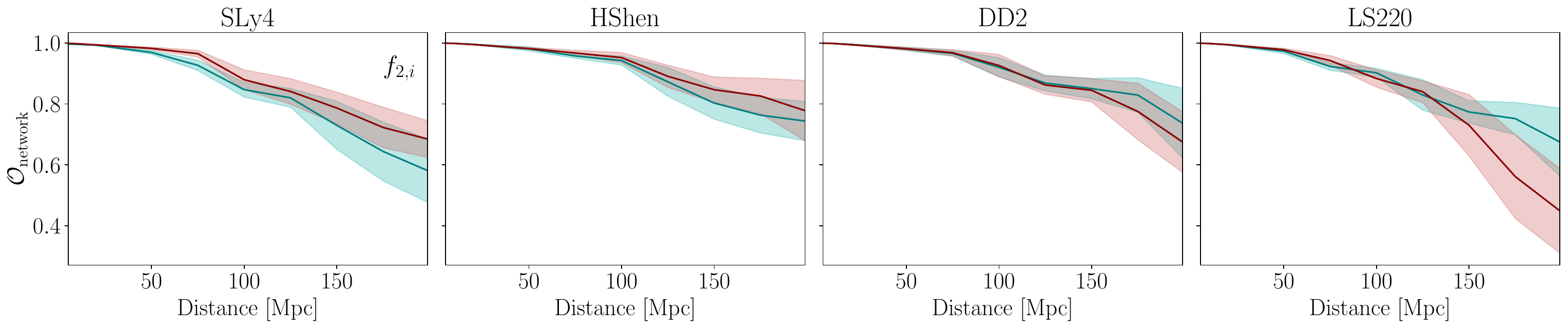}
    \includegraphics[width=\textwidth]{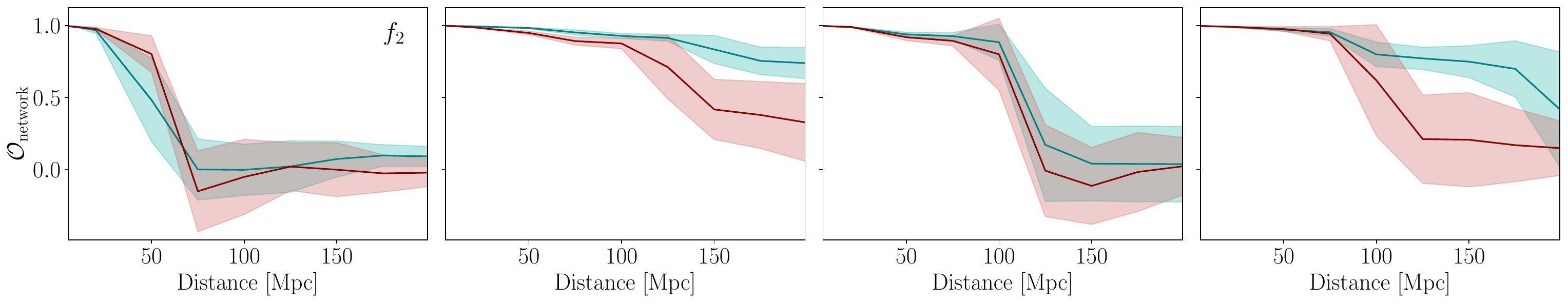}
    \includegraphics[width=\textwidth]{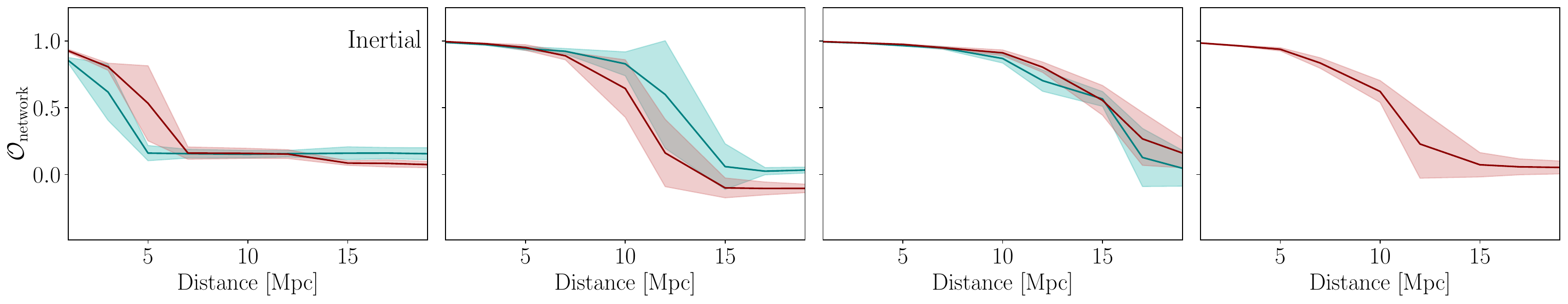}

    \caption{Detector network overlap between the injected and reconstructed signals as a function of the distance to the source. Top, middle, and bottom rows correspond to the $f_{2,i}$ modes, the $f_2$ modes, and the $f_{\rm inertial}$ modes, respectively. Each column corresponds to one EOS. Solid lines indicate the mean value over the waveform posterior distribution, and shaded areas are the standard deviations. The blue color corresponds to the hybrid version of the EOS and the red color to the tabulated version. Notice that the blue curve and shaded area are not shown for the {\tt LS220} EOS in the bottom-right plot, as the remnant for the hybrid version of this EOS collapses to a BH.}
    \label{fig:overlaps}
\end{figure*}

\begin{figure*}[t]
    \centering
    \includegraphics[width=\textwidth]{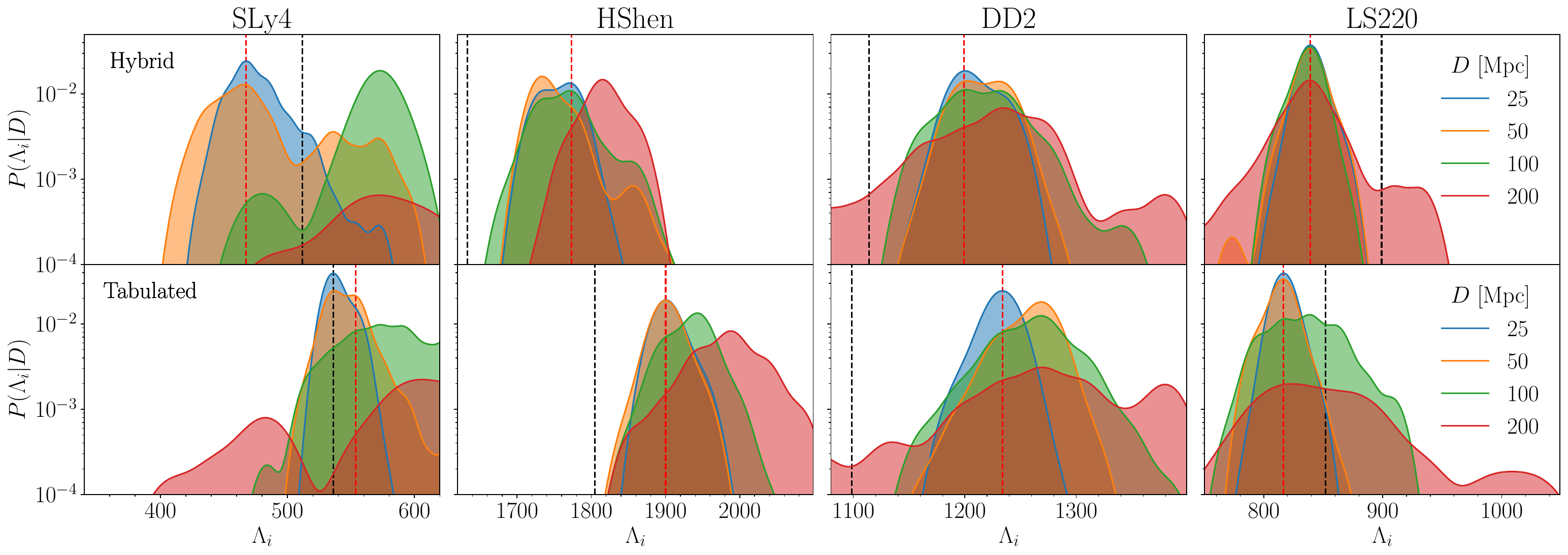}
    \includegraphics[width=0.765\textwidth,left]{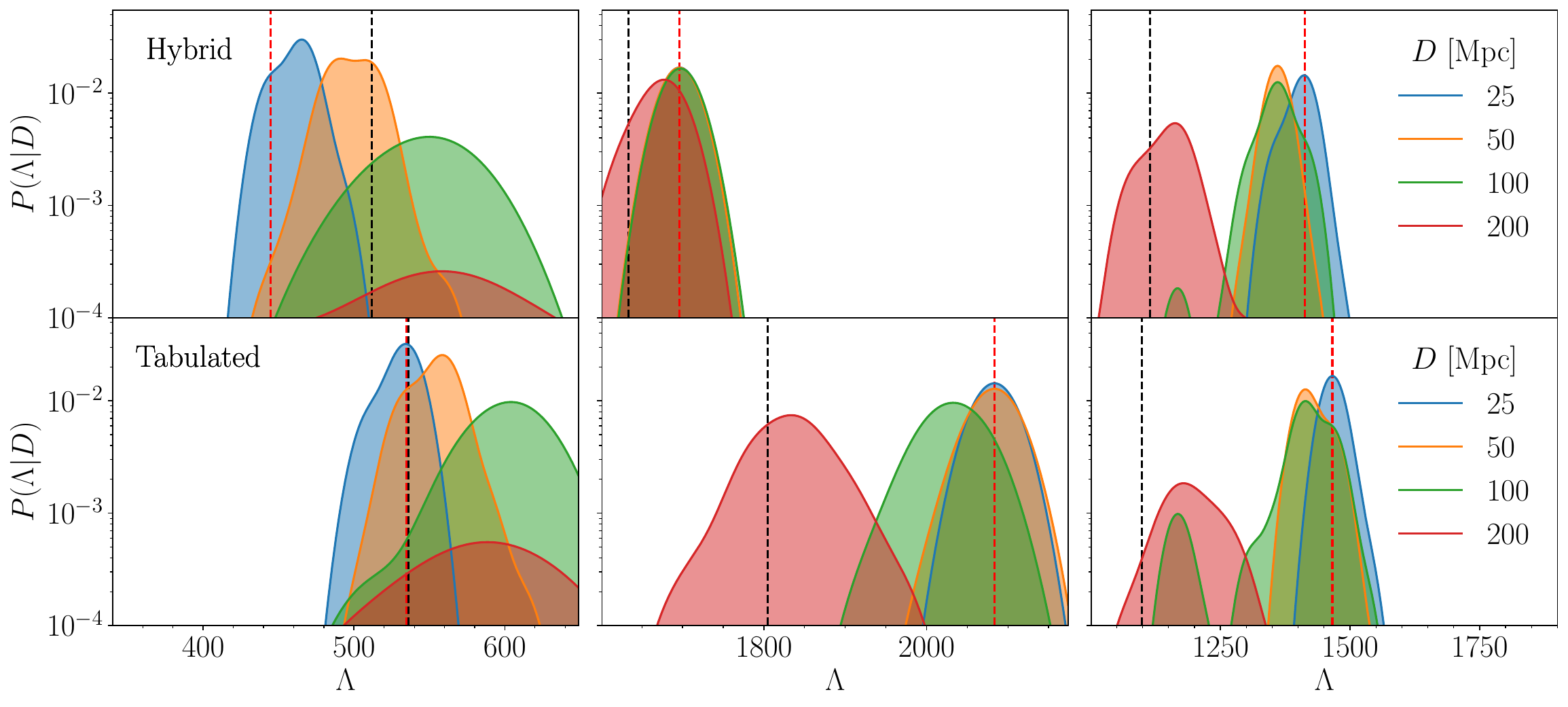}
    \caption{Posterior distributions of the tidal deformability parameter computed from the frequency peaks of the $f_{2,i}$ mode, $\Lambda_i$ (top row) and of the $f_2$ mode, $\Lambda$ (bottom row). Each column corresponds to a different EOS and different colors correspond to different distances to the source. The vertical red dashed line is the injected value of the parameter (obtained with the fit from~\cite{Guerra:2023}), and the vertical black dashed line is the true value shown in Table~\ref{table:Iparamenters} . We do not show the distributions for {\tt LS220} in the bottom row because the hybrid version collapses to a BH after the early postmerger phase. See the main text for details.}
    \label{fig:lambdas}
\end{figure*}

The middle panel of Fig.~\ref{fig:fpeaks} depicts the frequency peaks corresponding to the $f_2$ mode. These peaks are more difficult to recover than those of the $f_{2,i}$ mode, even though the differences between the hybrid and tabulated versions of the EOS are more prominent. For {\tt SLy4}, the recovery is inaccurate for distances $D \gtrsim 50$ Mpc, as the peaks of the posterior distributions are at significantly lower frequencies than the injected value, for both versions of the EOS. The corresponding $p$-values are 0.03 and 0.025 for the hybrid and tabulated versions, respectively, at $D = 100$ Mpc. On the other hand, {\tt HShen} is the EOS for which the peaks of the $f_2$ mode are best recovered, especially for the hybrid model, even at the largest distances considered. (This also holds for the case of the $f_{2,i}$ mode shown in the top panel.) For this EOS, there is a shift in the peak frequency of almost $200$ Hz between the hybrid and tabulated treatments of thermal effects, which corresponds to a difference of about $10\%$. This difference is large enough even for distances up to  $D \gtrsim 100 $ Mpc. As the distance to the source increases, the peaks for the tabulated version of the {\tt HShen} EOS are reconstructed at increasingly higher frequencies, to reach values that eventually almost overlap with the ones inferred for the hybrid case. The {\tt DD2} EOS also gives $f_2$ peaks at almost the same frequency for both versions of the EOS (only with a difference of $\approx 1.45\%$), as in the case of the $f_{2,i}$ mode shown in the top panel of Figure~\ref{fig:fpeaks}. However, the peaks of the $f_2$ mode are well recovered up to $D\approx 100$ Mpc,  lower than for the $f_{2,i}$ mode. Beyond this distance, {the mean of the distribution} starts differing more than 100 Hz from the injected value. For the tabulated version of the {\tt LS220} EOS, the $f_2$ mode frequency peak decreases only about 100 Hz compared to the $f_{2,i}$ mode. However, the hybrid version of this EOS displays a peak at a higher frequency. This can also be seen in the right panels in Fig.~\ref{fig:spectrograms}. This is due to the fact that the remnant  collapses to a BH only when the {\tt LS220} EOS implements a hybrid treatment of thermal effects. The signal amplitude of the $f_2$ mode looks large enough to be detectable for {\tt LS220} up to 200 Mpc.  Moreover, the posterior distributions for this mode do not overlap for distances of $D \leq 100$ Mpc.

In general, the hybrid and tabulated posterior distributions of the $f_2$ mode frequency peaks do not overlap as much as in the case  of the $f_{2,i}$ mode.This means that the \textit{hybrid} approach seems still valid at very early postmerger times. As the $f_2$ mode arises, the differences become larger and induce a significant bias between the approximate \textit{hybrid} approach and the self-consistent \textit{tabulated} one. However, since the signal amplitude decreases during this period, the uncertainty in the recovered peak frequency may hide the impact of this bias for some of the EOS considered. For example, while both posterior distributions of the $f_2$-mode frequency for the {\tt DD2} EOS still overlap even at $D = 25$ Mpc, the differences for the {\tt HShen} EOS are distinguishable up to $\sim 100$ Mpc. For {\tt SLy4}, the shift in the $f_2$-mode frequency is significant only for small distances to the source, since the signal amplitude is particularly lower for this EOS. Finally, the bias introduced by the \textit{hybrid} approach would also be detectable for {\tt LS220} up to $D \sim 100$ Mpc, as the peak of the posterior distribution for the tabulated version is over the left tail of the distribution of the hybrid version of this EOS. 

\subsubsection{\label{subsec:overlap_early} Overlap of the early postmerger phase }

The detector network overlap between the injected and reconstructed GW signals as a function of the distance to the source is displayed in Fig.~\ref{fig:overlaps} for our four EOSs. The top and middle rows in this figure correspond to the early postmerger phase (i.e.,~to the times reported in the first two columns in Table~\ref{table:time_windows}). The plots in the bottom row correspond to the late phase and will be discussed below. As noted in~\cite{Miravet:2023}, and in Appendix~\ref{sec::appendixB}, cases with an overlap value above~$\sim 0.75$ are considered faithful reconstructions, meaning the signal in these cases is detectable.

We start by considering the network overlap for the $f_{2,i}$ mode, shown in the top row of Fig.~\ref{fig:overlaps}. The blue colored line corresponds to the average value of the overlap for the hybrid version of the EOS and the red line to the tabulated version. The shaded regions are the standard deviations from the overlap posterior distributions. Each column corresponds to a different EOS. As expected, for lower distances the overlap is closer to one (perfect match). For all cases but the tabulated version of the {\tt LS220} EOS, the average value of the posterior distribution of the network overlap is over 0.5 up to distances of about 200 Mpc. The higher values of the overlap are obtained for the {\tt HShen} and for the {\tt DD2} EOS. No common trend for higher or lower values of the overlap depending on the treatment of thermal effects is observed across our EOS sample. 

The middle row of Fig.~\ref{fig:overlaps} depicts the corresponding network overlap for later postmerger times, in which the $f_2$ mode is dominant. In this case, the overlap at a given distance is lower than that achieved for the $f_{2,i}$ mode, for all EOSs. For the case of {\tt SLy4}, the average values of the posterior distributions fall abruptly below 0.5 for $D \gtrsim 50$ Mpc. As happens for the $f_{2,i}$ mode, {\tt HShen} also reaches the highest overlap for the $f_2$ mode, particularly for the hybrid version of this EOS, significantly larger than the value for the tabulated version. The latter reaches an overlap of about 0.5 at 150 Mpc. Regarding the {\tt DD2} EOS, values of the overlap higher than 0.75 are attained up to 100 Mpc. However, those values abruptly fall to zero overlap for larger distances. A similar trend is also observed for {\tt LS220} even though the hybrid version of this EOS yields higher overlap values for the $f_2$ mode than the tabulated version for significantly larger distances. For the $f_2$ mode, the network overlap seems larger for the hybrid case, as the signal amplitude is also slightly larger than for the tabulated approach. These findings suggest that, overall, the detectability prospects of the postmerger signal should be actually more conservative than those provided by the approximate hybrid approach.

\begin{figure*}[t]
\centering
   \includegraphics[width=0.49\textwidth]{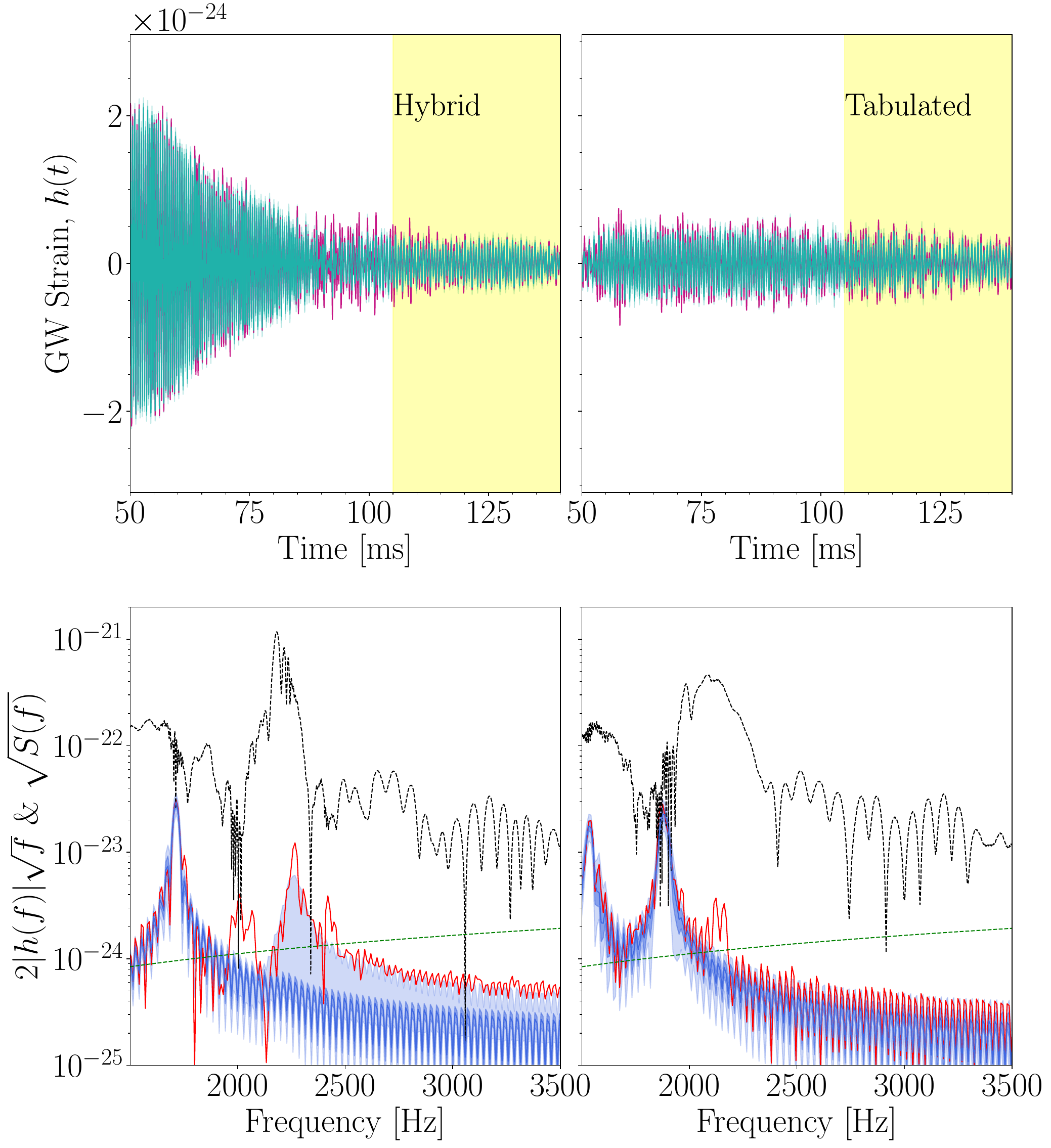}
   \includegraphics[width=0.49\textwidth]{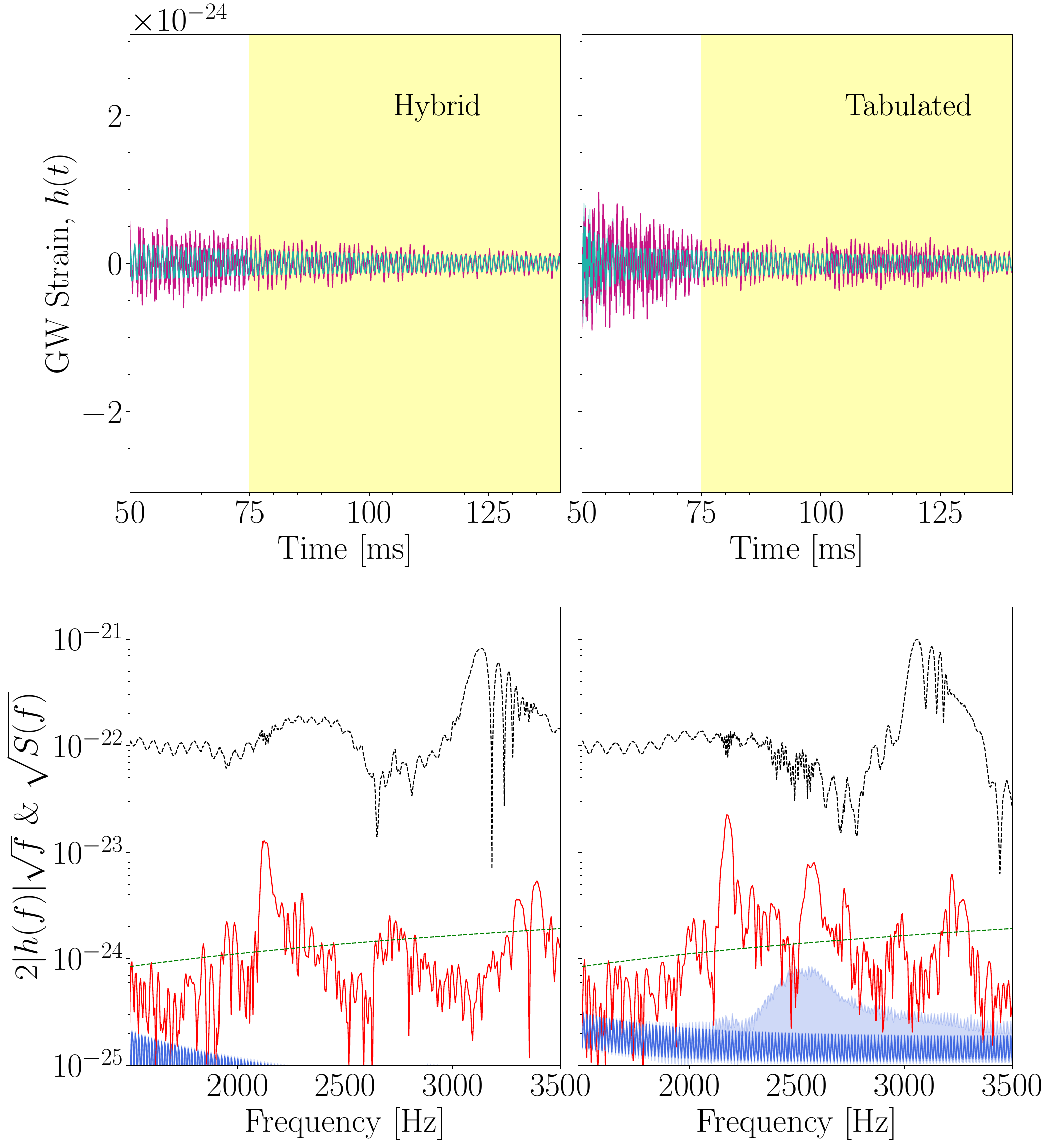}
   \caption{Waveforms (top row) and ASD (bottom row) of the injected (red) and reconstructed (blue) late postmerger GW signals for a source located at $D=7$ Mpc. The left panel corresponds to the {\tt HShen} EOS and the right panel to the {\tt SLy4} EOS. Within each panel, the left (right) column displays the hybrid (tabulated) version of the respective EOS. The black and green dotted lines in the bottom-row plots are the ASDs computed from the complete waveforms and the sensitivity curve of the ET detector, respectively.}
\label{fig:rec_HShen_SLy4_late}
\end{figure*}

\subsubsection{\label{subsec:tidal_def} Tidal deformability}

The frequency peaks of the early postmerger phase of the remnant can be used to infer properties of NSs by exploiting correlations with physical parameters through EOS-insensitive, quasiuniversal relations (see,~e.g.,~\cite{Read:2013, Bernuzzi:2014,Bauswein:2015,Takami:2015,Rezzolla:2016,Bauswein:2019, Bauswein:2019b,Soultanis:2022,Topolski:2023}). In particular, a number of empirical fits  between the frequencies of various modes (e.g.,~the peak frequency at merger, the $f_{2,i}$ mode, and the $f_2$ mode) and the tidal deformability parameter $\Lambda$ characterizing the quadrupole deformability of an isolated NS have been proposed (see~\cite{Soultanis:2022,Topolski:2023} and references therein for up-to-date revisions of existing literature). In~\cite{Guerra:2023}, we present new fits of the frequencies of the $f_{2,i}$ and $f_2$ modes to the tidal deformability parameter using our set of EOSs. We note that those quasiuniversal relations are built using hybrid EOSs only since the number of simulations with tabulated EOSs is not large enough to yield a meaningful fit.
However, their validity when applied to simulations with tabulated EOSs can be tested, using the standard deviation of the correlation for the hybrid EOS as a reference metric. For the $f_{2,i}$ and $f_2$ modes, the standard deviation is $67.54$ and $97.56$ Hz, respectively~\cite{Guerra:2023}. Using those fits we discuss here the posterior probability distributions of the tidal deformability parameter obtained from both the frequencies of the $f_{2,i}$ and $f_2$ modes and for the two different treatments of thermal effects.

In the top row of Fig.~\ref{fig:lambdas} we show the results for the $f_{2,i}$ mode, for different distances and all four EOSs. The distributions displayed are built using the empirical relations from~\cite{Guerra:2023}. The upper panels correspond to the hybrid version of the EOSs and the lower panels to the tabulated version. Since $\Lambda_i$ is directly calculated from $f_{2,i}$, the behavior of the posterior distributions of the two quantities with the distance is the same (cf.~uppermost row in Fig.~\ref{fig:fpeaks}). Our results indicate that $\Lambda_i$ could be reconstructed up to $D\approx 200$ Mpc for all EOS except for {\tt SLy4}, for which the reconstruction is acceptable only up to $D\lesssim 100$ Mpc. For the {\tt SLy4} EOS (first column), the distributions of $\Lambda_i$ are closer to the injected value (red vertical dashed line) for the tabulated version. On the other hand, {\tt HShen}, {\tt DD2} and {\tt LS220} yield a good recovery of $\Lambda_i$ for all the distances shown. Their $p$-values at 200 Mpc are above 0.15 for both versions of the EOS. 

The tidal deformability parameter can  also be computed using the frequency of the  $f_2$ mode. To do so, we use the fits for this mode presented in~\cite{Guerra:2023}. The {\tt LS220} EOS is discarded in this analysis because at the postmerger times considered the evolution of the remnant when using the hybrid version of this EOS already shows the formation of a BH. The differences on the distributions of $\Lambda$ between the hybrid and tabulated versions of the EOSs for the $f_2$ mode are displayed in the bottom row in Fig.~\ref{fig:lambdas}. As expected, the posterior distributions are similar to those obtained with the $f_{2}$ mode (middle row in Fig.~\ref{fig:fpeaks}). The most striking difference is that, for the tabulated version of the {\tt HShen} EOS, $\Lambda$ is inaccurately inferred at $D = 200$ Mpc. 

As stated in Sec.~\ref{sec:Numerical_set}, using tabulated or hybrid EOSs leads to slightly different initial NS configurations and, thus, to different values of the tidal deformability. This is why the vertical black lines displayed in Fig.~\ref{fig:lambdas}, corresponding to the values of $\Lambda$ reported in Table~\ref{table:Iparamenters}, are not the same for the two approaches for the EOS. This figure shows that the empirical fits of~\cite{Guerra:2023} can be also applied to simulations with tabulated EOS in most cases, as the variations in frequency are within 2$\sigma$ from the mean of the correlation. In general, we find that the fit for the $f_{2,i}$ mode (red vertical lines) is closest to the ``true'' value from the simulation. We observe that the tidal deformability might be detected up to several tens of Mpc for all EOSs (when computed from empirical fits built for hybrid models only).

\subsection{\label{sec:late_pm} Late postmerger phase: Inertial modes}

At later postmerger times than those considered in the preceding section ($t \gtrsim 50$ ms) the amplitude of the $f_2$ mode decreases and convective instabilities in the interior of the remnant set in (see~\cite{DePietri:2018,DePietri:2020,Guerra:2023}). Those trigger the excitation of inertial modes whose dynamics leave an imprint in the late postmerger signal.
Inertial modes attain smaller amplitudes than the modes from the early postmerger phase and their frequency peaks in the spectra are also lower than those of the $f_{2,i}$ and $f_2$ modes.

\subsubsection{\label{subsec:reco_late} Waveform reconstructions}

Figure~\ref{fig:rec_HShen_SLy4_late} shows the nonwhitened, time-domain signal
(top row) and the ASD (bottom
row) of the injected (red) and reconstructed (blue) late postmerger GW signals (with the detector ASD), for a source located at a distance of 7 Mpc. The left panel shows the waveforms and ASDs for the {\tt HShen} EOS while the right panel displays the corresponding quantities for the case of the {\tt SLy4} EOS. Within each panel, the left (right) column correspond to the hybrid (tabulated) version of the respective EOS. As before, the ASDs shown in the bottom row have been computed by Fourier transforming the waveforms in the time windows highlighted in yellow in the plots in the top row (see also Table~\ref{table:time_windows}). Likewise, the blue-shaded regions in the ASD
plots in both figures show the 50\% and 90\% CI of the distribution of the reconstructed waveforms.

At the distance considered and regardless of the treatment of thermal effects, the reconstructions of the late-time signals {\tt BayesWave} produces are accurate only for the {\tt HShen} EOS. This is due to the small amplitude of the late postmerger signal in the case of {\tt SLy4}. The frequency peak of the dominant inertial mode for this EOS is located around 2.2 kHz (see the ASD of the injected signal, colored in red). Despite the peak amplitude being above the sensitivity curve of the ET detector, {\tt BayesWave} cannot correctly capture it, as apparent from the CI of the reconstructed distributions. Regarding the {\tt HShen} EOS, the dominant frequency peak of the inertial modes is located below 2 kHz for both versions of the EOS. In the case of the tabulated version two peaks are visible around 1.5 and 2 kHz, while for the hybrid version of this EOS those two peaks appear at frequencies around 1.75 and 2.5 kHz. Notice that the peak located at around 2 kHz for the tabulated EOS is actually the $f_2$ mode. This mode is not yet completely damped at this late postmerger time (see the several frequencies that are excited at late times in Fig.~\ref{fig:spectrograms}). Therefore, we do not consider it when computing the posterior probability for the frequency peaks in the last row in Fig.~\ref{fig:fpeaks}. The same explanation holds for the peak at around 2.5 kHz in the hybrid case (see~\cite{Guerra:2023} for more details).

\subsubsection{\label{subsec:fp_late} Frequency peaks of the inertial modes}

The posterior distributions of the frequency peaks of the inertial modes are displayed in the bottom row in Fig.~\ref{fig:fpeaks} for all EOSs except {\tt LS220} (as the simulation with the hybrid version of this EOS collapses to a BH at early postmerger times). The upper (lower) rows in this figure represent the hybrid (tabulated)
versions of the corresponding EOS. The injections, whose frequencies are depicted by dashed vertical lines, are now performed at much shorter distances than we did for the $f_{2,i}$ and $f_2$ modes, due to the smaller amplitude of the inertial modes. The largest distance considered is now $D=15$ Mpc.

The characteristic peak frequencies of the inertial modes, $f_{\rm inertial}$, are lower than the ones from the fundamental modes, $f_{2,i}$ and $f_2$, as can be seen by direct comparison in Fig.~\ref{fig:fpeaks}. As for the quadrupolar modes, inertial modes also display a shift in frequencies depending on the particular treatment of thermal effects in the EOS. This shift appears to be sufficiently large only for the case of the {\tt HShen} EOS (middle panel) as the posterior distributions of the hybrid and the tabulated versions of the EOS do not overlap up to $D\approx 10$ Mpc. On the other hand, the small amplitude of the late postmerger signal for the {\tt SLy4} EOS is too low to yield a good reconstruction unless the source is located at a distance of less than 5 Mpc. Only for such short distances might the frequency shift in the posterior distributions be distinguished.

\subsubsection{\label{subsec:overlap_late} Overlap of the inertial modes}

As we did before for the $f_{2,i}$ and $f_2$ modes, we also use the network overlap function
to assess the reconstruction of the waveforms for the case of inertial modes. Those overlaps are shown in the bottom row in Fig.~\ref{fig:overlaps}. For the {\tt SLy4} EOS (first column), the overlap is above 0.5 for a distance to the source of less than 5 Mpc, with the tabulated version of the EOS attaining a higher value of the overlap for slightly larger distances. For both implementations of the thermal effects, the average overlap falls below 0.25 for distances above $\approx 7$ Mpc, which means that the injected and reconstructed waveforms differ considerably. Correspondingly, the waveform signals for {\tt HShen} and {\tt DD2} (second and third columns, respectively) are still reconstructed with a network overlap over 0.5 up to a distance of about 12 Mpc (15 Mpc for the case of {\tt DD2}). The tabulaled version of the {\tt HShen} EOS is more poorly recovered than its hybrid counterpart, with a smaller overlap at about 10 Mpc. Finally, in the case of the {\tt LS220} EOS (fourth column), the network overlap is above 0.5 for distances up to 10 Mpc. Notice that only the overlap of the tabulated version of {\tt LS220} is plotted in Fig.~\ref{fig:overlaps}, as the simulation with a hybrid EOS collapses to a BH before inertial modes have been excited.

\section{Conclusions}
\label{sec:discussion}

Numerical simulations of BNS mergers incorporate thermal effects in the EOS using two alternative approaches. The first one is a {\it hybrid} approach which assumes that the pressure and the internal energy are composed of two constituents, a cold, zero-temperature part described by a family of piecewise polytropes and a thermal part described by an ideal-gas-like EOS. The second approach employs {\it tabulated} representations of microphysical
finite-temperature EOSs, providing a self-consistent method to probe the impact of thermal effects in the merger dynamics. While the former is an approximation, it is nonetheless a widely employed approach as it reduces computational costs. Remarkably, the two procedures lead to measurable differences in the postmerger dynamics and GW emission, the latter standing out in the location of characteristic frequency peaks in the spectra (see,~e.g.,~\cite{Bauswein:2010dn, Guerra:2023}). Being the tabulated treatment of the EOS free (in principle) of any approximation, it is fair to associate those differences to the bias introduced by the hybrid approach.

In this paper, we have investigated the prospects for identifying such biases by reconstructing the GW signals of~\cite{Guerra:2023} using {\tt BayesWave}~\cite{Cornish:2015,Littenberg:2015}, building on our previous work in~\cite{Miravet:2023} where we focused on inertial modes only, excited in the very late part of the postmerger signal. Here, we have considered the entire postmerger signal, i.e.,~both its early part where the fundamental quadrupolar $f_2$ mode dominates the GW spectrum and its late part where inertial modes are excited. The time-domain waveforms of~\cite{Guerra:2023}, obtained through BNS merger simulations with four different EOSs, accounting for both descriptions of thermal effects, have been injected into Gaussian noise given by the sensitivity of the third-generation detector Einstein Telescope~\cite{ET:2010, Hild:2011}, selecting optimal sky location and inclination. Results for nonoptimal configurations are discussed in Appendix~\ref{sec:appendixA}. The capability of {\tt BayesWave} to reconstruct the injected signals has been assessed by computing the overlap function of the detector network. As the postmerger remnant evolves, the amplitude of the GW signal significantly decreases, resulting in a corresponding reduction of the overlap between injected and reconstructed waveforms. The same occurs as the distance to the source increases, irrespective of the portion of the postmerger signal being analyzed.

The two representations of thermal effects in the EOS result in frequency shifts of the dominant peaks in the GW spectra. In some cases, those differences are large enough to be told apart in the recovered signal, especially in the early postmerger phase, when the signal amplitude is the loudest, and at sufficiently small distances. They have been found to strongly depend on the EOS. Both the {\tt SLy4} EOS (at small enough distances) and the {\tt HShen} EOS (at significantly bigger distances) present large frequency shifts of the dominant $f_{2,i}$ and $f_2$ modes. 
On the other hand, for the {\tt DD2} and {\tt LS220} EOSs, no large enough frequency shifts between the hybrid and tabulated cases have been found to unambiguously differentiate with {\tt BayesWave} the treatment of thermal effects in the EOS, especially in the very early postmerger phase in which the $f_{2,i}$ mode is excited. The bias introduced by the hybrid approach seems to strongly depend on the  EOS. In some cases, this approximation may lead to incorrect guesses of the peak frequency even for source distances larger than $\sim 100$ Mpc.

Differences in the dominant peaks of the GW spectra are still present during the late postmerger phase, where the inertial modes dominate~\citep{Kastaun:2008,DePietri:2018,DePietri:2020,Guerra:2023}. These modes are associated with a part of the GW signal with a much lower amplitude than that of the $f_{2,i}$ and $f_2$ modes. Therefore, they are more difficult to detect~\cite{Miravet:2023}. Our results indicate that third-generation detectors such as ET may  be able to observe inertial modes  up to a distance of about 10 Mpc, depending on the EOS. For this late-time part of the signal, the shift in the peak frequency due to the different treatment of thermal effects can be above 200 Hz at most, for the case of the {\tt HShen} EOS. On the other hand, for the {\tt LS220} EOS, the difference is more obvious: The hybrid version of this EOS leads to the collapse of the remnant to a BH, as opposed to the tabulated version of the same EOS~\cite{Guerra:2023}. 
Notice that this frequency shift may potentially introduce a degeneracy with the EOS: Two different EOSs may potentially peak at the same frequency. Consequently, accurate modeling of thermal effects is essential for properly constraining the EOS.

The overlap function is also affected by the treatment of thermal effects. In most cases, the amplitude of the postmerger signal is lower for the tabulated approach, which leads to a worse signal reconstruction for a given source distance. This suggests that, broadly speaking, the range of detectability of the postmerger signal should be more conservative than that provided by the approximate hybrid approach. The latter overestimates the thermal pressure of the remnant, which leads to an artificially larger GW amplitude that can mistakenly translate into a wider detectability range.

Finally, we have also computed the tidal deformability from the frequency peaks of both the $f_{2,i}$ and $f_2$ modes and through the empirical fits presented in~\cite{Guerra:2023}. The differences in thermal effects between the hybrid and the tabulated EOS inferred through the analysis of the tidal deformability parameter are also more apparent for the $f_2$ mode, since the shift in the frequency peaks is more pronounced. However, due to the uncertainties arising from the fits, the recovered values of the tidal deformability from both approaches are compatible. 

The results of the work reported here are consistent with those recently presented by~\cite{Calderon-Bustillo:2023}, who employed Bayesian model selection to explore differences between the hybrid and the tabulated approaches for the same set of GW signals. The differences in the posterior distributions of the main frequency peaks in the early postmerger GW spectra in hybrid and tabulated models reported here may be resolved in third-generation detectors up to distances of about tens of Mpc, compatible with the values found by~\cite{Calderon-Bustillo:2023}. These results indicate that a self-consistent treatment of thermal effects in numerical-relativity simulations of BNS mergers seems mandatory to enhance the suitability of simulated postmerger waveforms in the detection prospects of future GW searches.

A few caveats remain. First, our binary models in Table~\ref{table:Iparamenters} are constructed to be as similar as numerically possible. However, there are inherent differences in the way the models are built (e.g.,~the minimum temperature value in the available tables is $T=0.01\rm\,MeV$, which affects the density distribution at low densities), making it impossible to create identical stars~(see Fig.~\ref{fig:MR}). Consequently, we generated initial data that are slightly different but minimized these differences as much as possible. We note that these configurations differ by less than $0.2\%$ on the gravitational mass, by less than
$2\%$ on the circumferential, and by $\lesssim 5\%$ on the tidal deformability number, among others.  These differences are comparable to the intrinsic error in the numerical simulations~\cite{Ruiz:2020via,Etienne:2011ea} and, thus, are unlikely to account for the differences observed in our studies. Second, recent efforts have focused on capturing the leading-order effects of degenerate matter on thermal pressure using phenomenological schemes (see,~e.g.,~\cite{Raithel:2019gws,Mroczek:2024sfp}). In particular, the approach reported in~\cite{Raithel:2019gws} reproduces thermal pressure within $30\%$ accuracy for some realistic EOSs at the typical densities in BNS mergers, while its temperature agrees with that from fully tabulated EOSs within $10\%$. Our numerical results indicate that, for certain EOSs, the hybrid approach may perform comparably to, or even as well as, these newer schemes. Therefore, further numerical studies are required to fully assess the advantage of these schemes over the hybrid approach.

This work has used the following open-source packages: \textsc{NumPy}~\cite{harris:2020}, \textsc{SciPy}~\cite{scipy:2020}, \textsc{Scikit-learn}~\cite{scikit-learn}, \textsc{Matplotlib}~\cite{Hunter:2007}, and \textsc{PyCBC}~\cite{pycbc}.

\begin{acknowledgments}

The authors thank Roberto De Pietri for a careful reading of the manuscript and Micaela Oertel for useful comments. This work has been supported by the Generalitat Valenciana through Grants No. CIDEGENT/2021/046 and Prometeo No. CIPROM/2022/49, by MCIN and Generalitat Valenciana with funding from European Union NextGenerationEU (PRTR-C17.I1, Grant ASFAE/2022/003), and by the Spanish Agencia Estatal de  Investigaci\'on through Grants No. PRE2019-087617 and No. PID2021-125485NB-C21 funded by MCIN/AEI/10.13039/501100011033 and ERDF A way of making Europe. Further support has been provided by the EU's Horizon 2020 Research and Innovation (RISE) program H2020-MSCA-RISE-2017 (FunFiCO-777740) and  by  the  EU  Staff  Exchange  (SE)  program HORIZON-MSCA-2021-SE-01 (NewFunFiCO-101086251). M.M.-T. acknowledges support from the Ministerio de Ciencia, Innovación y Universidades del Gobierno de España through the
“Ayuda para la Formación de Profesorado Universitario" (FPU) fellowship No.~FPU19/01750. D.G. acknowledges support from the Spanish Agencia Estatal de  Investigaci\'on through Grant No. PRE2019-087617.

The authors acknowledge the computational resources and technical support of the Spanish Supercomputing Network through the use of MareNostrum at the Barcelona Supercomputing Center (AECT-2023-1-0006) where the BNS merger simulations were performed, the computational
resources provided by the LIGO Laboratory and supported by
National Science Foundation Grants No. PHY-0757058 and No. PHY-0823459, and the resources from the Gravitational Wave Open Science Center, a service of the LIGO Laboratory, the LIGO Scientific Collaboration and the Virgo Collaboration. 
\end{acknowledgments}

\section*{DATA AVAILABILITY}

The data used in this paper can be made available on reasonable request to the authors.

\appendix

\section{Reconstruction of injections with nonoptimal sky location and orientation 
\label{sec:appendixA}}

The injections discussed in the main text of this paper were performed considering an optimal source inclination with respect to the ET detector $(\iota = 0)$ and an optimal sky location, with a right ascension of 2.9109 rad and a declination of 0.7627 rad. Therefore, the results represent the best-case scenario for a given source distance. However, in reality the source can be anywhere in the sky and have an arbitrary declination. Hence, the effective distance to the source can be actually larger. This possibility is briefly discussed in this appendix.

\begin{center}
  \begin{table}[h]
  \caption{Overlap functions for different inclinations and sky locations for a distance $D = 150$ Mpc to the source. The first (second) row corresponds to the mode  $f_{2,i}$ ($f_2$). In parentheses we show the percentage value with respect to the overlap for the optimal case.
      \label{table:non_opt}}
    \begin{tabular}{ccccc}
        \hline
        \hline
       Mode & Optimal $\iota$    & Nonoptimal $\iota$   & Optimal $\iota$  & Nonoptimal $\iota$ \\
       & Optimal & Optimal & Nonoptimal & Nonoptimal  \\
       & sky loc & sky loc  &  sky loc &  sky loc \\
      \hline
        $f_{2,i}$    &   0.803    &  0.650 (81.0 \%)   &  0.656 (81.7 \%) & 0.630 (78.5 \%)  \\
        $f_2$         &  0.835    &  0.791 (94.7 \%)  & 0.793 (95.0 \%)  &  0.728 (87.2 \%)  \\
        \hline
        \hline
    \end{tabular}
  \end{table}
\end{center}

In Table~\ref{table:non_opt} we report the value of the overlap function for the fundamental quadrupolar frequency peaks for a source at a distance $D = 150$ Mpc and for different combinations of sky locations and inclinations. We consider only the hybrid version of the {\tt HShen} EOS as this is the one yielding the best detectability prospects in the optimal case. The nonoptimal inclination is set to $\iota = 0.5585$ rad and the right ascension and declination in the sky for a representative nonoptimal case are chosen to be 3.4462 and 0.45 rad, respectively. As expected, the overlap function decreases with respect to the optimal case. The lowest values found are $78.5 \%$ (with respect to the optimal case) and $87.2\%$ for the $f_{2,i}$ and $f_2$ modes, respectively. Therefore, for signals coming from a nonoptimal sky location and/or from a source with a nonoptimal inclination, the effective distance will not be much larger than the optimal case. Furthermore, we note that the effect of the actual sky position of the source will become less of a concern if a network of detectors built in different locations is used.

\section{Signal-to-noise ratio of the injected signals}
\label{sec::appendixB}

\begin{figure}[t]
    \centering 
    \includegraphics[width=\linewidth]{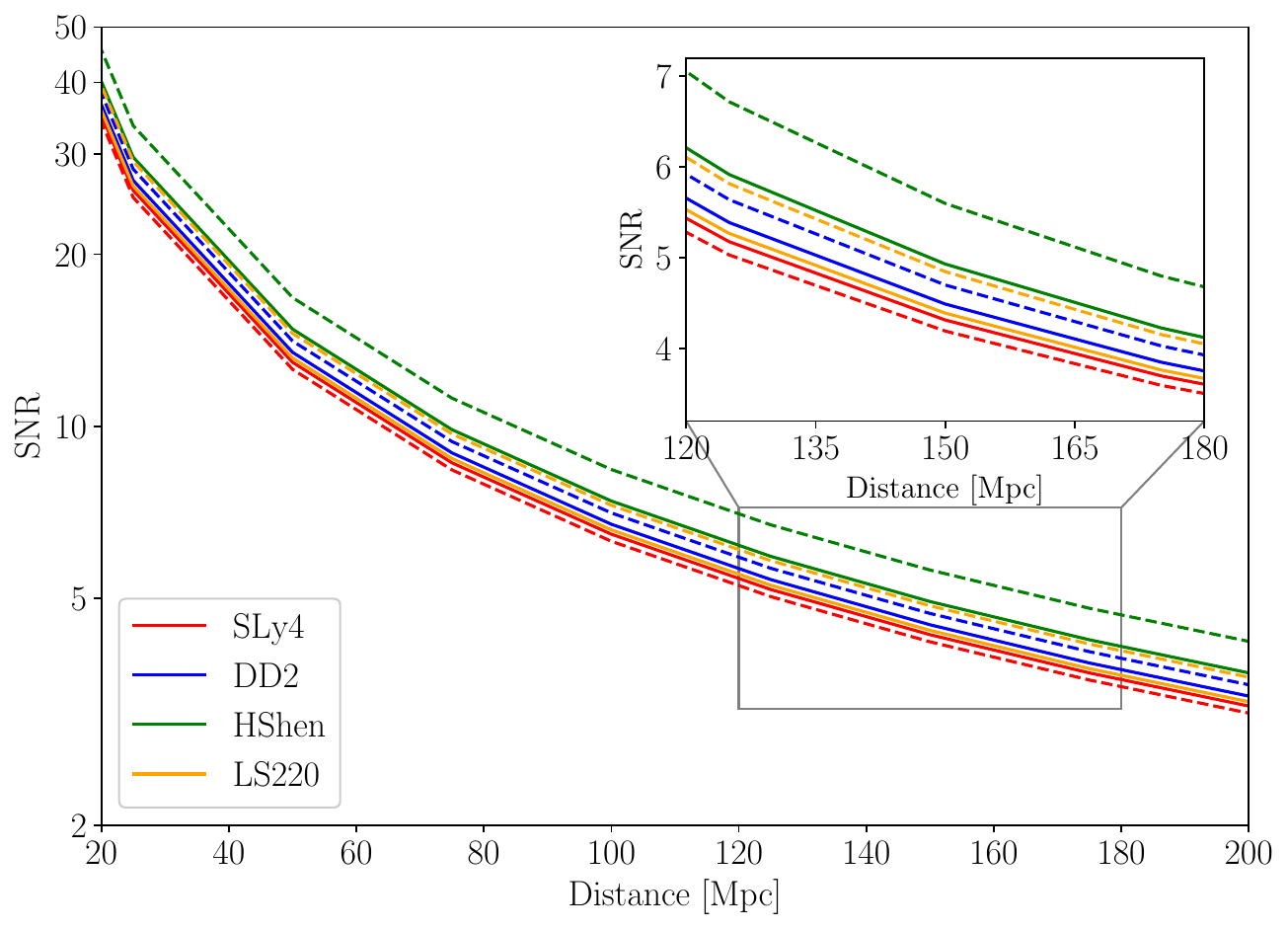}
    \caption{Values of the SNR for high-frequency signals injected into the ET detector as a function of the distance to the GW source, which is optimally oriented. The different colors represent different EOSs. The solid lines correspond to the tabulated versions, whereas the dashed curves refer to the hybrid cases. As distance increases, the SNR decreases, and there is a slight dependence on the EOS considered.}
    \label{fig:snr_app}
\end{figure}

\begin{figure*}[t]
    \centering 
    \includegraphics[width=\textwidth]{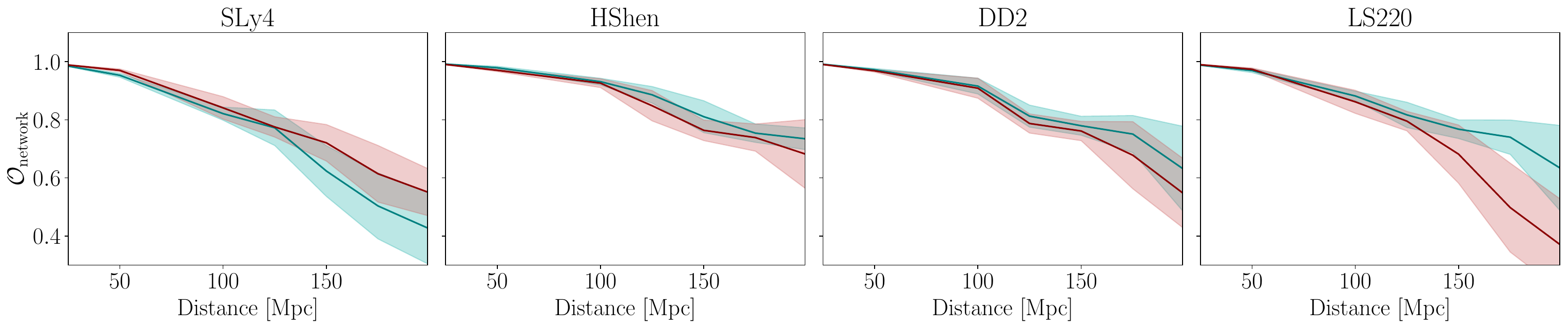}
    \caption{The same as Fig~\ref{fig:overlaps}, but the network overlap function corresponds to high-frequency signals injected into the ET detector, from 10 ms before merger to $\sim$30 ms after merger.} 
    \label{fig:ov_app}
\end{figure*}

The signal-to-noise ratio (SNR) is a quantity widely used to measure the excess power of the signal with respect to the detector noise. The squared value can be expressed as the inner product of the FFT of the strain [see Eq.~\eqref{inner_prod}]:
\begin{equation}\label{snr}
    \rho^2 = \langle \Tilde{h}(f) | \Tilde{h}(f)\rangle\,.
\end{equation}

In this appendix, we compute this quantity corresponding to each distance to the GW source. We compute the SNR for the ET detector using the last 10 ms before merger and up to $\sim$30 ms after merger, capturing both the $f_{2,i}$ and the $f_2$ modes. We put a low-frequency threshold of 1096 Hz to remove the contribution of the inspiral phase. In Fig.~\ref{fig:snr_app}, we depict the resulting SNR for different distances $D$ to the optimally oriented source. The values of the SNR are consistent with our findings. The signals from the {\tt HShen} EOS have a higher SNR, and their detection range is found to be larger. In contrast, the lower SNR of {\tt SLy4} is also consistent with our results. 

We also compute the network overlap $\mathcal{O}_{\rm network}$ for this time window, depicted in Fig.~\ref{fig:ov_app}. The EOSs with a higher overlap also possess a higher SNR. When the overlap function goes below the threshold value ($\sim0.75$), the SNR of the signal is $\sim4$. For lower values of the SNR, the reconstruction performance fails abruptly.  In all cases, we reach an overlap above $\sim0.9$ for values of the SNR above $\sim5-6$, as in~\cite{Chatz:2017}.

\clearpage
\bibliography{refs}

\begin{thebibliography}{106}
\expandafter\ifx\csname natexlab\endcsname\relax\def\natexlab#1{#1}\fi
\expandafter\ifx\csname bibnamefont\endcsname\relax
  \def\bibnamefont#1{#1}\fi
\expandafter\ifx\csname bibfnamefont\endcsname\relax
  \def\bibfnamefont#1{#1}\fi
\expandafter\ifx\csname citenamefont\endcsname\relax
  \def\citenamefont#1{#1}\fi
\expandafter\ifx\csname url\endcsname\relax
  \def\url#1{\texttt{#1}}\fi
\expandafter\ifx\csname urlprefix\endcsname\relax\def\urlprefix{URL }\fi
\providecommand{\bibinfo}[2]{#2}
\providecommand{\eprint}[2][]{\url{#2}}

\bibitem[{\citenamefont{{Abbott}
  et~al.}(2019{\natexlab{a}})\citenamefont{{Abbott}, {Abbott}, {Abbott},
  {Abraham}, {Acernese}, {Ackley}, {Adams}, {Adhikari}, {Adya}, {Affeldt}
  et~al.}}]{GWTC-1}
\bibinfo{author}{\bibfnamefont{B.~P.} \bibnamefont{{Abbott}}},
  \bibinfo{author}{\bibfnamefont{R.}~\bibnamefont{{Abbott}}},
  \bibinfo{author}{\bibfnamefont{T.~D.} \bibnamefont{{Abbott}}},
  \bibinfo{author}{\bibfnamefont{S.}~\bibnamefont{{Abraham}}},
  \bibinfo{author}{\bibfnamefont{F.}~\bibnamefont{{Acernese}}},
  \bibinfo{author}{\bibfnamefont{K.}~\bibnamefont{{Ackley}}},
  \bibinfo{author}{\bibfnamefont{C.}~\bibnamefont{{Adams}}},
  \bibinfo{author}{\bibfnamefont{R.~X.} \bibnamefont{{Adhikari}}},
  \bibinfo{author}{\bibfnamefont{V.~B.} \bibnamefont{{Adya}}},
  \bibinfo{author}{\bibfnamefont{C.}~\bibnamefont{{Affeldt}}},
  \bibnamefont{et~al.}, \bibinfo{journal}{Physical Review X}
  \textbf{\bibinfo{volume}{9}}, \bibinfo{eid}{031040}
  (\bibinfo{year}{2019}{\natexlab{a}}), \eprint{1811.12907}.

\bibitem[{\citenamefont{{Abbott} et~al.}(2021)\citenamefont{{Abbott}, {Abbott},
  {Abraham}, {Acernese}, {Ackley}, {Adams}, {Adams}, {Adhikari}, {Adya},
  {Affeldt} et~al.}}]{GWTC-2}
\bibinfo{author}{\bibfnamefont{R.}~\bibnamefont{{Abbott}}},
  \bibinfo{author}{\bibfnamefont{T.~D.} \bibnamefont{{Abbott}}},
  \bibinfo{author}{\bibfnamefont{S.}~\bibnamefont{{Abraham}}},
  \bibinfo{author}{\bibfnamefont{F.}~\bibnamefont{{Acernese}}},
  \bibinfo{author}{\bibfnamefont{K.}~\bibnamefont{{Ackley}}},
  \bibinfo{author}{\bibfnamefont{A.}~\bibnamefont{{Adams}}},
  \bibinfo{author}{\bibfnamefont{C.}~\bibnamefont{{Adams}}},
  \bibinfo{author}{\bibfnamefont{R.~X.} \bibnamefont{{Adhikari}}},
  \bibinfo{author}{\bibfnamefont{V.~B.} \bibnamefont{{Adya}}},
  \bibinfo{author}{\bibfnamefont{C.}~\bibnamefont{{Affeldt}}},
  \bibnamefont{et~al.}, \bibinfo{journal}{Physical Review X}
  \textbf{\bibinfo{volume}{11}}, \bibinfo{eid}{021053} (\bibinfo{year}{2021}),
  \eprint{2010.14527}.

\bibitem[{\citenamefont{{The LIGO Scientific Collaboration}
  et~al.}(2021)\citenamefont{{The LIGO Scientific Collaboration}, {the Virgo
  Collaboration}, {Abbott}, {Abbott}, {Acernese}, {Ackley}, {Adams},
  {Adhikari}, {Adhikari}, {Adya} et~al.}}]{GWTC-2.1}
\bibinfo{author}{\bibnamefont{{The LIGO Scientific Collaboration}}},
  \bibinfo{author}{\bibnamefont{{the Virgo Collaboration}}},
  \bibinfo{author}{\bibfnamefont{R.}~\bibnamefont{{Abbott}}},
  \bibinfo{author}{\bibfnamefont{T.~D.} \bibnamefont{{Abbott}}},
  \bibinfo{author}{\bibfnamefont{F.}~\bibnamefont{{Acernese}}},
  \bibinfo{author}{\bibfnamefont{K.}~\bibnamefont{{Ackley}}},
  \bibinfo{author}{\bibfnamefont{C.}~\bibnamefont{{Adams}}},
  \bibinfo{author}{\bibfnamefont{N.}~\bibnamefont{{Adhikari}}},
  \bibinfo{author}{\bibfnamefont{R.~X.} \bibnamefont{{Adhikari}}},
  \bibinfo{author}{\bibfnamefont{V.~B.} \bibnamefont{{Adya}}},
  \bibnamefont{et~al.}, \bibinfo{journal}{arXiv e-prints}
  \bibinfo{eid}{arXiv:2108.01045} (\bibinfo{year}{2021}), \eprint{2108.01045}.

\bibitem[{\citenamefont{{Abbott} et~al.}(2023)\citenamefont{{Abbott}, {Abbott},
  {Acernese}, {Ackley}, {Adams}, {Adhikari}, {Adhikari}, {Adya}, {Affeldt},
  {Agarwal} et~al.}}]{GWTC-3}
\bibinfo{author}{\bibfnamefont{R.}~\bibnamefont{{Abbott}}},
  \bibinfo{author}{\bibfnamefont{T.~D.} \bibnamefont{{Abbott}}},
  \bibinfo{author}{\bibfnamefont{F.}~\bibnamefont{{Acernese}}},
  \bibinfo{author}{\bibfnamefont{K.}~\bibnamefont{{Ackley}}},
  \bibinfo{author}{\bibfnamefont{C.}~\bibnamefont{{Adams}}},
  \bibinfo{author}{\bibfnamefont{N.}~\bibnamefont{{Adhikari}}},
  \bibinfo{author}{\bibfnamefont{R.~X.} \bibnamefont{{Adhikari}}},
  \bibinfo{author}{\bibfnamefont{V.~B.} \bibnamefont{{Adya}}},
  \bibinfo{author}{\bibfnamefont{C.}~\bibnamefont{{Affeldt}}},
  \bibinfo{author}{\bibfnamefont{D.}~\bibnamefont{{Agarwal}}},
  \bibnamefont{et~al.}, \bibinfo{journal}{Physical Review X}
  \textbf{\bibinfo{volume}{13}}, \bibinfo{eid}{011048} (\bibinfo{year}{2023}),
  \eprint{2111.03634}.

\bibitem[{\citenamefont{Abbott
  et~al.}(2017{\natexlab{a}})}]{LIGOScientific:2017pwl}
\bibinfo{author}{\bibfnamefont{B.~P.} \bibnamefont{Abbott}}
  \bibnamefont{et~al.} (\bibinfo{collaboration}{LIGO Scientific, Virgo}),
  \bibinfo{journal}{Astrophys. J. Lett.} \textbf{\bibinfo{volume}{850}},
  \bibinfo{pages}{L39} (\bibinfo{year}{2017}{\natexlab{a}}),
  \eprint{1710.05836}.

\bibitem[{\citenamefont{Abbott et~al.}(2017{\natexlab{b}})}]{GBM:2017lvd}
\bibinfo{author}{\bibfnamefont{B.~P.} \bibnamefont{Abbott}}
  \bibnamefont{et~al.}, \bibinfo{journal}{Astrophys. J.}
  \textbf{\bibinfo{volume}{848}}, \bibinfo{pages}{L12}
  (\bibinfo{year}{2017}{\natexlab{b}}), \eprint{1710.05833}.

\bibitem[{\citenamefont{Abbott
  et~al.}(2017{\natexlab{c}})}]{LIGOScientific:2017zic}
\bibinfo{author}{\bibfnamefont{B.~P.} \bibnamefont{Abbott}}
  \bibnamefont{et~al.} (\bibinfo{collaboration}{LIGO Scientific, Virgo,
  Fermi-GBM, INTEGRAL}), \bibinfo{journal}{Astrophys. J. Lett.}
  \textbf{\bibinfo{volume}{848}}, \bibinfo{pages}{L13}
  (\bibinfo{year}{2017}{\natexlab{c}}), \eprint{1710.05834}.

\bibitem[{\citenamefont{Li and Paczynski}(1998)}]{Li:1998bw}
\bibinfo{author}{\bibfnamefont{L.-X.} \bibnamefont{Li}} \bibnamefont{and}
  \bibinfo{author}{\bibfnamefont{B.}~\bibnamefont{Paczynski}},
  \bibinfo{journal}{Astrophys. J. Lett.} \textbf{\bibinfo{volume}{507}},
  \bibinfo{pages}{L59} (\bibinfo{year}{1998}), \eprint{astro-ph/9807272}.

\bibitem[{\citenamefont{Metzger}(2017)}]{Metzger:2016pju}
\bibinfo{author}{\bibfnamefont{B.~D.} \bibnamefont{Metzger}},
  \bibinfo{journal}{Living Rev. Rel.} \textbf{\bibinfo{volume}{20}},
  \bibinfo{pages}{3} (\bibinfo{year}{2017}), \eprint{1610.09381}.

\bibitem[{\citenamefont{{Troja} et~al.}(2017)\citenamefont{{Troja}, {Piro},
  {van Eerten}, {Wollaeger}, {Im}, {Fox}, {Butler}, {Cenko}, {Sakamoto},
  {Fryer} et~al.}}]{Troja:2017}
\bibinfo{author}{\bibfnamefont{E.}~\bibnamefont{{Troja}}},
  \bibinfo{author}{\bibfnamefont{L.}~\bibnamefont{{Piro}}},
  \bibinfo{author}{\bibfnamefont{H.}~\bibnamefont{{van Eerten}}},
  \bibinfo{author}{\bibfnamefont{R.~T.} \bibnamefont{{Wollaeger}}},
  \bibinfo{author}{\bibfnamefont{M.}~\bibnamefont{{Im}}},
  \bibinfo{author}{\bibfnamefont{O.~D.} \bibnamefont{{Fox}}},
  \bibinfo{author}{\bibfnamefont{N.~R.} \bibnamefont{{Butler}}},
  \bibinfo{author}{\bibfnamefont{S.~B.} \bibnamefont{{Cenko}}},
  \bibinfo{author}{\bibfnamefont{T.}~\bibnamefont{{Sakamoto}}},
  \bibinfo{author}{\bibfnamefont{C.~L.} \bibnamefont{{Fryer}}},
  \bibnamefont{et~al.}, \bibinfo{journal}{\nat} \textbf{\bibinfo{volume}{551}},
  \bibinfo{pages}{71} (\bibinfo{year}{2017}), \eprint{1710.05433}.

\bibitem[{\citenamefont{{Kasen} et~al.}(2017)\citenamefont{{Kasen}, {Metzger},
  {Barnes}, {Quataert}, and {Ramirez-Ruiz}}}]{Kasen:2017}
\bibinfo{author}{\bibfnamefont{D.}~\bibnamefont{{Kasen}}},
  \bibinfo{author}{\bibfnamefont{B.}~\bibnamefont{{Metzger}}},
  \bibinfo{author}{\bibfnamefont{J.}~\bibnamefont{{Barnes}}},
  \bibinfo{author}{\bibfnamefont{E.}~\bibnamefont{{Quataert}}},
  \bibnamefont{and}
  \bibinfo{author}{\bibfnamefont{E.}~\bibnamefont{{Ramirez-Ruiz}}},
  \bibinfo{journal}{\nat} \textbf{\bibinfo{volume}{551}}, \bibinfo{pages}{80}
  (\bibinfo{year}{2017}), \eprint{1710.05463}.

\bibitem[{\citenamefont{{Abbott}
  et~al.}(2017{\natexlab{a}})\citenamefont{{Abbott}, {Abbott}, {Abbott},
  {Acernese}, {Ackley}, {Adams}, {Adams}, {Addesso}, {Adhikari}, {Adya}
  et~al.}}]{lvk_hubble:2017}
\bibinfo{author}{\bibfnamefont{B.~P.} \bibnamefont{{Abbott}}},
  \bibinfo{author}{\bibfnamefont{R.}~\bibnamefont{{Abbott}}},
  \bibinfo{author}{\bibfnamefont{T.~D.} \bibnamefont{{Abbott}}},
  \bibinfo{author}{\bibfnamefont{F.}~\bibnamefont{{Acernese}}},
  \bibinfo{author}{\bibfnamefont{K.}~\bibnamefont{{Ackley}}},
  \bibinfo{author}{\bibfnamefont{C.}~\bibnamefont{{Adams}}},
  \bibinfo{author}{\bibfnamefont{T.}~\bibnamefont{{Adams}}},
  \bibinfo{author}{\bibfnamefont{P.}~\bibnamefont{{Addesso}}},
  \bibinfo{author}{\bibfnamefont{R.~X.} \bibnamefont{{Adhikari}}},
  \bibinfo{author}{\bibfnamefont{V.~B.} \bibnamefont{{Adya}}},
  \bibnamefont{et~al.}, \bibinfo{journal}{\nat} \textbf{\bibinfo{volume}{551}},
  \bibinfo{pages}{85} (\bibinfo{year}{2017}{\natexlab{a}}),
  \eprint{1710.05835}.

\bibitem[{\citenamefont{Dietrich et~al.}(2020)\citenamefont{Dietrich, Coughlin,
  Pang, Bulla, Heinzel, Issa, Tews, and Antier}}]{Dietrich:2020efo}
\bibinfo{author}{\bibfnamefont{T.}~\bibnamefont{Dietrich}},
  \bibinfo{author}{\bibfnamefont{M.~W.} \bibnamefont{Coughlin}},
  \bibinfo{author}{\bibfnamefont{P.~T.~H.} \bibnamefont{Pang}},
  \bibinfo{author}{\bibfnamefont{M.}~\bibnamefont{Bulla}},
  \bibinfo{author}{\bibfnamefont{J.}~\bibnamefont{Heinzel}},
  \bibinfo{author}{\bibfnamefont{L.}~\bibnamefont{Issa}},
  \bibinfo{author}{\bibfnamefont{I.}~\bibnamefont{Tews}}, \bibnamefont{and}
  \bibinfo{author}{\bibfnamefont{S.}~\bibnamefont{Antier}},
  \bibinfo{journal}{Science} \textbf{\bibinfo{volume}{370}},
  \bibinfo{pages}{1450} (\bibinfo{year}{2020}), \eprint{2002.11355}.

\bibitem[{\citenamefont{Rezzolla et~al.}(2018)\citenamefont{Rezzolla, Most, and
  Weih}}]{Rezzolla:2017aly}
\bibinfo{author}{\bibfnamefont{L.}~\bibnamefont{Rezzolla}},
  \bibinfo{author}{\bibfnamefont{E.~R.} \bibnamefont{Most}}, \bibnamefont{and}
  \bibinfo{author}{\bibfnamefont{L.~R.} \bibnamefont{Weih}},
  \bibinfo{journal}{Astrophys. J. Lett.} \textbf{\bibinfo{volume}{852}},
  \bibinfo{pages}{L25} (\bibinfo{year}{2018}), \eprint{1711.00314}.

\bibitem[{\citenamefont{Ruiz et~al.}(2018)\citenamefont{Ruiz, Shapiro, and
  Tsokaros}}]{Ruiz:2017due}
\bibinfo{author}{\bibfnamefont{M.}~\bibnamefont{Ruiz}},
  \bibinfo{author}{\bibfnamefont{S.~L.} \bibnamefont{Shapiro}},
  \bibnamefont{and} \bibinfo{author}{\bibfnamefont{A.}~\bibnamefont{Tsokaros}},
  \bibinfo{journal}{Phys. Rev. D} \textbf{\bibinfo{volume}{97}},
  \bibinfo{pages}{021501} (\bibinfo{year}{2018}), \eprint{1711.00473}.

\bibitem[{\citenamefont{Shibata et~al.}(2017)\citenamefont{Shibata,
  Fujibayashi, Hotokezaka, Kiuchi, Kyutoku, Sekiguchi, and
  Tanaka}}]{Shibata:2017xdx}
\bibinfo{author}{\bibfnamefont{M.}~\bibnamefont{Shibata}},
  \bibinfo{author}{\bibfnamefont{S.}~\bibnamefont{Fujibayashi}},
  \bibinfo{author}{\bibfnamefont{K.}~\bibnamefont{Hotokezaka}},
  \bibinfo{author}{\bibfnamefont{K.}~\bibnamefont{Kiuchi}},
  \bibinfo{author}{\bibfnamefont{K.}~\bibnamefont{Kyutoku}},
  \bibinfo{author}{\bibfnamefont{Y.}~\bibnamefont{Sekiguchi}},
  \bibnamefont{and} \bibinfo{author}{\bibfnamefont{M.}~\bibnamefont{Tanaka}},
  \bibinfo{journal}{Phys. Rev. D} \textbf{\bibinfo{volume}{96}},
  \bibinfo{pages}{123012} (\bibinfo{year}{2017}), \eprint{1710.07579}.

\bibitem[{\citenamefont{Margalit and Metzger}(2017)}]{Margalit:2017dij}
\bibinfo{author}{\bibfnamefont{B.}~\bibnamefont{Margalit}} \bibnamefont{and}
  \bibinfo{author}{\bibfnamefont{B.~D.} \bibnamefont{Metzger}},
  \bibinfo{journal}{Astrophys. J. Lett.} \textbf{\bibinfo{volume}{850}},
  \bibinfo{pages}{L19} (\bibinfo{year}{2017}), \eprint{1710.05938}.

\bibitem[{\citenamefont{{Abbott} et~al.}(2018)\citenamefont{{Abbott}, {Abbott},
  {Abbott}, {Acernese}, {Ackley}, {Adams}, {Adams}, {Addesso}, {Adhikari},
  {Adya} et~al.}}]{lvk_eos:2018}
\bibinfo{author}{\bibfnamefont{B.~P.} \bibnamefont{{Abbott}}},
  \bibinfo{author}{\bibfnamefont{R.}~\bibnamefont{{Abbott}}},
  \bibinfo{author}{\bibfnamefont{T.~D.} \bibnamefont{{Abbott}}},
  \bibinfo{author}{\bibfnamefont{F.}~\bibnamefont{{Acernese}}},
  \bibinfo{author}{\bibfnamefont{K.}~\bibnamefont{{Ackley}}},
  \bibinfo{author}{\bibfnamefont{C.}~\bibnamefont{{Adams}}},
  \bibinfo{author}{\bibfnamefont{T.}~\bibnamefont{{Adams}}},
  \bibinfo{author}{\bibfnamefont{P.}~\bibnamefont{{Addesso}}},
  \bibinfo{author}{\bibfnamefont{R.~X.} \bibnamefont{{Adhikari}}},
  \bibinfo{author}{\bibfnamefont{V.~B.} \bibnamefont{{Adya}}},
  \bibnamefont{et~al.}, \bibinfo{journal}{\prl} \textbf{\bibinfo{volume}{121}},
  \bibinfo{eid}{161101} (\bibinfo{year}{2018}), \eprint{1805.11581}.

\bibitem[{\citenamefont{{Abbott}
  et~al.}(2019{\natexlab{b}})\citenamefont{{Abbott}, {Abbott}, {Abbott},
  {Acernese}, {Ackley}, {Adams}, {Adams}, {Addesso}, {Adhikari}, {Adya}
  et~al.}}]{lvk_eos:2019}
\bibinfo{author}{\bibfnamefont{B.~P.} \bibnamefont{{Abbott}}},
  \bibinfo{author}{\bibfnamefont{R.}~\bibnamefont{{Abbott}}},
  \bibinfo{author}{\bibfnamefont{T.~D.} \bibnamefont{{Abbott}}},
  \bibinfo{author}{\bibfnamefont{F.}~\bibnamefont{{Acernese}}},
  \bibinfo{author}{\bibfnamefont{K.}~\bibnamefont{{Ackley}}},
  \bibinfo{author}{\bibfnamefont{C.}~\bibnamefont{{Adams}}},
  \bibinfo{author}{\bibfnamefont{T.}~\bibnamefont{{Adams}}},
  \bibinfo{author}{\bibfnamefont{P.}~\bibnamefont{{Addesso}}},
  \bibinfo{author}{\bibfnamefont{R.~X.} \bibnamefont{{Adhikari}}},
  \bibinfo{author}{\bibfnamefont{V.~B.} \bibnamefont{{Adya}}},
  \bibnamefont{et~al.}, \bibinfo{journal}{Physical Review X}
  \textbf{\bibinfo{volume}{9}}, \bibinfo{eid}{011001}
  (\bibinfo{year}{2019}{\natexlab{b}}), \eprint{1805.11579}.

\bibitem[{\citenamefont{{Abbott}
  et~al.}(2017{\natexlab{b}})\citenamefont{{Abbott}, {Abbott}, {Abbott},
  {Acernese}, {Ackley}, {Adams}, {Adams}, {Addesso}, {Adhikari}, {Adya}
  et~al.}}]{lvk_pm:2017}
\bibinfo{author}{\bibfnamefont{B.~P.} \bibnamefont{{Abbott}}},
  \bibinfo{author}{\bibfnamefont{R.}~\bibnamefont{{Abbott}}},
  \bibinfo{author}{\bibfnamefont{T.~D.} \bibnamefont{{Abbott}}},
  \bibinfo{author}{\bibfnamefont{F.}~\bibnamefont{{Acernese}}},
  \bibinfo{author}{\bibfnamefont{K.}~\bibnamefont{{Ackley}}},
  \bibinfo{author}{\bibfnamefont{C.}~\bibnamefont{{Adams}}},
  \bibinfo{author}{\bibfnamefont{T.}~\bibnamefont{{Adams}}},
  \bibinfo{author}{\bibfnamefont{P.}~\bibnamefont{{Addesso}}},
  \bibinfo{author}{\bibfnamefont{R.~X.} \bibnamefont{{Adhikari}}},
  \bibinfo{author}{\bibfnamefont{V.~B.} \bibnamefont{{Adya}}},
  \bibnamefont{et~al.}, \bibinfo{journal}{ApJL} \textbf{\bibinfo{volume}{851}},
  \bibinfo{eid}{L16} (\bibinfo{year}{2017}{\natexlab{b}}), \eprint{1710.09320}.

\bibitem[{\citenamefont{Lackey et~al.}(2017)\citenamefont{Lackey, Bernuzzi,
  Galley, Meidam, and Van Den~Broeck}}]{Lackey:2016krb}
\bibinfo{author}{\bibfnamefont{B.~D.} \bibnamefont{Lackey}},
  \bibinfo{author}{\bibfnamefont{S.}~\bibnamefont{Bernuzzi}},
  \bibinfo{author}{\bibfnamefont{C.~R.} \bibnamefont{Galley}},
  \bibinfo{author}{\bibfnamefont{J.}~\bibnamefont{Meidam}}, \bibnamefont{and}
  \bibinfo{author}{\bibfnamefont{C.}~\bibnamefont{Van Den~Broeck}},
  \bibinfo{journal}{Phys. Rev. D} \textbf{\bibinfo{volume}{95}},
  \bibinfo{pages}{104036} (\bibinfo{year}{2017}), \eprint{1610.04742}.

\bibitem[{\citenamefont{Narikawa and Uchikata}(2022)}]{Narikawa:2022saj}
\bibinfo{author}{\bibfnamefont{T.}~\bibnamefont{Narikawa}} \bibnamefont{and}
  \bibinfo{author}{\bibfnamefont{N.}~\bibnamefont{Uchikata}},
  \bibinfo{journal}{Phys. Rev. D} \textbf{\bibinfo{volume}{106}},
  \bibinfo{pages}{103006} (\bibinfo{year}{2022}), \eprint{2205.06023}.

\bibitem[{\citenamefont{Sun et~al.}(2022)\citenamefont{Sun, Ruiz, Shapiro, and
  Tsokaros}}]{Sun:2022vri}
\bibinfo{author}{\bibfnamefont{L.}~\bibnamefont{Sun}},
  \bibinfo{author}{\bibfnamefont{M.}~\bibnamefont{Ruiz}},
  \bibinfo{author}{\bibfnamefont{S.~L.} \bibnamefont{Shapiro}},
  \bibnamefont{and} \bibinfo{author}{\bibfnamefont{A.}~\bibnamefont{Tsokaros}},
  \bibinfo{journal}{Phys. Rev. D} \textbf{\bibinfo{volume}{105}},
  \bibinfo{pages}{104028} (\bibinfo{year}{2022}), \eprint{2202.12901}.

\bibitem[{\citenamefont{Foucart et~al.}(2023)\citenamefont{Foucart, Duez, Haas,
  Kidder, Pfeiffer, Scheel, and Spira-Savett}}]{Foucart:2022kon}
\bibinfo{author}{\bibfnamefont{F.}~\bibnamefont{Foucart}},
  \bibinfo{author}{\bibfnamefont{M.~D.} \bibnamefont{Duez}},
  \bibinfo{author}{\bibfnamefont{R.}~\bibnamefont{Haas}},
  \bibinfo{author}{\bibfnamefont{L.~E.} \bibnamefont{Kidder}},
  \bibinfo{author}{\bibfnamefont{H.~P.} \bibnamefont{Pfeiffer}},
  \bibinfo{author}{\bibfnamefont{M.~A.} \bibnamefont{Scheel}},
  \bibnamefont{and}
  \bibinfo{author}{\bibfnamefont{E.}~\bibnamefont{Spira-Savett}},
  \bibinfo{journal}{Phys. Rev. D} \textbf{\bibinfo{volume}{107}},
  \bibinfo{pages}{103055} (\bibinfo{year}{2023}), \eprint{2210.05670}.

\bibitem[{\citenamefont{Foucart et~al.}(2020)\citenamefont{Foucart, Duez,
  Hebert, Kidder, Pfeiffer, and Scheel}}]{Foucart:2020qjb}
\bibinfo{author}{\bibfnamefont{F.}~\bibnamefont{Foucart}},
  \bibinfo{author}{\bibfnamefont{M.~D.} \bibnamefont{Duez}},
  \bibinfo{author}{\bibfnamefont{F.}~\bibnamefont{Hebert}},
  \bibinfo{author}{\bibfnamefont{L.~E.} \bibnamefont{Kidder}},
  \bibinfo{author}{\bibfnamefont{H.~P.} \bibnamefont{Pfeiffer}},
  \bibnamefont{and} \bibinfo{author}{\bibfnamefont{M.~A.}
  \bibnamefont{Scheel}}, \bibinfo{journal}{Astrophys. J. Lett.}
  \textbf{\bibinfo{volume}{902}}, \bibinfo{pages}{L27} (\bibinfo{year}{2020}),
  \eprint{2008.08089}.

\bibitem[{\citenamefont{Gieg et~al.}(2022)\citenamefont{Gieg, Schianchi,
  Dietrich, and Ujevic}}]{Gieg:2022mut}
\bibinfo{author}{\bibfnamefont{H.}~\bibnamefont{Gieg}},
  \bibinfo{author}{\bibfnamefont{F.}~\bibnamefont{Schianchi}},
  \bibinfo{author}{\bibfnamefont{T.}~\bibnamefont{Dietrich}}, \bibnamefont{and}
  \bibinfo{author}{\bibfnamefont{M.}~\bibnamefont{Ujevic}},
  \bibinfo{journal}{Universe} \textbf{\bibinfo{volume}{8}},
  \bibinfo{pages}{370} (\bibinfo{year}{2022}), \eprint{2206.01337}.

\bibitem[{\citenamefont{Hayashi et~al.}(2022)\citenamefont{Hayashi,
  Fujibayashi, Kiuchi, Kyutoku, Sekiguchi, and Shibata}}]{Hayashi:2021oxy}
\bibinfo{author}{\bibfnamefont{K.}~\bibnamefont{Hayashi}},
  \bibinfo{author}{\bibfnamefont{S.}~\bibnamefont{Fujibayashi}},
  \bibinfo{author}{\bibfnamefont{K.}~\bibnamefont{Kiuchi}},
  \bibinfo{author}{\bibfnamefont{K.}~\bibnamefont{Kyutoku}},
  \bibinfo{author}{\bibfnamefont{Y.}~\bibnamefont{Sekiguchi}},
  \bibnamefont{and} \bibinfo{author}{\bibfnamefont{M.}~\bibnamefont{Shibata}},
  \bibinfo{journal}{Phys. Rev. D} \textbf{\bibinfo{volume}{106}},
  \bibinfo{pages}{023008} (\bibinfo{year}{2022}), \eprint{2111.04621}.

\bibitem[{\citenamefont{Radice et~al.}(2022)\citenamefont{Radice, Bernuzzi,
  Perego, and Haas}}]{Radice:2021jtw}
\bibinfo{author}{\bibfnamefont{D.}~\bibnamefont{Radice}},
  \bibinfo{author}{\bibfnamefont{S.}~\bibnamefont{Bernuzzi}},
  \bibinfo{author}{\bibfnamefont{A.}~\bibnamefont{Perego}}, \bibnamefont{and}
  \bibinfo{author}{\bibfnamefont{R.}~\bibnamefont{Haas}},
  \bibinfo{journal}{Mon. Not. Roy. Astron. Soc.}
  \textbf{\bibinfo{volume}{512}}, \bibinfo{pages}{1499} (\bibinfo{year}{2022}),
  \eprint{2111.14858}.

\bibitem[{\citenamefont{{Janka} et~al.}(1993)\citenamefont{{Janka}, {Zwerger},
  and {Moenchmeyer}}}]{1993A&A...268..360J}
\bibinfo{author}{\bibfnamefont{H.~T.} \bibnamefont{{Janka}}},
  \bibinfo{author}{\bibfnamefont{T.}~\bibnamefont{{Zwerger}}},
  \bibnamefont{and}
  \bibinfo{author}{\bibfnamefont{R.}~\bibnamefont{{Moenchmeyer}}},
  \bibinfo{journal}{"Astronomy and Astrophysics"}
  \textbf{\bibinfo{volume}{268}}, \bibinfo{pages}{360} (\bibinfo{year}{1993}).

\bibitem[{\citenamefont{Dimmelmeier et~al.}(2002)\citenamefont{Dimmelmeier,
  Font, and Muller}}]{Dimmelmeier:2002bk}
\bibinfo{author}{\bibfnamefont{H.}~\bibnamefont{Dimmelmeier}},
  \bibinfo{author}{\bibfnamefont{J.~A.} \bibnamefont{Font}}, \bibnamefont{and}
  \bibinfo{author}{\bibfnamefont{E.}~\bibnamefont{Muller}},
  \bibinfo{journal}{Astron. Astrophys.} \textbf{\bibinfo{volume}{388}},
  \bibinfo{pages}{917} (\bibinfo{year}{2002}), \eprint{astro-ph/0204288}.

\bibitem[{\citenamefont{Shibata et~al.}(2005)\citenamefont{Shibata, Taniguchi,
  and Uryu}}]{shibata:2005ss}
\bibinfo{author}{\bibfnamefont{M.}~\bibnamefont{Shibata}},
  \bibinfo{author}{\bibfnamefont{K.}~\bibnamefont{Taniguchi}},
  \bibnamefont{and} \bibinfo{author}{\bibfnamefont{K.}~\bibnamefont{Uryu}},
  \bibinfo{journal}{Phys. Rev. D} \textbf{\bibinfo{volume}{71}},
  \bibinfo{pages}{084021} (\bibinfo{year}{2005}), \eprint{gr-qc/0503119}.

\bibitem[{\citenamefont{Constantinou et~al.}(2015)\citenamefont{Constantinou,
  Muccioli, Prakash, and Lattimer}}]{Constantinou:2015mna}
\bibinfo{author}{\bibfnamefont{C.}~\bibnamefont{Constantinou}},
  \bibinfo{author}{\bibfnamefont{B.}~\bibnamefont{Muccioli}},
  \bibinfo{author}{\bibfnamefont{M.}~\bibnamefont{Prakash}}, \bibnamefont{and}
  \bibinfo{author}{\bibfnamefont{J.~M.} \bibnamefont{Lattimer}},
  \bibinfo{journal}{Phys. Rev. C} \textbf{\bibinfo{volume}{92}},
  \bibinfo{pages}{025801} (\bibinfo{year}{2015}), \eprint{1504.03982}.

\bibitem[{\citenamefont{Takami et~al.}(2015)\citenamefont{Takami, Rezzolla, and
  Baiotti}}]{Takami:2014tva}
\bibinfo{author}{\bibfnamefont{K.}~\bibnamefont{Takami}},
  \bibinfo{author}{\bibfnamefont{L.}~\bibnamefont{Rezzolla}}, \bibnamefont{and}
  \bibinfo{author}{\bibfnamefont{L.}~\bibnamefont{Baiotti}},
  \bibinfo{journal}{Phys. Rev. D} \textbf{\bibinfo{volume}{91}},
  \bibinfo{pages}{064001} (\bibinfo{year}{2015}), \eprint{1412.3240}.

\bibitem[{\citenamefont{Lim and Holt}(2019)}]{Lim:2019ozm}
\bibinfo{author}{\bibfnamefont{Y.}~\bibnamefont{Lim}} \bibnamefont{and}
  \bibinfo{author}{\bibfnamefont{J.~W.} \bibnamefont{Holt}}
  (\bibinfo{year}{2019}), \eprint{1909.09089}.

\bibitem[{\citenamefont{Raithel et~al.}(2021)\citenamefont{Raithel,
  Paschalidis, and \"Ozel}}]{Raithel:2021hye}
\bibinfo{author}{\bibfnamefont{C.}~\bibnamefont{Raithel}},
  \bibinfo{author}{\bibfnamefont{V.}~\bibnamefont{Paschalidis}},
  \bibnamefont{and} \bibinfo{author}{\bibfnamefont{F.}~\bibnamefont{\"Ozel}},
  \bibinfo{journal}{Phys. Rev. D} \textbf{\bibinfo{volume}{104}},
  \bibinfo{pages}{063016} (\bibinfo{year}{2021}), \eprint{2104.07226}.

\bibitem[{\citenamefont{Bauswein et~al.}(2010)\citenamefont{Bauswein, Janka,
  and Oechslin}}]{Bauswein:2010dn}
\bibinfo{author}{\bibfnamefont{A.}~\bibnamefont{Bauswein}},
  \bibinfo{author}{\bibfnamefont{H.~T.} \bibnamefont{Janka}}, \bibnamefont{and}
  \bibinfo{author}{\bibfnamefont{R.}~\bibnamefont{Oechslin}},
  \bibinfo{journal}{Phys. Rev. D} \textbf{\bibinfo{volume}{82}},
  \bibinfo{pages}{084043} (\bibinfo{year}{2010}), \eprint{1006.3315}.

\bibitem[{\citenamefont{Figura et~al.}(2021)\citenamefont{Figura, Li, Lu,
  Burgio, Li, and Schulze}}]{Figura:2021bcn}
\bibinfo{author}{\bibfnamefont{A.}~\bibnamefont{Figura}},
  \bibinfo{author}{\bibfnamefont{F.}~\bibnamefont{Li}},
  \bibinfo{author}{\bibfnamefont{J.-J.} \bibnamefont{Lu}},
  \bibinfo{author}{\bibfnamefont{G.~F.} \bibnamefont{Burgio}},
  \bibinfo{author}{\bibfnamefont{Z.-H.} \bibnamefont{Li}}, \bibnamefont{and}
  \bibinfo{author}{\bibfnamefont{H.~J.} \bibnamefont{Schulze}},
  \bibinfo{journal}{Phys. Rev. D} \textbf{\bibinfo{volume}{103}},
  \bibinfo{pages}{083012} (\bibinfo{year}{2021}), \eprint{2103.02365}.

\bibitem[{\citenamefont{Oechslin et~al.}(2007)\citenamefont{Oechslin, Janka,
  and Marek}}]{Oechslin:2006uk}
\bibinfo{author}{\bibfnamefont{R.}~\bibnamefont{Oechslin}},
  \bibinfo{author}{\bibfnamefont{H.~T.} \bibnamefont{Janka}}, \bibnamefont{and}
  \bibinfo{author}{\bibfnamefont{A.}~\bibnamefont{Marek}},
  \bibinfo{journal}{Astron. Astrophys.} \textbf{\bibinfo{volume}{467}},
  \bibinfo{pages}{395} (\bibinfo{year}{2007}), \eprint{astro-ph/0611047}.

\bibitem[{\citenamefont{Sekiguchi et~al.}(2011)\citenamefont{Sekiguchi, Kiuchi,
  Kyutoku, and Shibata}}]{Sekiguchi:2011zd}
\bibinfo{author}{\bibfnamefont{Y.}~\bibnamefont{Sekiguchi}},
  \bibinfo{author}{\bibfnamefont{K.}~\bibnamefont{Kiuchi}},
  \bibinfo{author}{\bibfnamefont{K.}~\bibnamefont{Kyutoku}}, \bibnamefont{and}
  \bibinfo{author}{\bibfnamefont{M.}~\bibnamefont{Shibata}},
  \bibinfo{journal}{Phys. Rev. Lett.} \textbf{\bibinfo{volume}{107}},
  \bibinfo{pages}{051102} (\bibinfo{year}{2011}), \eprint{1105.2125}.

\bibitem[{\citenamefont{Fields et~al.}(2023)\citenamefont{Fields, Prakash,
  Breschi, Radice, Bernuzzi, and Schneider}}]{Fields:2023bhs}
\bibinfo{author}{\bibfnamefont{J.}~\bibnamefont{Fields}},
  \bibinfo{author}{\bibfnamefont{A.}~\bibnamefont{Prakash}},
  \bibinfo{author}{\bibfnamefont{M.}~\bibnamefont{Breschi}},
  \bibinfo{author}{\bibfnamefont{D.}~\bibnamefont{Radice}},
  \bibinfo{author}{\bibfnamefont{S.}~\bibnamefont{Bernuzzi}}, \bibnamefont{and}
  \bibinfo{author}{\bibfnamefont{A.~d.~S.} \bibnamefont{Schneider}}
  (\bibinfo{year}{2023}), \eprint{2302.11359}.

\bibitem[{\citenamefont{Espino et~al.}(2022)\citenamefont{Espino, Bozzola, and
  Paschalidis}}]{Espino:2022mtb}
\bibinfo{author}{\bibfnamefont{P.~L.} \bibnamefont{Espino}},
  \bibinfo{author}{\bibfnamefont{G.}~\bibnamefont{Bozzola}}, \bibnamefont{and}
  \bibinfo{author}{\bibfnamefont{V.}~\bibnamefont{Paschalidis}}
  (\bibinfo{year}{2022}), \eprint{2210.13481}.

\bibitem[{\citenamefont{Werneck et~al.}(2023)}]{Werneck:2022exo}
\bibinfo{author}{\bibfnamefont{L.~R.} \bibnamefont{Werneck}}
  \bibnamefont{et~al.}, \bibinfo{journal}{Phys. Rev. D}
  \textbf{\bibinfo{volume}{107}}, \bibinfo{pages}{044037}
  (\bibinfo{year}{2023}), \eprint{2208.14487}.

\bibitem[{\citenamefont{{Guerra} et~al.}(2024)\citenamefont{{Guerra}, {Ruiz},
  {Pasquali}, {Cerd{\'a}-Duran}, {Font}, and {Rios}}}]{Guerra:2023}
\bibinfo{author}{\bibfnamefont{D.}~\bibnamefont{{Guerra}}},
  \bibinfo{author}{\bibfnamefont{M.}~\bibnamefont{{Ruiz}}},
  \bibinfo{author}{\bibfnamefont{M.}~\bibnamefont{{Pasquali}}},
  \bibinfo{author}{\bibfnamefont{P.}~\bibnamefont{{Cerd{\'a}-Duran}}},
  \bibinfo{author}{\bibfnamefont{J.~A.} \bibnamefont{{Font}}},
  \bibnamefont{and} \bibinfo{author}{\bibfnamefont{A.}~\bibnamefont{{Rios}}},
  \bibinfo{journal}{in preparation}  (\bibinfo{year}{2024}).

\bibitem[{\citenamefont{{Stergioulas} et~al.}(2011)\citenamefont{{Stergioulas},
  {Bauswein}, {Zagkouris}, and {Janka}}}]{Stergioulas:2011}
\bibinfo{author}{\bibfnamefont{N.}~\bibnamefont{{Stergioulas}}},
  \bibinfo{author}{\bibfnamefont{A.}~\bibnamefont{{Bauswein}}},
  \bibinfo{author}{\bibfnamefont{K.}~\bibnamefont{{Zagkouris}}},
  \bibnamefont{and} \bibinfo{author}{\bibfnamefont{H.-T.}
  \bibnamefont{{Janka}}}, \bibinfo{journal}{Mon.~Not.~Roy.~Astron.~Soc.}
  \textbf{\bibinfo{volume}{418}}, \bibinfo{pages}{427} (\bibinfo{year}{2011}),
  \eprint{1105.0368}.

\bibitem[{\citenamefont{{Hotokezaka} et~al.}(2013)\citenamefont{{Hotokezaka},
  {Kiuchi}, {Kyutoku}, {Muranushi}, {Sekiguchi}, {Shibata}, and
  {Taniguchi}}}]{Hotokezaka:2013}
\bibinfo{author}{\bibfnamefont{K.}~\bibnamefont{{Hotokezaka}}},
  \bibinfo{author}{\bibfnamefont{K.}~\bibnamefont{{Kiuchi}}},
  \bibinfo{author}{\bibfnamefont{K.}~\bibnamefont{{Kyutoku}}},
  \bibinfo{author}{\bibfnamefont{T.}~\bibnamefont{{Muranushi}}},
  \bibinfo{author}{\bibfnamefont{Y.-i.} \bibnamefont{{Sekiguchi}}},
  \bibinfo{author}{\bibfnamefont{M.}~\bibnamefont{{Shibata}}},
  \bibnamefont{and}
  \bibinfo{author}{\bibfnamefont{K.}~\bibnamefont{{Taniguchi}}},
  \bibinfo{journal}{\prd} \textbf{\bibinfo{volume}{88}}, \bibinfo{eid}{044026}
  (\bibinfo{year}{2013}), \eprint{1307.5888}.

\bibitem[{\citenamefont{{Bauswein} and {Stergioulas}}(2015)}]{Bauswein:2015}
\bibinfo{author}{\bibfnamefont{A.}~\bibnamefont{{Bauswein}}} \bibnamefont{and}
  \bibinfo{author}{\bibfnamefont{N.}~\bibnamefont{{Stergioulas}}},
  \bibinfo{journal}{\prd} \textbf{\bibinfo{volume}{91}}, \bibinfo{eid}{124056}
  (\bibinfo{year}{2015}), \eprint{1502.03176}.

\bibitem[{\citenamefont{{Takami} et~al.}(2015)\citenamefont{{Takami},
  {Rezzolla}, and {Baiotti}}}]{Takami:2015}
\bibinfo{author}{\bibfnamefont{K.}~\bibnamefont{{Takami}}},
  \bibinfo{author}{\bibfnamefont{L.}~\bibnamefont{{Rezzolla}}},
  \bibnamefont{and}
  \bibinfo{author}{\bibfnamefont{L.}~\bibnamefont{{Baiotti}}},
  \bibinfo{journal}{\prd} \textbf{\bibinfo{volume}{91}}, \bibinfo{eid}{064001}
  (\bibinfo{year}{2015}), \eprint{1412.3240}.

\bibitem[{\citenamefont{{Bauswein} et~al.}(2016)\citenamefont{{Bauswein},
  {Stergioulas}, and {Janka}}}]{Bauswein:2016}
\bibinfo{author}{\bibfnamefont{A.}~\bibnamefont{{Bauswein}}},
  \bibinfo{author}{\bibfnamefont{N.}~\bibnamefont{{Stergioulas}}},
  \bibnamefont{and} \bibinfo{author}{\bibfnamefont{H.-T.}
  \bibnamefont{{Janka}}}, \bibinfo{journal}{European Physical Journal A}
  \textbf{\bibinfo{volume}{52}}, \bibinfo{eid}{56} (\bibinfo{year}{2016}),
  \eprint{1508.05493}.

\bibitem[{\citenamefont{{Bauswein} and {Stergioulas}}(2019)}]{Bauswein:2019}
\bibinfo{author}{\bibfnamefont{A.}~\bibnamefont{{Bauswein}}} \bibnamefont{and}
  \bibinfo{author}{\bibfnamefont{N.}~\bibnamefont{{Stergioulas}}},
  \bibinfo{journal}{Journal of Physics G Nuclear Physics}
  \textbf{\bibinfo{volume}{46}}, \bibinfo{eid}{113002} (\bibinfo{year}{2019}),
  \eprint{1901.06969}.

\bibitem[{\citenamefont{{Rezzolla} and {Takami}}(2016)}]{Rezzolla:2016}
\bibinfo{author}{\bibfnamefont{L.}~\bibnamefont{{Rezzolla}}} \bibnamefont{and}
  \bibinfo{author}{\bibfnamefont{K.}~\bibnamefont{{Takami}}},
  \bibinfo{journal}{\prd} \textbf{\bibinfo{volume}{93}}, \bibinfo{eid}{124051}
  (\bibinfo{year}{2016}), \eprint{1604.00246}.

\bibitem[{\citenamefont{{Read} et~al.}(2013)\citenamefont{{Read}, {Baiotti},
  {Creighton}, {Friedman}, {Giacomazzo}, {Kyutoku}, {Markakis}, {Rezzolla},
  {Shibata}, and {Taniguchi}}}]{Read:2013}
\bibinfo{author}{\bibfnamefont{J.~S.} \bibnamefont{{Read}}},
  \bibinfo{author}{\bibfnamefont{L.}~\bibnamefont{{Baiotti}}},
  \bibinfo{author}{\bibfnamefont{J.~D.~E.} \bibnamefont{{Creighton}}},
  \bibinfo{author}{\bibfnamefont{J.~L.} \bibnamefont{{Friedman}}},
  \bibinfo{author}{\bibfnamefont{B.}~\bibnamefont{{Giacomazzo}}},
  \bibinfo{author}{\bibfnamefont{K.}~\bibnamefont{{Kyutoku}}},
  \bibinfo{author}{\bibfnamefont{C.}~\bibnamefont{{Markakis}}},
  \bibinfo{author}{\bibfnamefont{L.}~\bibnamefont{{Rezzolla}}},
  \bibinfo{author}{\bibfnamefont{M.}~\bibnamefont{{Shibata}}},
  \bibnamefont{and}
  \bibinfo{author}{\bibfnamefont{K.}~\bibnamefont{{Taniguchi}}},
  \bibinfo{journal}{\prd} \textbf{\bibinfo{volume}{88}}, \bibinfo{eid}{044042}
  (\bibinfo{year}{2013}), \eprint{1306.4065}.

\bibitem[{\citenamefont{{Topolski} et~al.}(2024)\citenamefont{{Topolski},
  {Tootle}, and {Rezzolla}}}]{Topolski:2023}
\bibinfo{author}{\bibfnamefont{K.}~\bibnamefont{{Topolski}}},
  \bibinfo{author}{\bibfnamefont{S.~D.} \bibnamefont{{Tootle}}},
  \bibnamefont{and}
  \bibinfo{author}{\bibfnamefont{L.}~\bibnamefont{{Rezzolla}}},
  \bibinfo{journal}{\apj} \textbf{\bibinfo{volume}{960}}, \bibinfo{eid}{86}
  (\bibinfo{year}{2024}), \eprint{2310.10728}.

\bibitem[{\citenamefont{{De Pietri} et~al.}(2018)\citenamefont{{De Pietri},
  {Feo}, {Font}, {L{\"o}ffler}, {Maione}, {Pasquali}, and
  {Stergioulas}}}]{DePietri:2018}
\bibinfo{author}{\bibfnamefont{R.}~\bibnamefont{{De Pietri}}},
  \bibinfo{author}{\bibfnamefont{A.}~\bibnamefont{{Feo}}},
  \bibinfo{author}{\bibfnamefont{J.~A.} \bibnamefont{{Font}}},
  \bibinfo{author}{\bibfnamefont{F.}~\bibnamefont{{L{\"o}ffler}}},
  \bibinfo{author}{\bibfnamefont{F.}~\bibnamefont{{Maione}}},
  \bibinfo{author}{\bibfnamefont{M.}~\bibnamefont{{Pasquali}}},
  \bibnamefont{and}
  \bibinfo{author}{\bibfnamefont{N.}~\bibnamefont{{Stergioulas}}},
  \bibinfo{journal}{\prl} \textbf{\bibinfo{volume}{120}}, \bibinfo{eid}{221101}
  (\bibinfo{year}{2018}), \eprint{1802.03288}.

\bibitem[{\citenamefont{{De Pietri} et~al.}(2020)\citenamefont{{De Pietri},
  {Feo}, {Font}, {L{\"o}ffler}, {Pasquali}, and {Stergioulas}}}]{DePietri:2020}
\bibinfo{author}{\bibfnamefont{R.}~\bibnamefont{{De Pietri}}},
  \bibinfo{author}{\bibfnamefont{A.}~\bibnamefont{{Feo}}},
  \bibinfo{author}{\bibfnamefont{J.~A.} \bibnamefont{{Font}}},
  \bibinfo{author}{\bibfnamefont{F.}~\bibnamefont{{L{\"o}ffler}}},
  \bibinfo{author}{\bibfnamefont{M.}~\bibnamefont{{Pasquali}}},
  \bibnamefont{and}
  \bibinfo{author}{\bibfnamefont{N.}~\bibnamefont{{Stergioulas}}},
  \bibinfo{journal}{\prd} \textbf{\bibinfo{volume}{101}}, \bibinfo{eid}{064052}
  (\bibinfo{year}{2020}), \eprint{1910.04036}.

\bibitem[{\citenamefont{{Kastaun}}(2008)}]{Kastaun:2008}
\bibinfo{author}{\bibfnamefont{W.}~\bibnamefont{{Kastaun}}},
  \bibinfo{journal}{\prd} \textbf{\bibinfo{volume}{77}}, \bibinfo{eid}{124019}
  (\bibinfo{year}{2008}), \eprint{0804.1151}.

\bibitem[{\citenamefont{{Cornish} and {Littenberg}}(2015)}]{Cornish:2015}
\bibinfo{author}{\bibfnamefont{N.~J.} \bibnamefont{{Cornish}}}
  \bibnamefont{and} \bibinfo{author}{\bibfnamefont{T.~B.}
  \bibnamefont{{Littenberg}}}, \bibinfo{journal}{Classical and Quantum Gravity}
  \textbf{\bibinfo{volume}{32}}, \bibinfo{eid}{135012} (\bibinfo{year}{2015}),
  \eprint{1410.3835}.

\bibitem[{\citenamefont{{Littenberg} and {Cornish}}(2015)}]{Littenberg:2015}
\bibinfo{author}{\bibfnamefont{T.~B.} \bibnamefont{{Littenberg}}}
  \bibnamefont{and} \bibinfo{author}{\bibfnamefont{N.~J.}
  \bibnamefont{{Cornish}}}, \bibinfo{journal}{\prd}
  \textbf{\bibinfo{volume}{91}}, \bibinfo{eid}{084034} (\bibinfo{year}{2015}),
  \eprint{1410.3852}.

\bibitem[{\citenamefont{Ruiz et~al.}(2021)\citenamefont{Ruiz, Tsokaros, and
  Shapiro}}]{Ruiz:2021qmm}
\bibinfo{author}{\bibfnamefont{M.}~\bibnamefont{Ruiz}},
  \bibinfo{author}{\bibfnamefont{A.}~\bibnamefont{Tsokaros}}, \bibnamefont{and}
  \bibinfo{author}{\bibfnamefont{S.~L.} \bibnamefont{Shapiro}},
  \bibinfo{journal}{Phys. Rev. D} \textbf{\bibinfo{volume}{104}},
  \bibinfo{pages}{124049} (\bibinfo{year}{2021}), \eprint{2110.11968}.

\bibitem[{\citenamefont{Foucart et~al.}(2016)\citenamefont{Foucart, Haas, Duez,
  O'Connor, Ott, Roberts, Kidder, Lippuner, Pfeiffer, and
  Scheel}}]{Foucart:2015gaa}
\bibinfo{author}{\bibfnamefont{F.}~\bibnamefont{Foucart}},
  \bibinfo{author}{\bibfnamefont{R.}~\bibnamefont{Haas}},
  \bibinfo{author}{\bibfnamefont{M.~D.} \bibnamefont{Duez}},
  \bibinfo{author}{\bibfnamefont{E.}~\bibnamefont{O'Connor}},
  \bibinfo{author}{\bibfnamefont{C.~D.} \bibnamefont{Ott}},
  \bibinfo{author}{\bibfnamefont{L.}~\bibnamefont{Roberts}},
  \bibinfo{author}{\bibfnamefont{L.~E.} \bibnamefont{Kidder}},
  \bibinfo{author}{\bibfnamefont{J.}~\bibnamefont{Lippuner}},
  \bibinfo{author}{\bibfnamefont{H.~P.} \bibnamefont{Pfeiffer}},
  \bibnamefont{and} \bibinfo{author}{\bibfnamefont{M.~A.}
  \bibnamefont{Scheel}}, \bibinfo{journal}{Phys. Rev. D}
  \textbf{\bibinfo{volume}{93}}, \bibinfo{pages}{044019}
  (\bibinfo{year}{2016}), \eprint{1510.06398}.

\bibitem[{\citenamefont{Blacker et~al.}(2020)\citenamefont{Blacker, Bastian,
  Bauswein, Blaschke, Fischer, Oertel, Soultanis, and Typel}}]{Blacker:2020nlq}
\bibinfo{author}{\bibfnamefont{S.}~\bibnamefont{Blacker}},
  \bibinfo{author}{\bibfnamefont{N.-U.~F.} \bibnamefont{Bastian}},
  \bibinfo{author}{\bibfnamefont{A.}~\bibnamefont{Bauswein}},
  \bibinfo{author}{\bibfnamefont{D.~B.} \bibnamefont{Blaschke}},
  \bibinfo{author}{\bibfnamefont{T.}~\bibnamefont{Fischer}},
  \bibinfo{author}{\bibfnamefont{M.}~\bibnamefont{Oertel}},
  \bibinfo{author}{\bibfnamefont{T.}~\bibnamefont{Soultanis}},
  \bibnamefont{and} \bibinfo{author}{\bibfnamefont{S.}~\bibnamefont{Typel}},
  \bibinfo{journal}{Phys. Rev. D} \textbf{\bibinfo{volume}{102}},
  \bibinfo{pages}{123023} (\bibinfo{year}{2020}), \eprint{2006.03789}.

\bibitem[{\citenamefont{Rivieccio et~al.}(2024)\citenamefont{Rivieccio, Guerra,
  Ruiz, and Font}}]{Rivieccio:2024sfm}
\bibinfo{author}{\bibfnamefont{G.}~\bibnamefont{Rivieccio}},
  \bibinfo{author}{\bibfnamefont{D.}~\bibnamefont{Guerra}},
  \bibinfo{author}{\bibfnamefont{M.}~\bibnamefont{Ruiz}}, \bibnamefont{and}
  \bibinfo{author}{\bibfnamefont{J.~A.} \bibnamefont{Font}},
  \bibinfo{journal}{Phys. Rev. D} \textbf{\bibinfo{volume}{109}},
  \bibinfo{pages}{064032} (\bibinfo{year}{2024}), \eprint{2401.06849}.

\bibitem[{\citenamefont{{Miravet-Ten{\'e}s}
  et~al.}(2023)\citenamefont{{Miravet-Ten{\'e}s}, {Castillo}, {De Pietri},
  {Cerd{\'a}-Dur{\'a}n}, and {Font}}}]{Miravet:2023}
\bibinfo{author}{\bibfnamefont{M.}~\bibnamefont{{Miravet-Ten{\'e}s}}},
  \bibinfo{author}{\bibfnamefont{F.~L.} \bibnamefont{{Castillo}}},
  \bibinfo{author}{\bibfnamefont{R.}~\bibnamefont{{De Pietri}}},
  \bibinfo{author}{\bibfnamefont{P.}~\bibnamefont{{Cerd{\'a}-Dur{\'a}n}}},
  \bibnamefont{and} \bibinfo{author}{\bibfnamefont{J.~A.}
  \bibnamefont{{Font}}}, \bibinfo{journal}{\prd}
  \textbf{\bibinfo{volume}{107}}, \bibinfo{eid}{103053} (\bibinfo{year}{2023}),
  \eprint{2302.04553}.

\bibitem[{\citenamefont{{Punturo} et~al.}(2010)\citenamefont{{Punturo},
  {Abernathy}, {Acernese}, {Allen}, {Andersson}, {Arun}, {Barone}, {Barr},
  {Barsuglia}, {Beker} et~al.}}]{ET:2010}
\bibinfo{author}{\bibfnamefont{M.}~\bibnamefont{{Punturo}}},
  \bibinfo{author}{\bibfnamefont{M.}~\bibnamefont{{Abernathy}}},
  \bibinfo{author}{\bibfnamefont{F.}~\bibnamefont{{Acernese}}},
  \bibinfo{author}{\bibfnamefont{B.}~\bibnamefont{{Allen}}},
  \bibinfo{author}{\bibfnamefont{N.}~\bibnamefont{{Andersson}}},
  \bibinfo{author}{\bibfnamefont{K.}~\bibnamefont{{Arun}}},
  \bibinfo{author}{\bibfnamefont{F.}~\bibnamefont{{Barone}}},
  \bibinfo{author}{\bibfnamefont{B.}~\bibnamefont{{Barr}}},
  \bibinfo{author}{\bibfnamefont{M.}~\bibnamefont{{Barsuglia}}},
  \bibinfo{author}{\bibfnamefont{M.}~\bibnamefont{{Beker}}},
  \bibnamefont{et~al.}, \bibinfo{journal}{Classical and Quantum Gravity}
  \textbf{\bibinfo{volume}{27}}, \bibinfo{eid}{194002} (\bibinfo{year}{2010}).

\bibitem[{\citenamefont{{Hild} et~al.}(2011)\citenamefont{{Hild}, {Abernathy},
  {Acernese}, {Amaro-Seoane}, {Andersson}, {Arun}, {Barone}, {Barr},
  {Barsuglia}, {Beker} et~al.}}]{Hild:2011}
\bibinfo{author}{\bibfnamefont{S.}~\bibnamefont{{Hild}}},
  \bibinfo{author}{\bibfnamefont{M.}~\bibnamefont{{Abernathy}}},
  \bibinfo{author}{\bibfnamefont{F.}~\bibnamefont{{Acernese}}},
  \bibinfo{author}{\bibfnamefont{P.}~\bibnamefont{{Amaro-Seoane}}},
  \bibinfo{author}{\bibfnamefont{N.}~\bibnamefont{{Andersson}}},
  \bibinfo{author}{\bibfnamefont{K.}~\bibnamefont{{Arun}}},
  \bibinfo{author}{\bibfnamefont{F.}~\bibnamefont{{Barone}}},
  \bibinfo{author}{\bibfnamefont{B.}~\bibnamefont{{Barr}}},
  \bibinfo{author}{\bibfnamefont{M.}~\bibnamefont{{Barsuglia}}},
  \bibinfo{author}{\bibfnamefont{M.}~\bibnamefont{{Beker}}},
  \bibnamefont{et~al.}, \bibinfo{journal}{Classical and Quantum Gravity}
  \textbf{\bibinfo{volume}{28}}, \bibinfo{eid}{094013} (\bibinfo{year}{2011}),
  \eprint{1012.0908}.

\bibitem[{\citenamefont{{Maggiore} et~al.}(2020)\citenamefont{{Maggiore}, {Van
  Den Broeck}, {Bartolo}, {Belgacem}, {Bertacca}, {Bizouard}, {Branchesi},
  {Clesse}, {Foffa}, {Garc{\'\i}a-Bellido} et~al.}}]{Science_case_ET}
\bibinfo{author}{\bibfnamefont{M.}~\bibnamefont{{Maggiore}}},
  \bibinfo{author}{\bibfnamefont{C.}~\bibnamefont{{Van Den Broeck}}},
  \bibinfo{author}{\bibfnamefont{N.}~\bibnamefont{{Bartolo}}},
  \bibinfo{author}{\bibfnamefont{E.}~\bibnamefont{{Belgacem}}},
  \bibinfo{author}{\bibfnamefont{D.}~\bibnamefont{{Bertacca}}},
  \bibinfo{author}{\bibfnamefont{M.~A.} \bibnamefont{{Bizouard}}},
  \bibinfo{author}{\bibfnamefont{M.}~\bibnamefont{{Branchesi}}},
  \bibinfo{author}{\bibfnamefont{S.}~\bibnamefont{{Clesse}}},
  \bibinfo{author}{\bibfnamefont{S.}~\bibnamefont{{Foffa}}},
  \bibinfo{author}{\bibfnamefont{J.}~\bibnamefont{{Garc{\'\i}a-Bellido}}},
  \bibnamefont{et~al.}, \bibinfo{journal}{JCAP}
  \textbf{\bibinfo{volume}{2020}}, \bibinfo{eid}{050} (\bibinfo{year}{2020}),
  \eprint{1912.02622}.

\bibitem[{\citenamefont{{Branchesi} et~al.}(2023)\citenamefont{{Branchesi},
  {Maggiore}, {Alonso}, {Badger}, {Banerjee}, {Beirnaert}, {Belgacem},
  {Bhagwat}, {Boileau}, {Borhanian} et~al.}}]{COBA_study}
\bibinfo{author}{\bibfnamefont{M.}~\bibnamefont{{Branchesi}}},
  \bibinfo{author}{\bibfnamefont{M.}~\bibnamefont{{Maggiore}}},
  \bibinfo{author}{\bibfnamefont{D.}~\bibnamefont{{Alonso}}},
  \bibinfo{author}{\bibfnamefont{C.}~\bibnamefont{{Badger}}},
  \bibinfo{author}{\bibfnamefont{B.}~\bibnamefont{{Banerjee}}},
  \bibinfo{author}{\bibfnamefont{F.}~\bibnamefont{{Beirnaert}}},
  \bibinfo{author}{\bibfnamefont{E.}~\bibnamefont{{Belgacem}}},
  \bibinfo{author}{\bibfnamefont{S.}~\bibnamefont{{Bhagwat}}},
  \bibinfo{author}{\bibfnamefont{G.}~\bibnamefont{{Boileau}}},
  \bibinfo{author}{\bibfnamefont{S.}~\bibnamefont{{Borhanian}}},
  \bibnamefont{et~al.}, \bibinfo{journal}{JCAP}
  \textbf{\bibinfo{volume}{2023}}, \bibinfo{eid}{068} (\bibinfo{year}{2023}),
  \eprint{2303.15923}.

\bibitem[{\citenamefont{{Villa-Ortega}
  et~al.}(2023)\citenamefont{{Villa-Ortega}, {Lorenzo-Medina}, {Calder{\'o}n
  Bustillo}, {Ruiz}, {Guerra}, {Cerd{\'a}-Duran}, and
  {Font}}}]{Calderon-Bustillo:2023}
\bibinfo{author}{\bibfnamefont{V.}~\bibnamefont{{Villa-Ortega}}},
  \bibinfo{author}{\bibfnamefont{A.}~\bibnamefont{{Lorenzo-Medina}}},
  \bibinfo{author}{\bibfnamefont{J.}~\bibnamefont{{Calder{\'o}n Bustillo}}},
  \bibinfo{author}{\bibfnamefont{M.}~\bibnamefont{{Ruiz}}},
  \bibinfo{author}{\bibfnamefont{D.}~\bibnamefont{{Guerra}}},
  \bibinfo{author}{\bibfnamefont{P.}~\bibnamefont{{Cerd{\'a}-Duran}}},
  \bibnamefont{and} \bibinfo{author}{\bibfnamefont{J.~A.}
  \bibnamefont{{Font}}}, \bibinfo{journal}{arXiv e-prints}
  \bibinfo{eid}{arXiv:2310.20378} (\bibinfo{year}{2023}), \eprint{2310.20378}.

\bibitem[{\citenamefont{{Raithel} and {Paschalidis}}(2023)}]{Raithel:2023}
\bibinfo{author}{\bibfnamefont{C.~A.} \bibnamefont{{Raithel}}}
  \bibnamefont{and}
  \bibinfo{author}{\bibfnamefont{V.}~\bibnamefont{{Paschalidis}}},
  \bibinfo{journal}{arXiv e-prints} \bibinfo{eid}{arXiv:2312.14046}
  (\bibinfo{year}{2023}), \eprint{2312.14046}.

\bibitem[{\citenamefont{Typel et~al.}(2010)\citenamefont{Typel, Ropke, Klahn,
  Blaschke, and Wolter}}]{Typel:2009sy}
\bibinfo{author}{\bibfnamefont{S.}~\bibnamefont{Typel}},
  \bibinfo{author}{\bibfnamefont{G.}~\bibnamefont{Ropke}},
  \bibinfo{author}{\bibfnamefont{T.}~\bibnamefont{Klahn}},
  \bibinfo{author}{\bibfnamefont{D.}~\bibnamefont{Blaschke}}, \bibnamefont{and}
  \bibinfo{author}{\bibfnamefont{H.~H.} \bibnamefont{Wolter}},
  \bibinfo{journal}{Phys. Rev. C} \textbf{\bibinfo{volume}{81}},
  \bibinfo{pages}{015803} (\bibinfo{year}{2010}), \eprint{0908.2344}.

\bibitem[{\citenamefont{Shen et~al.}(2011)\citenamefont{Shen, Toki, Oyamatsu,
  and Sumiyoshi}}]{Shen:2011qu}
\bibinfo{author}{\bibfnamefont{H.}~\bibnamefont{Shen}},
  \bibinfo{author}{\bibfnamefont{H.}~\bibnamefont{Toki}},
  \bibinfo{author}{\bibfnamefont{K.}~\bibnamefont{Oyamatsu}}, \bibnamefont{and}
  \bibinfo{author}{\bibfnamefont{K.}~\bibnamefont{Sumiyoshi}},
  \bibinfo{journal}{Astrophys. J. Suppl.} \textbf{\bibinfo{volume}{197}},
  \bibinfo{pages}{20} (\bibinfo{year}{2011}), \eprint{1105.1666}.

\bibitem[{\citenamefont{James M.~Lattimer}(1991)}]{Lattimer91}
\bibinfo{author}{\bibfnamefont{F.~D.~S.} \bibnamefont{James M.~Lattimer}},
  \bibinfo{journal}{Nucl. Phys. A} \textbf{\bibinfo{volume}{535}},
  \bibinfo{pages}{331} (\bibinfo{year}{1991}).

\bibitem[{\citenamefont{Chabanat et~al.}(1998)\citenamefont{Chabanat, Bonche,
  Haensel, Meyer, and Shaeffer}}]{chabanat98}
\bibinfo{author}{\bibfnamefont{E.}~\bibnamefont{Chabanat}},
  \bibinfo{author}{\bibfnamefont{P.}~\bibnamefont{Bonche}},
  \bibinfo{author}{\bibfnamefont{P.}~\bibnamefont{Haensel}},
  \bibinfo{author}{\bibfnamefont{J.}~\bibnamefont{Meyer}}, \bibnamefont{and}
  \bibinfo{author}{\bibfnamefont{R.}~\bibnamefont{Shaeffer}},
  \bibinfo{journal}{Nucl. Phys. A} \textbf{\bibinfo{volume}{635}},
  \bibinfo{pages}{231} (\bibinfo{year}{1998}).

\bibitem[{\citenamefont{Gourgoulhon et~al.}(2001)\citenamefont{Gourgoulhon,
  Grandclement, Taniguchi, Marck, and Bonazzola}}]{Gourgoulhon:2000nn}
\bibinfo{author}{\bibfnamefont{E.}~\bibnamefont{Gourgoulhon}},
  \bibinfo{author}{\bibfnamefont{P.}~\bibnamefont{Grandclement}},
  \bibinfo{author}{\bibfnamefont{K.}~\bibnamefont{Taniguchi}},
  \bibinfo{author}{\bibfnamefont{J.-A.} \bibnamefont{Marck}}, \bibnamefont{and}
  \bibinfo{author}{\bibfnamefont{S.}~\bibnamefont{Bonazzola}},
  \bibinfo{journal}{Phys. Rev. D} \textbf{\bibinfo{volume}{63}},
  \bibinfo{pages}{064029} (\bibinfo{year}{2001}), \eprint{gr-qc/0007028}.

\bibitem[{\citenamefont{{Taniguchi} and {Gourgoulhon}}(2002)}]{tg02}
\bibinfo{author}{\bibfnamefont{K.}~\bibnamefont{{Taniguchi}}} \bibnamefont{and}
  \bibinfo{author}{\bibfnamefont{E.}~\bibnamefont{{Gourgoulhon}}},
  \bibinfo{journal}{Phys. Rev. D} \textbf{\bibinfo{volume}{66}},
  \bibinfo{eid}{104019} (\bibinfo{year}{2002}).

\bibitem[{Lor()}]{Lorene}
\bibinfo{note}{{\tt http://www.lorene.obspm.fr/}}.

\bibitem[{\citenamefont{Schneider et~al.}(2017)\citenamefont{Schneider,
  Roberts, and Ott}}]{Schneider:2017tfi}
\bibinfo{author}{\bibfnamefont{A.~S.} \bibnamefont{Schneider}},
  \bibinfo{author}{\bibfnamefont{L.~F.} \bibnamefont{Roberts}},
  \bibnamefont{and} \bibinfo{author}{\bibfnamefont{C.~D.} \bibnamefont{Ott}},
  \bibinfo{journal}{Phys. Rev. C} \textbf{\bibinfo{volume}{96}},
  \bibinfo{pages}{065802} (\bibinfo{year}{2017}), \eprint{1707.01527}.

\bibitem[{ste()}]{stellarcollapse}
\bibinfo{note}{{\tt https://stellarcollapse.org/SROEOS}}.

\bibitem[{\citenamefont{{Pilgrim}}(2021)}]{2021JOSS....6.3859P}
\bibinfo{author}{\bibfnamefont{C.}~\bibnamefont{{Pilgrim}}},
  \bibinfo{journal}{The Journal of Open Source Software}
  \textbf{\bibinfo{volume}{6}}, \bibinfo{eid}{3859} (\bibinfo{year}{2021}).

\bibitem[{\citenamefont{Read et~al.}(2009)\citenamefont{Read, Lackey, Owen, and
  Friedman}}]{Read:2008iy}
\bibinfo{author}{\bibfnamefont{J.~S.} \bibnamefont{Read}},
  \bibinfo{author}{\bibfnamefont{B.~D.} \bibnamefont{Lackey}},
  \bibinfo{author}{\bibfnamefont{B.~J.} \bibnamefont{Owen}}, \bibnamefont{and}
  \bibinfo{author}{\bibfnamefont{J.~L.} \bibnamefont{Friedman}},
  \bibinfo{journal}{Phys. Rev.} \textbf{\bibinfo{volume}{D79}},
  \bibinfo{pages}{124032} (\bibinfo{year}{2009}).

\bibitem[{\citenamefont{Etienne et~al.}(2015)\citenamefont{Etienne,
  Paschalidis, Haas, M\"osta, and Shapiro}}]{Etienne:2015cea}
\bibinfo{author}{\bibfnamefont{Z.~B.} \bibnamefont{Etienne}},
  \bibinfo{author}{\bibfnamefont{V.}~\bibnamefont{Paschalidis}},
  \bibinfo{author}{\bibfnamefont{R.}~\bibnamefont{Haas}},
  \bibinfo{author}{\bibfnamefont{P.}~\bibnamefont{M\"osta}}, \bibnamefont{and}
  \bibinfo{author}{\bibfnamefont{S.~L.} \bibnamefont{Shapiro}},
  \bibinfo{journal}{Class. Quant. Grav.} \textbf{\bibinfo{volume}{32}},
  \bibinfo{pages}{175009} (\bibinfo{year}{2015}), \eprint{1501.07276}.

\bibitem[{\citenamefont{Loffler et~al.}(2012)}]{Loffler:2011ay}
\bibinfo{author}{\bibfnamefont{F.}~\bibnamefont{Loffler}} \bibnamefont{et~al.},
  \bibinfo{journal}{Class. Quant. Grav.} \textbf{\bibinfo{volume}{29}},
  \bibinfo{pages}{115001} (\bibinfo{year}{2012}), \eprint{1111.3344}.

\bibitem[{\citenamefont{Noble et~al.}(2006)\citenamefont{Noble, Gammie,
  McKinney, and Del~Zanna}}]{Noble:2005gf}
\bibinfo{author}{\bibfnamefont{S.~C.} \bibnamefont{Noble}},
  \bibinfo{author}{\bibfnamefont{C.~F.} \bibnamefont{Gammie}},
  \bibinfo{author}{\bibfnamefont{J.~C.} \bibnamefont{McKinney}},
  \bibnamefont{and}
  \bibinfo{author}{\bibfnamefont{L.}~\bibnamefont{Del~Zanna}},
  \bibinfo{journal}{Astrophys. J.} \textbf{\bibinfo{volume}{641}},
  \bibinfo{pages}{626} (\bibinfo{year}{2006}), \eprint{astro-ph/0512420}.

\bibitem[{\citenamefont{{Baumgarte} and {Shapiro}}(1998)}]{Baumgarte:1998}
\bibinfo{author}{\bibfnamefont{T.~W.} \bibnamefont{{Baumgarte}}}
  \bibnamefont{and} \bibinfo{author}{\bibfnamefont{S.~L.}
  \bibnamefont{{Shapiro}}}, \bibinfo{journal}{\prd}
  \textbf{\bibinfo{volume}{59}}, \bibinfo{eid}{024007} (\bibinfo{year}{1998}),
  \eprint{gr-qc/9810065}.

\bibitem[{\citenamefont{Shibata and Nakamura}(1995)}]{Shibata:1995}
\bibinfo{author}{\bibfnamefont{M.}~\bibnamefont{Shibata}} \bibnamefont{and}
  \bibinfo{author}{\bibfnamefont{T.}~\bibnamefont{Nakamura}},
  \bibinfo{journal}{Phys. Rev. D} \textbf{\bibinfo{volume}{52}},
  \bibinfo{pages}{5428} (\bibinfo{year}{1995}).

\bibitem[{\citenamefont{{Banyuls} et~al.}(1997)\citenamefont{{Banyuls}, {Font},
  {Ib{\'a}{\~n}ez}, {Mart{\'\i}}, and {Miralles}}}]{Banyuls:1997}
\bibinfo{author}{\bibfnamefont{F.}~\bibnamefont{{Banyuls}}},
  \bibinfo{author}{\bibfnamefont{J.~A.} \bibnamefont{{Font}}},
  \bibinfo{author}{\bibfnamefont{J.~M.} \bibnamefont{{Ib{\'a}{\~n}ez}}},
  \bibinfo{author}{\bibfnamefont{J.~M.} \bibnamefont{{Mart{\'\i}}}},
  \bibnamefont{and} \bibinfo{author}{\bibfnamefont{J.~A.}
  \bibnamefont{{Miralles}}}, \bibinfo{journal}{\apj}
  \textbf{\bibinfo{volume}{476}}, \bibinfo{pages}{221} (\bibinfo{year}{1997}).

\bibitem[{\citenamefont{{Font}}(2008)}]{Font:2008}
\bibinfo{author}{\bibfnamefont{J.~A.} \bibnamefont{{Font}}},
  \bibinfo{journal}{Living Reviews in Relativity}
  \textbf{\bibinfo{volume}{11}}, \bibinfo{eid}{7} (\bibinfo{year}{2008}).

\bibitem[{\citenamefont{{B{\'e}csy} et~al.}(2017)\citenamefont{{B{\'e}csy},
  {Raffai}, {Cornish}, {Essick}, {Kanner}, {Katsavounidis}, {Littenberg},
  {Millhouse}, and {Vitale}}}]{BECSY:2017}
\bibinfo{author}{\bibfnamefont{B.}~\bibnamefont{{B{\'e}csy}}},
  \bibinfo{author}{\bibfnamefont{P.}~\bibnamefont{{Raffai}}},
  \bibinfo{author}{\bibfnamefont{N.~J.} \bibnamefont{{Cornish}}},
  \bibinfo{author}{\bibfnamefont{R.}~\bibnamefont{{Essick}}},
  \bibinfo{author}{\bibfnamefont{J.}~\bibnamefont{{Kanner}}},
  \bibinfo{author}{\bibfnamefont{E.}~\bibnamefont{{Katsavounidis}}},
  \bibinfo{author}{\bibfnamefont{T.~B.} \bibnamefont{{Littenberg}}},
  \bibinfo{author}{\bibfnamefont{M.}~\bibnamefont{{Millhouse}}},
  \bibnamefont{and} \bibinfo{author}{\bibfnamefont{S.}~\bibnamefont{{Vitale}}},
  \bibinfo{journal}{\apj} \textbf{\bibinfo{volume}{839}}, \bibinfo{eid}{15}
  (\bibinfo{year}{2017}), \eprint{1612.02003}.

\bibitem[{\citenamefont{Cooley and Tukey}(1965)}]{Cooley:1965}
\bibinfo{author}{\bibfnamefont{J.~W.} \bibnamefont{Cooley}} \bibnamefont{and}
  \bibinfo{author}{\bibfnamefont{J.~W.} \bibnamefont{Tukey}},
  \bibinfo{journal}{Mathematics of Computation} \textbf{\bibinfo{volume}{19}},
  \bibinfo{pages}{297} (\bibinfo{year}{1965}).

\bibitem[{\citenamefont{{Kastaun}}(2021)}]{pycactus:2021}
\bibinfo{author}{\bibfnamefont{W.}~\bibnamefont{{Kastaun}}},
  \emph{\bibinfo{title}{{PyCactus: Post-processing tools for Cactus
  computational toolkit simulation data}}}, \bibinfo{howpublished}{Astrophysics
  Source Code Library, record ascl:2107.017} (\bibinfo{year}{2021}),
  \eprint{2107.017}.

\bibitem[{\citenamefont{Chatziioannou et~al.}(2017)\citenamefont{Chatziioannou,
  Clark, Bauswein, Millhouse, Littenberg, and Cornish}}]{Chatz:2017}
\bibinfo{author}{\bibfnamefont{K.}~\bibnamefont{Chatziioannou}},
  \bibinfo{author}{\bibfnamefont{J.~A.} \bibnamefont{Clark}},
  \bibinfo{author}{\bibfnamefont{A.}~\bibnamefont{Bauswein}},
  \bibinfo{author}{\bibfnamefont{M.}~\bibnamefont{Millhouse}},
  \bibinfo{author}{\bibfnamefont{T.~B.} \bibnamefont{Littenberg}},
  \bibnamefont{and} \bibinfo{author}{\bibfnamefont{N.}~\bibnamefont{Cornish}},
  \bibinfo{journal}{\prd} \textbf{\bibinfo{volume}{96}},
  \bibinfo{pages}{124035} (\bibinfo{year}{2017}).

\bibitem[{\citenamefont{{Evans} et~al.}(2021)\citenamefont{{Evans}, {Adhikari},
  {Afle}, {Ballmer}, {Biscoveanu}, {Borhanian}, {Brown}, {Chen}, {Eisenstein},
  {Gruson} et~al.}}]{Cosmic_Explorer}
\bibinfo{author}{\bibfnamefont{M.}~\bibnamefont{{Evans}}},
  \bibinfo{author}{\bibfnamefont{R.~X.} \bibnamefont{{Adhikari}}},
  \bibinfo{author}{\bibfnamefont{C.}~\bibnamefont{{Afle}}},
  \bibinfo{author}{\bibfnamefont{S.~W.} \bibnamefont{{Ballmer}}},
  \bibinfo{author}{\bibfnamefont{S.}~\bibnamefont{{Biscoveanu}}},
  \bibinfo{author}{\bibfnamefont{S.}~\bibnamefont{{Borhanian}}},
  \bibinfo{author}{\bibfnamefont{D.~A.} \bibnamefont{{Brown}}},
  \bibinfo{author}{\bibfnamefont{Y.}~\bibnamefont{{Chen}}},
  \bibinfo{author}{\bibfnamefont{R.}~\bibnamefont{{Eisenstein}}},
  \bibinfo{author}{\bibfnamefont{A.}~\bibnamefont{{Gruson}}},
  \bibnamefont{et~al.}, \bibinfo{journal}{arXiv e-prints}
  \bibinfo{eid}{arXiv:2109.09882} (\bibinfo{year}{2021}), \eprint{2109.09882}.

\bibitem[{\citenamefont{{Blackburn} et~al.}(2008)\citenamefont{{Blackburn},
  {Cadonati}, {Caride}, {Caudill}, {Chatterji}, {Christensen}, {Dalrymple},
  {Desai}, {Di Credico}, {Ely} et~al.}}]{Blackburn:2008}
\bibinfo{author}{\bibfnamefont{L.}~\bibnamefont{{Blackburn}}},
  \bibinfo{author}{\bibfnamefont{L.}~\bibnamefont{{Cadonati}}},
  \bibinfo{author}{\bibfnamefont{S.}~\bibnamefont{{Caride}}},
  \bibinfo{author}{\bibfnamefont{S.}~\bibnamefont{{Caudill}}},
  \bibinfo{author}{\bibfnamefont{S.}~\bibnamefont{{Chatterji}}},
  \bibinfo{author}{\bibfnamefont{N.}~\bibnamefont{{Christensen}}},
  \bibinfo{author}{\bibfnamefont{J.}~\bibnamefont{{Dalrymple}}},
  \bibinfo{author}{\bibfnamefont{S.}~\bibnamefont{{Desai}}},
  \bibinfo{author}{\bibfnamefont{A.}~\bibnamefont{{Di Credico}}},
  \bibinfo{author}{\bibfnamefont{G.}~\bibnamefont{{Ely}}},
  \bibnamefont{et~al.}, \bibinfo{journal}{Classical and Quantum Gravity}
  \textbf{\bibinfo{volume}{25}}, \bibinfo{eid}{184004} (\bibinfo{year}{2008}),
  \eprint{0804.0800}.

\bibitem[{\citenamefont{{Abbott} et~al.}(2009)\citenamefont{{Abbott}, {Abbott},
  {Adhikari}, {Ajith}, {Allen}, {Allen}, {Amin}, {Anderson}, {Anderson},
  {Arain} et~al.}}]{Abbott:2009}
\bibinfo{author}{\bibfnamefont{B.~P.} \bibnamefont{{Abbott}}},
  \bibinfo{author}{\bibfnamefont{R.}~\bibnamefont{{Abbott}}},
  \bibinfo{author}{\bibfnamefont{R.}~\bibnamefont{{Adhikari}}},
  \bibinfo{author}{\bibfnamefont{P.}~\bibnamefont{{Ajith}}},
  \bibinfo{author}{\bibfnamefont{B.}~\bibnamefont{{Allen}}},
  \bibinfo{author}{\bibfnamefont{G.}~\bibnamefont{{Allen}}},
  \bibinfo{author}{\bibfnamefont{R.~S.} \bibnamefont{{Amin}}},
  \bibinfo{author}{\bibfnamefont{S.~B.} \bibnamefont{{Anderson}}},
  \bibinfo{author}{\bibfnamefont{W.~G.} \bibnamefont{{Anderson}}},
  \bibinfo{author}{\bibfnamefont{M.~A.} \bibnamefont{{Arain}}},
  \bibnamefont{et~al.}, \bibinfo{journal}{Reports on Progress in Physics}
  \textbf{\bibinfo{volume}{72}}, \bibinfo{eid}{076901} (\bibinfo{year}{2009}),
  \eprint{0711.3041}.

\bibitem[{\citenamefont{{Aasi} et~al.}(2012)\citenamefont{{Aasi}, {Abadie},
  {Abbott}, {Abbott}, {Abbott}, {Abernathy}, {Accadia}, {Acernese}, {Adams},
  {Adams} et~al.}}]{Aasi:2012}
\bibinfo{author}{\bibfnamefont{J.}~\bibnamefont{{Aasi}}},
  \bibinfo{author}{\bibfnamefont{J.}~\bibnamefont{{Abadie}}},
  \bibinfo{author}{\bibfnamefont{B.~P.} \bibnamefont{{Abbott}}},
  \bibinfo{author}{\bibfnamefont{R.}~\bibnamefont{{Abbott}}},
  \bibinfo{author}{\bibfnamefont{T.~D.} \bibnamefont{{Abbott}}},
  \bibinfo{author}{\bibfnamefont{M.}~\bibnamefont{{Abernathy}}},
  \bibinfo{author}{\bibfnamefont{T.}~\bibnamefont{{Accadia}}},
  \bibinfo{author}{\bibfnamefont{F.}~\bibnamefont{{Acernese}}},
  \bibinfo{author}{\bibfnamefont{C.}~\bibnamefont{{Adams}}},
  \bibinfo{author}{\bibfnamefont{T.}~\bibnamefont{{Adams}}},
  \bibnamefont{et~al.}, \bibinfo{journal}{Classical and Quantum Gravity}
  \textbf{\bibinfo{volume}{29}}, \bibinfo{eid}{155002} (\bibinfo{year}{2012}),
  \eprint{1203.5613}.

\bibitem[{\citenamefont{{Bernuzzi} et~al.}(2014)\citenamefont{{Bernuzzi},
  {Nagar}, {Balmelli}, {Dietrich}, and {Ujevic}}}]{Bernuzzi:2014}
\bibinfo{author}{\bibfnamefont{S.}~\bibnamefont{{Bernuzzi}}},
  \bibinfo{author}{\bibfnamefont{A.}~\bibnamefont{{Nagar}}},
  \bibinfo{author}{\bibfnamefont{S.}~\bibnamefont{{Balmelli}}},
  \bibinfo{author}{\bibfnamefont{T.}~\bibnamefont{{Dietrich}}},
  \bibnamefont{and} \bibinfo{author}{\bibfnamefont{M.}~\bibnamefont{{Ujevic}}},
  \bibinfo{journal}{\prl} \textbf{\bibinfo{volume}{112}}, \bibinfo{eid}{201101}
  (\bibinfo{year}{2014}), \eprint{1402.6244}.

\bibitem[{\citenamefont{{Bauswein} et~al.}(2019)\citenamefont{{Bauswein},
  {Bastian}, {Blaschke}, {Chatziioannou}, {Clark}, {Fischer}, and
  {Oertel}}}]{Bauswein:2019b}
\bibinfo{author}{\bibfnamefont{A.}~\bibnamefont{{Bauswein}}},
  \bibinfo{author}{\bibfnamefont{N.-U.~F.} \bibnamefont{{Bastian}}},
  \bibinfo{author}{\bibfnamefont{D.~B.} \bibnamefont{{Blaschke}}},
  \bibinfo{author}{\bibfnamefont{K.}~\bibnamefont{{Chatziioannou}}},
  \bibinfo{author}{\bibfnamefont{J.~A.} \bibnamefont{{Clark}}},
  \bibinfo{author}{\bibfnamefont{T.}~\bibnamefont{{Fischer}}},
  \bibnamefont{and} \bibinfo{author}{\bibfnamefont{M.}~\bibnamefont{{Oertel}}},
  \bibinfo{journal}{\prl} \textbf{\bibinfo{volume}{122}}, \bibinfo{eid}{061102}
  (\bibinfo{year}{2019}), \eprint{1809.01116}.

\bibitem[{\citenamefont{{Soultanis} et~al.}(2022)\citenamefont{{Soultanis},
  {Bauswein}, and {Stergioulas}}}]{Soultanis:2022}
\bibinfo{author}{\bibfnamefont{T.}~\bibnamefont{{Soultanis}}},
  \bibinfo{author}{\bibfnamefont{A.}~\bibnamefont{{Bauswein}}},
  \bibnamefont{and}
  \bibinfo{author}{\bibfnamefont{N.}~\bibnamefont{{Stergioulas}}},
  \bibinfo{journal}{\prd} \textbf{\bibinfo{volume}{105}}, \bibinfo{eid}{043020}
  (\bibinfo{year}{2022}), \eprint{2111.08353}.

\bibitem[{\citenamefont{Ruiz et~al.}(2020)\citenamefont{Ruiz, Tsokaros, and
  Shapiro}}]{Ruiz:2020via}
\bibinfo{author}{\bibfnamefont{M.}~\bibnamefont{Ruiz}},
  \bibinfo{author}{\bibfnamefont{A.}~\bibnamefont{Tsokaros}}, \bibnamefont{and}
  \bibinfo{author}{\bibfnamefont{S.~L.} \bibnamefont{Shapiro}},
  \bibinfo{journal}{Phys. Rev. D} \textbf{\bibinfo{volume}{101}},
  \bibinfo{pages}{064042} (\bibinfo{year}{2020}), \eprint{2001.09153}.

\bibitem[{\citenamefont{Etienne et~al.}(2012)\citenamefont{Etienne, Liu,
  Paschalidis, and Shapiro}}]{Etienne:2011ea}
\bibinfo{author}{\bibfnamefont{Z.~B.} \bibnamefont{Etienne}},
  \bibinfo{author}{\bibfnamefont{Y.~T.} \bibnamefont{Liu}},
  \bibinfo{author}{\bibfnamefont{V.}~\bibnamefont{Paschalidis}},
  \bibnamefont{and} \bibinfo{author}{\bibfnamefont{S.~L.}
  \bibnamefont{Shapiro}}, \bibinfo{journal}{Phys. Rev. D}
  \textbf{\bibinfo{volume}{85}}, \bibinfo{pages}{064029}
  (\bibinfo{year}{2012}), \eprint{1112.0568}.

\bibitem[{\citenamefont{Raithel et~al.}(2019)\citenamefont{Raithel, Ozel, and
  Psaltis}}]{Raithel:2019gws}
\bibinfo{author}{\bibfnamefont{C.~A.} \bibnamefont{Raithel}},
  \bibinfo{author}{\bibfnamefont{F.}~\bibnamefont{Ozel}}, \bibnamefont{and}
  \bibinfo{author}{\bibfnamefont{D.}~\bibnamefont{Psaltis}},
  \bibinfo{journal}{Astrophys. J.} \textbf{\bibinfo{volume}{875}},
  \bibinfo{pages}{12} (\bibinfo{year}{2019}), \eprint{1902.10735}.

\bibitem[{\citenamefont{Mroczek et~al.}(2024)\citenamefont{Mroczek, Yao, Zine,
  Noronha-Hostler, Dexheimer, Haber, and Most}}]{Mroczek:2024sfp}
\bibinfo{author}{\bibfnamefont{D.}~\bibnamefont{Mroczek}},
  \bibinfo{author}{\bibfnamefont{N.}~\bibnamefont{Yao}},
  \bibinfo{author}{\bibfnamefont{K.}~\bibnamefont{Zine}},
  \bibinfo{author}{\bibfnamefont{J.}~\bibnamefont{Noronha-Hostler}},
  \bibinfo{author}{\bibfnamefont{V.}~\bibnamefont{Dexheimer}},
  \bibinfo{author}{\bibfnamefont{A.}~\bibnamefont{Haber}}, \bibnamefont{and}
  \bibinfo{author}{\bibfnamefont{E.~R.} \bibnamefont{Most}}
  (\bibinfo{year}{2024}), \eprint{2404.01658}.

\bibitem[{\citenamefont{Harris et~al.}(2020)\citenamefont{Harris, Millman,
  van~der Walt, Gommers, Virtanen, Cournapeau, Wieser, Taylor, Berg, Smith
  et~al.}}]{harris:2020}
\bibinfo{author}{\bibfnamefont{C.~R.} \bibnamefont{Harris}},
  \bibinfo{author}{\bibfnamefont{K.~J.} \bibnamefont{Millman}},
  \bibinfo{author}{\bibfnamefont{S.~J.} \bibnamefont{van~der Walt}},
  \bibinfo{author}{\bibfnamefont{R.}~\bibnamefont{Gommers}},
  \bibinfo{author}{\bibfnamefont{P.}~\bibnamefont{Virtanen}},
  \bibinfo{author}{\bibfnamefont{D.}~\bibnamefont{Cournapeau}},
  \bibinfo{author}{\bibfnamefont{E.}~\bibnamefont{Wieser}},
  \bibinfo{author}{\bibfnamefont{J.}~\bibnamefont{Taylor}},
  \bibinfo{author}{\bibfnamefont{S.}~\bibnamefont{Berg}},
  \bibinfo{author}{\bibfnamefont{N.~J.} \bibnamefont{Smith}},
  \bibnamefont{et~al.}, \bibinfo{journal}{Nature}
  \textbf{\bibinfo{volume}{585}}, \bibinfo{pages}{357} (\bibinfo{year}{2020}).

\bibitem[{\citenamefont{Virtanen et~al.}(2020)\citenamefont{Virtanen, Gommers,
  Oliphant, Haberland, Reddy, Cournapeau, Burovski, Peterson, Weckesser, Bright
  et~al.}}]{scipy:2020}
\bibinfo{author}{\bibfnamefont{P.}~\bibnamefont{Virtanen}},
  \bibinfo{author}{\bibfnamefont{R.}~\bibnamefont{Gommers}},
  \bibinfo{author}{\bibfnamefont{T.~E.} \bibnamefont{Oliphant}},
  \bibinfo{author}{\bibfnamefont{M.}~\bibnamefont{Haberland}},
  \bibinfo{author}{\bibfnamefont{T.}~\bibnamefont{Reddy}},
  \bibinfo{author}{\bibfnamefont{D.}~\bibnamefont{Cournapeau}},
  \bibinfo{author}{\bibfnamefont{E.}~\bibnamefont{Burovski}},
  \bibinfo{author}{\bibfnamefont{P.}~\bibnamefont{Peterson}},
  \bibinfo{author}{\bibfnamefont{W.}~\bibnamefont{Weckesser}},
  \bibinfo{author}{\bibfnamefont{J.}~\bibnamefont{Bright}},
  \bibnamefont{et~al.}, \bibinfo{journal}{Nature Methods}
  \textbf{\bibinfo{volume}{17}}, \bibinfo{pages}{261} (\bibinfo{year}{2020}).

\bibitem[{\citenamefont{Pedregosa et~al.}(2011)\citenamefont{Pedregosa,
  Varoquaux, Gramfort, Michel, Thirion, Grisel, Blondel, Prettenhofer, Weiss,
  Dubourg et~al.}}]{scikit-learn}
\bibinfo{author}{\bibfnamefont{F.}~\bibnamefont{Pedregosa}},
  \bibinfo{author}{\bibfnamefont{G.}~\bibnamefont{Varoquaux}},
  \bibinfo{author}{\bibfnamefont{A.}~\bibnamefont{Gramfort}},
  \bibinfo{author}{\bibfnamefont{V.}~\bibnamefont{Michel}},
  \bibinfo{author}{\bibfnamefont{B.}~\bibnamefont{Thirion}},
  \bibinfo{author}{\bibfnamefont{O.}~\bibnamefont{Grisel}},
  \bibinfo{author}{\bibfnamefont{M.}~\bibnamefont{Blondel}},
  \bibinfo{author}{\bibfnamefont{P.}~\bibnamefont{Prettenhofer}},
  \bibinfo{author}{\bibfnamefont{R.}~\bibnamefont{Weiss}},
  \bibinfo{author}{\bibfnamefont{V.}~\bibnamefont{Dubourg}},
  \bibnamefont{et~al.}, \bibinfo{journal}{Journal of Machine Learning Research}
  \textbf{\bibinfo{volume}{12}}, \bibinfo{pages}{2825} (\bibinfo{year}{2011}).

\bibitem[{\citenamefont{Hunter}(2007)}]{Hunter:2007}
\bibinfo{author}{\bibfnamefont{J.~D.} \bibnamefont{Hunter}},
  \bibinfo{journal}{Computing in Science \& Engineering}
  \textbf{\bibinfo{volume}{9}}, \bibinfo{pages}{90} (\bibinfo{year}{2007}).

\bibitem[{\citenamefont{Nitz et~al.}(2024)\citenamefont{Nitz, Harry, Brown,
  Biwer, Willis, Canton, Capano, Dent, Pekowsky, Davies et~al.}}]{pycbc}
\bibinfo{author}{\bibfnamefont{A.}~\bibnamefont{Nitz}},
  \bibinfo{author}{\bibfnamefont{I.}~\bibnamefont{Harry}},
  \bibinfo{author}{\bibfnamefont{D.}~\bibnamefont{Brown}},
  \bibinfo{author}{\bibfnamefont{C.~M.} \bibnamefont{Biwer}},
  \bibinfo{author}{\bibfnamefont{J.}~\bibnamefont{Willis}},
  \bibinfo{author}{\bibfnamefont{T.~D.} \bibnamefont{Canton}},
  \bibinfo{author}{\bibfnamefont{C.}~\bibnamefont{Capano}},
  \bibinfo{author}{\bibfnamefont{T.}~\bibnamefont{Dent}},
  \bibinfo{author}{\bibfnamefont{L.}~\bibnamefont{Pekowsky}},
  \bibinfo{author}{\bibfnamefont{G.~S.~C.} \bibnamefont{Davies}},
  \bibnamefont{et~al.}, \emph{\bibinfo{title}{gwastro/pycbc: v2.3.3 release of
  pycbc}} (\bibinfo{year}{2024}).

\end{thebibliography}

\end{document}